\documentclass[reprint,onecolumn, amsmath,amssymb,aps,floatfix, superscriptaddress]{revtex4-2}
\usepackage{upgreek}             
\usepackage{wrapfig}
\usepackage{fancyhdr}
\usepackage{graphicx}
\usepackage[export]{adjustbox} 
\usepackage{enumitem}
\usepackage{xcolor}
\usepackage{multirow}
\usepackage{colortbl}
\usepackage{boldline}
\usepackage{soul}
\usepackage{hyperref}
\newcommand{\micron}{$\upmu$m}
\def\degree{$^{\circ}$}
\def\Micro-X{\mbox{Micro-X}}

\newcommand{\angstrom}{\mbox{\normalfont\AA}}

\begin{document}

%\tableofcontents
%\newpage
\title{First Flight Performance of the Micro-X Microcalorimeter X-Ray Sounding Rocket}
\newcommand{\gsfc}{\affiliation{NASA Goddard Space Flight Center, Greenbelt, MD, USA}}
\newcommand{\nwu}{\affiliation{Department of Physics and Astronomy, Northwestern University, Evanston, IL, USA}}
\newcommand{\kavli}{\affiliation{MIT Kavli Institute, Massachusetts Institute of Technology, Cambridge, MA, USA}}
\newcommand{\nist}{\affiliation{National Institute of Standards and Technology, Boulder, CO, USA}}
\newcommand{\llnl}{\affiliation{Lawrence Livermore National Laboratory, Livermore, CA, USA}}

\author{Joseph~S.~Adams}
\gsfc
\author{Robert~Baker}
\gsfc
\author{Simon~R.~Bandler}
\gsfc
\author{No\"emie~Bastidon}
\nwu
\affiliation{Current address: The Australian National University, Canberra, Australia}
\author{Daniel~Castro}
\affiliation{Center for Astrophysics $\vert$ Harvard \& Smithsonian, Cambridge, MA, USA}
\author{Meredith~E.~Danowksi}
\nwu
\affiliation{Current address: Ball Aerospace, Boulder, CO, USA}
\author{William~B.~Doriese}
\nist
\author{Megan~E.~Eckart}
\llnl
\author{Enectal\'i~Figueroa-Feliciano}
\nwu
\author{Joshua~Fuhrman}
\nwu
\author{David~C.~Goldfinger}
\nwu
\affiliation{Current address: Center for Astrophysics $\vert$ Harvard \& Smithsonian, Cambridge, MA, USA}
\author{Sarah~N.T.~Heine}
\kavli
\author{Gene~Hilton}
\nist
\author{Antonia~J.F.~Hubbard}
\email[Corresponding author: ]{hubbard8@llnl.gov}
\llnl
\author{Daniel~Jardin}
\nwu
\author{Richard~L.~Kelley}
\gsfc
\author{Caroline~A.~Kilbourne}
\gsfc
\author{Steven~W.~Leman}
\kavli
\author{Ren\'ee~E.~Manzagol-Harwood}
\nwu
\author{Dan~McCammon}
\affiliation{University of Wisconsin, Madison, WI, USA}
\author{Philip~H.H.~Oakley}
\affiliation{Current address: Ball Aerospace, Boulder, CO, USA}
\kavli
\author{Takashi~Okajima}
\gsfc
\author{Frederick~Scott~Porter}
\gsfc
\author{Carl~D.~Reintsema}
\nist
\author{John~Rutherford}
\kavli
\author{Tarek~Saab}
\affiliation{Department of Physics, University of Florida, Gainesville, FL, USA}
\author{Kosuke~Sato}
\kavli
\affiliation{Current address: Department of Physics, Saitama University, Saitama, Japan}
\author{Peter~Serlemitsos}
\gsfc
\author{Stephen~J.~Smith}
\gsfc
\author{Yang~Soong}
\gsfc
\author{Patrick~Wikus}
%\altaffiliation[Current address: ]{Bruker BioSpin AG, Fllanden, Switzerland}
\kavli
\affiliation{Current address: Bruker BioSpin AG, Fallanden, Switzerland}

\collaboration{The Micro-X collaboration}
\date{\today}

\begin{abstract}
The flight of the Micro-X sounding rocket on July 22, 2018 marked the first operation of Transition-Edge Sensors and their SQUID readouts in space. The instrument combines the microcalorimeter array with an imaging mirror to take high-resolution spectra from extended X-ray sources. The first flight target was the Cassiopeia~A Supernova Remnant. While a rocket pointing malfunction led to no time on-target, data from the flight was used to evaluate the performance of the instrument and demonstrate the flight viability of the payload. The instrument successfully achieved a stable cryogenic environment, executed all flight operations, and observed X-rays from the on-board calibration source. The flight environment did not significantly affect the performance of the detectors compared to ground operation. The flight provided an invaluable test of the impact of external magnetic fields and the instrument configuration on detector performance. This flight provides a milestone in the flight readiness of these detector and readout technologies, both of which have been selected for future X-ray observatories. 
\end{abstract}

\maketitle
%%%%%%%%%%%%%%%%%%%%%%%%%%%%%%%%%
% !TEX root = ../instrument_paper_revB.tex
\section{Introduction}
\label{sec:Intro}

Decades of development of spacecraft instrumentation and microcalorimeter detector technology are converging to provide a new class of instruments with unprecedented sensitivity to the X-ray sky. High-spectral-resolution X-ray spectroscopy has, with few exceptions, traditionally been confined to point sources or bright, compact features within extended sources. By contrast, microcalorimeters can provide high-spectral-resolution observations of extended sources. The two missions that have flown microcalorimeters, the X-Ray Quantum Calorimeter (XQC) sounding rocket~\cite{mccammon2002} and the Soft X-ray Spectrometer (SXS)~\cite{HitomiSXS} aboard Hitomi~\cite{Hitomi_2018}, have demonstrated the value of flying such high-resolution detectors. Both of these instruments used silicon thermistors, as will the Resolve instrument~\cite{Resolve} aboard the X-Ray Imaging and Spectroscopy Mission (XRISM) satellite~\cite{XRISM}. 

The development of the next generation of microcalorimeters, Transition-Edge Sensors (TESs), has been a major research effort over the past two decades~\cite{Ullom_2015}.With its first flight on July 22, 2018, Micro-X became the first mission to fly TES microcalorimeters in space. This was also the first space operation of a Time-Division Multiplexing Superconducting QUantum Interference Device (TDM SQUID) readout. This new flight heritage marks a milestone in the maturity of these technologies, which have been selected or proposed for the next generation of X-ray observatories. Notably, both TES microcalorimeters and TDM SQUID readouts are baselined for the X-ray Integral Field Unit (X-IFU)~\cite{XIFU_2018,XIFU_2018b} on the Athena X-ray observatory~\cite{Athena_2013,Athena_2014}, and TES microcalorimeters are baselined for the Lynx X-ray Microcalorimeter (LXM) instrument~\citep{LXM} on the Lynx observatory~\cite{Lynx_2019,Lynx_2018}. Demonstrating these technologies on a sounding rocket allows them to be flight-tested on a lower-cost, recoverable instrument before satellite operation. 

The Micro-X instrument is designed to take high-resolution spectra of extended X-ray sources. It combines an imaging mirror with a microcalorimeter array to provide both spectral and spatial data for X-rays in the 0.3-2.5~keV range. The goals of the program are both scientific and technical. Scientifically, the mission is designed to observe Supernova Remnants (SNRs) with high energy resolution and medium angular resolution. The instrument can also be modified for galactic dark matter searches. Technically, the mission is designed to advance the flight readiness of TES microcalorimeters and their TDM SQUID readouts for future missions. 

This paper provides a description of the instrument and its first flight performance. Section~\ref{sec: Mission Overview} provides an overview of the mission, including the science goals and the fundamentals of microcalorimeters and sounding rockets. Section~\ref{sec:Instrument} describes the instrument. Section~\ref{sec:Flight Performance} analyzes the instrument performance in the first flight, and Section~\ref{sec:Reflight} summarizes the modifications implemented for the second flight on August~21, 2022. Data and results from the second flight will be the subject of a future publication.
%%%%%%%%%%%%%%%%%%%%%%%%%%%%%%%%%
% !TEX root = ../instrument_paper_revB.tex
\section{Mission Overview}
\label{sec: Mission Overview}

\subsection{Astrophysics Program}

The Micro-X science program focuses on observations of SNRs and indirect detection of galactic dark matter. The primary target for SNR observations is the Bright Eastern Knot (BEK) of the Puppis~A SNR, which exploded in a core-collapse supernova approximately 4000~years ago~\citep{Becker:2012, Zavlin:1999}. The BEK is the interaction site of the SNR with a dense ambient cloud~\cite{Hwang:2005}. Observations of the BEK by Chandra and XMM-Newton reveal the X-ray emission from this region to be dominated by a thermal plasma rich in O, Ne, Mg, Si, and Fe. All of these elements have strong emission lines that can be well resolved spectrally by Micro-X, and the main morphological features are well-matched to the Point Spread Function (PSF) of the Micro-X optics \cite{Hwang:2005}. The primary science goals of the Micro-X BEK observation are to understand the importance of charge exchange (CX) reactions and to study particle acceleration at the shock front, as described in \cite{Katsuda:2012}. 

For launches occurring when Puppis~A is not visible from the launch site, the secondary SNR target is Casseiopeia~A. The goals of that observation are to: 1)~measure line fluxes and map emission lines to estimate elemental abundances, 2)~compare plasma diagnostics between elements, 3)~determine system dynamics from Doppler shifts, 4)~constrain nucleosynthesis models. Detailed projections for this observation are described in~\cite{Rutherford:2013}. 

The dark matter observation would use a modified version of the instrument, built up in a large field-of-view (FOV) configuration to search for a diffuse galactic dark matter signal, as described in \cite{microxDM:2015, HubbardLTD}. Dark matter detection in the X-ray band is particularly compelling because several X-ray satellites report an excess at 3.5~keV that may be a product of a dark matter interaction~\cite{keVwhitepaper:2017, Kev_DM, Boyarsky:2019, Bulbul:2014, Boyarsky2018}. The primary dark matter target is a quiet X-ray region just off the galactic plane, and the secondary target is the Galactic Center. The background spectrum at the primary target is a nearly flat continuum with 0.6~counts/flight/2.5~eV~\cite{microxDM:2015}. Using the flux derived from~\cite{BoyarskyA:2014}, 20.3$\pm$4.5 signal events are expected in a single flight, corresponding to $>$5$\sigma$ significance~\cite{HubbardLTD}. Micro-X will be sensitive to both the reported 3.5~keV excess and to any X-ray signature of dark matter in this energy band. If a line is detected, Micro-X can discern between dark matter and emission of atomic origin by mapping the Doppler shift of the line across the Galaxy with multiple flights~\cite{Doppler1, Doppler2}. The Micro-X observations are statistics-limited, so additional flights will further increase the sensitivity of the results. 

%%%%%%%%%%%%%%%%%%%%%%%%%%%%%%%%%%%
\subsection{TES Microcalorimeters}
\label{sec:TES_intro}

\begin{figure}[htb]
	\centering
	\includegraphics[width=\textwidth]{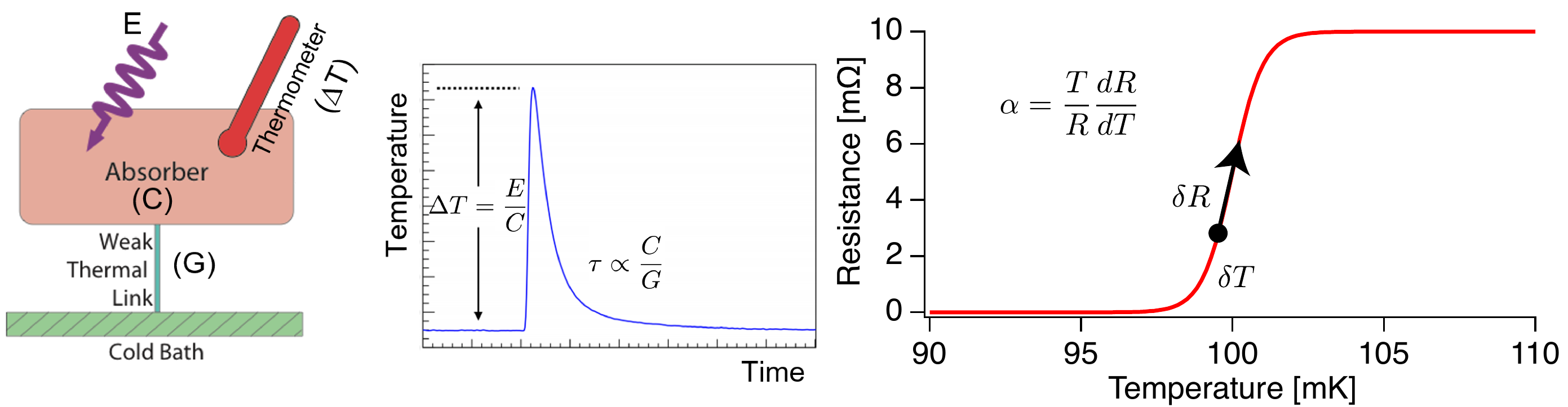}
	\caption{When an X-ray is incident on a TES microcalorimeter (left), the temperature of the absorber rises, then returns to its base temperature as heat dissipates through the weak thermal link. The size and shape of the pulse (center) observed by the TES thermometer depend on the energy of the X-ray (E), the heat capacity of the absorber (C) and the thermal conductivity of the thermal link (G). The resistance of the TES sharply changes with the temperature in the absorber (right). The size of the response depends on where the TES is biased in its superconducting transition and the slope of the transition.} %($\alpha$) took alpha out since it is not strictly the slope of the transition
	\label{fig:TES}
	\vspace{-5pt}
\end{figure}

The X-ray detectors used for Micro-X are cryogenic microcalorimeters that have been engineered to provide excellent energy resolution in the keV regime. They consist of an X-ray absorber, a thermometer, and a weak thermal link to connect the absorber with a heat sink, as shown in Figure~\ref{fig:TES} (left). When an X-ray of energy $E$ is incident on the pixel, the temperature of the absorber rises by $\Delta T \approx \frac{E}{C}$, where  $C$ is its heat capacity. As the deposited heat dissipates into the heat sink, the absorber cools back down with the time constant %proportional to the natural time constant,
$\tau \propto \tau_o = \frac{C}{G}$, where $G$ is the thermal conductivity of the weak thermal link. This creates the characteristic ``pulse" observed by the thermometer, shown in Figure~\ref{fig:TES} (center). Additional information on microcalorimeters can be found in~\cite{McCammon2005}. 

Micro-X uses TESs as its microcalorimeter thermometers. A TES is a superconducting film operating at a stable, cryogenic temperature that places it between its superconducting and normal resistance ranges. Once biased into this transition regime, shown in Figure~\ref{fig:TES} (right), the small temperature change from an absorbed X-ray produces a large change in the TES resistance, and the energy of the X-ray can thus be reconstructed to high precision. The increased temperature also produces Joule power dissipation in the TES, providing a negative electrothermal feedback mechanism than maintains stable operation and appropriate biasing within the narrow superconducting transition \cite{Irwin:2005}. The slope of the TES transition is given by the dimensionless parameter $\alpha = \frac{T}{R} \frac{dR}{dT}$. TES microcalorimeters have achieved energy resolutions down to 1.6~$\pm$~0.1~eV  for 5.89~keV X-rays in a stable laboratory environment~\citep{Smith:2012, Minuissi:2018}. The Micro-X readout scheme is described in \S\ref{sec:FEA}. 
%TESs are conventionally read out with SQUIDs, which serve as amplifiers. Micro-X uses a TDM SQUID readout scheme that is specifically described in \S\ref{sec:FEA}.  

%%%%%%%%%%%%%%%%%%%%%%%%%%%%%%
\subsection{Sounding Rockets}

Sounding rockets missions use suborbital flights to provide approximately five minutes of observation above the atmosphere~\citep{SRHB}. X-rays in the Micro-X signal bandpass are attenuated by the atmosphere, so they can only be observed above 160~km altitude. Sounding rocket payloads can be recovered after flight with the ability to re-fly, making them a particularly good testbed for new technologies, as with Micro-X. Specialized engineering is required to successfully operate sensitive detectors, like microcalorimeters, in the physically extreme and time-constrained conditions of a sounding rocket flight. The instrument design must account for intense vibrational loads, the zero-gravity environment of space, significant changes in environmental pressure and temperature, and the mechanical shock of impact, among other considerations. 
%\com{PW: You could point out that the flight is not only vibrationally violent - there are also large "steady" g-loads, zero-g (which requires special engineering like the porous plug), thermal challenges, etc.}

\begin{figure}[htb]
	\begin{center}
		\includegraphics[width=\textwidth]{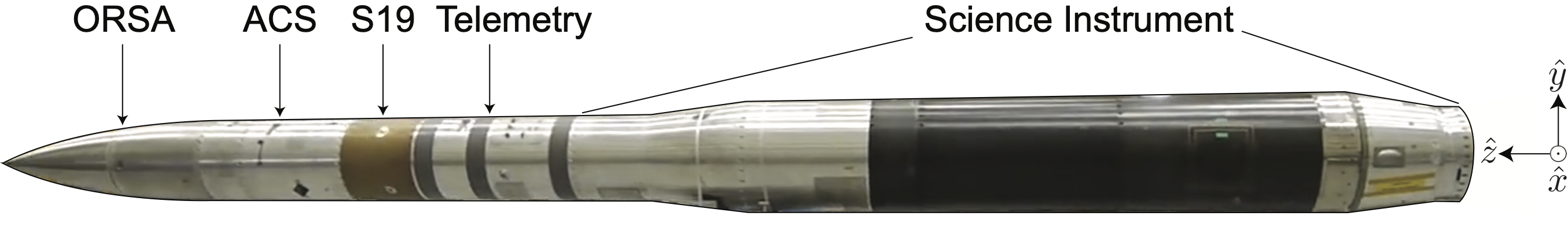}
	\end{center}
	\caption{The Micro-X payload includes the science instrument and standard rocket systems for flight operations, including data transmission (telemetry), trajectory control (S-19), pointing (ACS), and a parachute for the descent (ORSA). The rocket motors are not shown.}
	\label{fig:Payload}
\end{figure}

%The first Micro-X flight launched from the White Sands Missile Range (WSMR) in New Mexico on July 22, 2018.
The payload, shown in Figure~\ref{fig:Payload}, includes the science instrument (\S\ref{sec:Instrument}) and several standard payload systems that are supplied by the NASA Sounding Rocket Operations Contract (NSROC)~\cite{SRHB}. These systems include an Ogive Recovery System Assembly (ORSA) to control the descent, a celestial attitude control system (ACS) for pointing, a S-19 boost guidance system for trajectory control, and three telemetry links for data transmission. The telemetry antennae include one 5~W and two 10~W (minimum) transmitters. The 5~W transmitter has a carrier frequency of 2279.5~MHz, with the 10~W transmitters at 2235.5~MHz and 2382.5~MHz. The payload flies on a two-stage Black Brant IX Mk4, with a Terrier Mk70 first stage booster. 

%%%%%%%%%%%%%%%%%%%%%%%%%%%%%%%
% !TEX root = ../instrument_paper_revB.tex
\section{The Micro-X Microcalorimeter Sounding Rocket Instrument}
\label{sec:Instrument}

\begin{figure}[htb]
	\begin{center}
		\includegraphics[width=\textwidth]{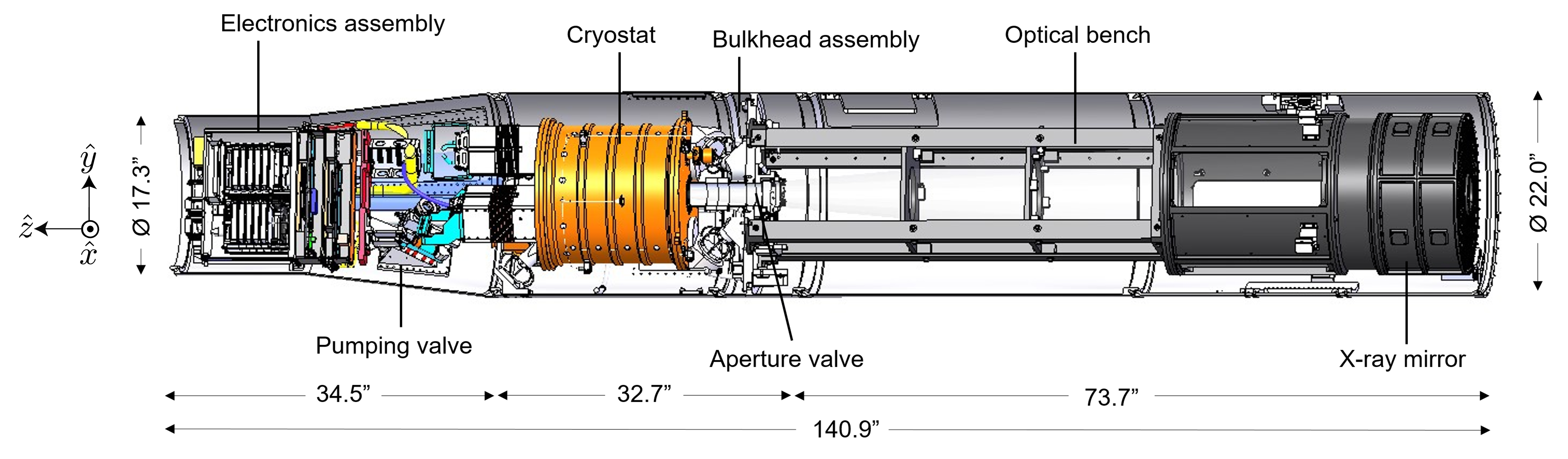}
	\end{center}
	\caption{X-rays enter the Micro-X science instrument from the aft side (right in figure). They are focused by the X-ray optics to pass through the aperture valve and onto the detector array. The array sits inside the cryostat, with temperature control maintained by a LHe tank (pumped through the pumping valve) and an ADR. On-board electronics read out data to telemetry and control the detector, ADR, and valve systems. } 
	\label{InstrumentImage}
\end{figure}

X-rays from the astronomical target enter the Micro-X science instrument, shown in Figure~\ref{InstrumentImage}, through the optics section (\S\ref{sec:Optics}), where they are focused by the mirror onto the detector array (\S\ref{sec:Detectors}). The array sits in the vacuum space of the cryostat. Its temperature is controlled by an Adiabatic Demagnetization Refrigerator (ADR) (\S\ref{sec:Cryostat}), backed by a pumped liquid helium (LHe) system. An X-ray calibration source illuminates the detector array throughout the entire flight to characterize the detector response. Key instrument parameters are shown in Table \ref{table:DetectorSpecs}. %A set of filters, used to block optical and infrared light, sit between the detectors and the gate valve that opens to the optics section (the ``aperture valve"). 

%\begin{wraptable}{r}{0 pt}
\begin{table}[htb]
\hspace{-0.1 in}
\vspace{-10pt}
    \begin{tabular}{ll}
    \hline
		\textbf{Parameter} & \textbf{Specification}\\
		\hline
		Array size & 7.2$\times$7.2~mm$^2$\\
		Absorber size & 590$\times$590~$\upmu$m$^2$\\
		TES size & 140$\times$140~$\upmu$m$^2$\\
		Number of pixels & 128\\
		Fill fraction & 97\%\\
		Instrument effective area (1~keV) & $>$ 300~cm$^2$\\
		Target energy waveband & 0.3 - 2.5~keV \\
		Detector quantum efficiency & $\sim$100\%\\
		Spectral resolution & Goal: 5~eV\\
		TES transition temperature & 120~mK\\
	   	ADR operating temperature & 75~mK $\pm$ 10~$\upmu$K\\
		Field of view & 11.8'\\
		Field of view / pixel & 0.98' \\
		Angular resolution (HPD) & 2.4 arcminutes\\
	   	\hline
    \end{tabular}
	\caption{The Micro-X instrument parameters are optimized for high-spectral-resolution observations of extended X-ray sources. Flight performance is discussed in \S\ref{sec:Flight Performance}. The detector quantum efficiency and spectral resolution parameters listed correspond to the target energy waveband. Fig.\ref{fig:optics_eff_area} shows the total effective area.}
	\label{table:DetectorSpecs}
%\end{wraptable}
\end{table}

The instrument electronics include two science readout chains, ADR control, and valve controls. There is no uplink; all flight commands are pre-programmed on timers (\S\ref{sec:Operations}). The science chains (\S\ref{sec:Science Chain Electronics}) read out the detector data and control the SQUIDs. The ADR controller (\S\ref{sec:ADR Electronics}) maintains stable thermal conditions at the detector array. The valve controllers actuate the gate valves that open the detectors to the optics (``aperture valve") and the LHe tank to the vacuum of space (``LHe pumping valve") (\S\ref{sec:ADR Electronics}). The instrument electronics interface with the telemetry system, which supplies power to the instrument, controls the event timers, and encodes and transmits flight data. A WFF93 encoder (5.0~Mbit/s) reads in data to monitor the status and health of the instrument, and two MV encoders (20~Mbit/s) read in detector data, one for each science chain. The transmitted data is supplied to the ground support electronics, which control the instrument before flight and record telemetered data during flight. 

Once the payload passes 89~km in altitude, a hermetic door at the aft end of the payload (the ``shutter door") opens to expose the optics to space. At this point, the celestial ACS housed in the central bore of the X-ray mirror uses the stars in its field of view to point the payload at the science target. At an altitude of 160~km, the aperture valve opens, and X-rays from the science target can reach the detector array. This is the start of the science observation. The observation continues until the aperture valve and the shutter door close before the payload re-enters the atmosphere and descends on a parachute. %\com{PW: To the "uninitiated", it could be a bit unclear what the shutter door is. You could show the shutter door in Figure 3, or explain it in more detail in the text.}

%%% DETECTORS + SQUIDS %%%
% !TEX root = ../instrument_paper_revC.tex
\subsection{Detectors and Science Readout}
\subsubsection{Detector Array}
\label{sec:Detectors}

\begin{figure}[htb]
\begin{center}
    \includegraphics[height=1.8in]{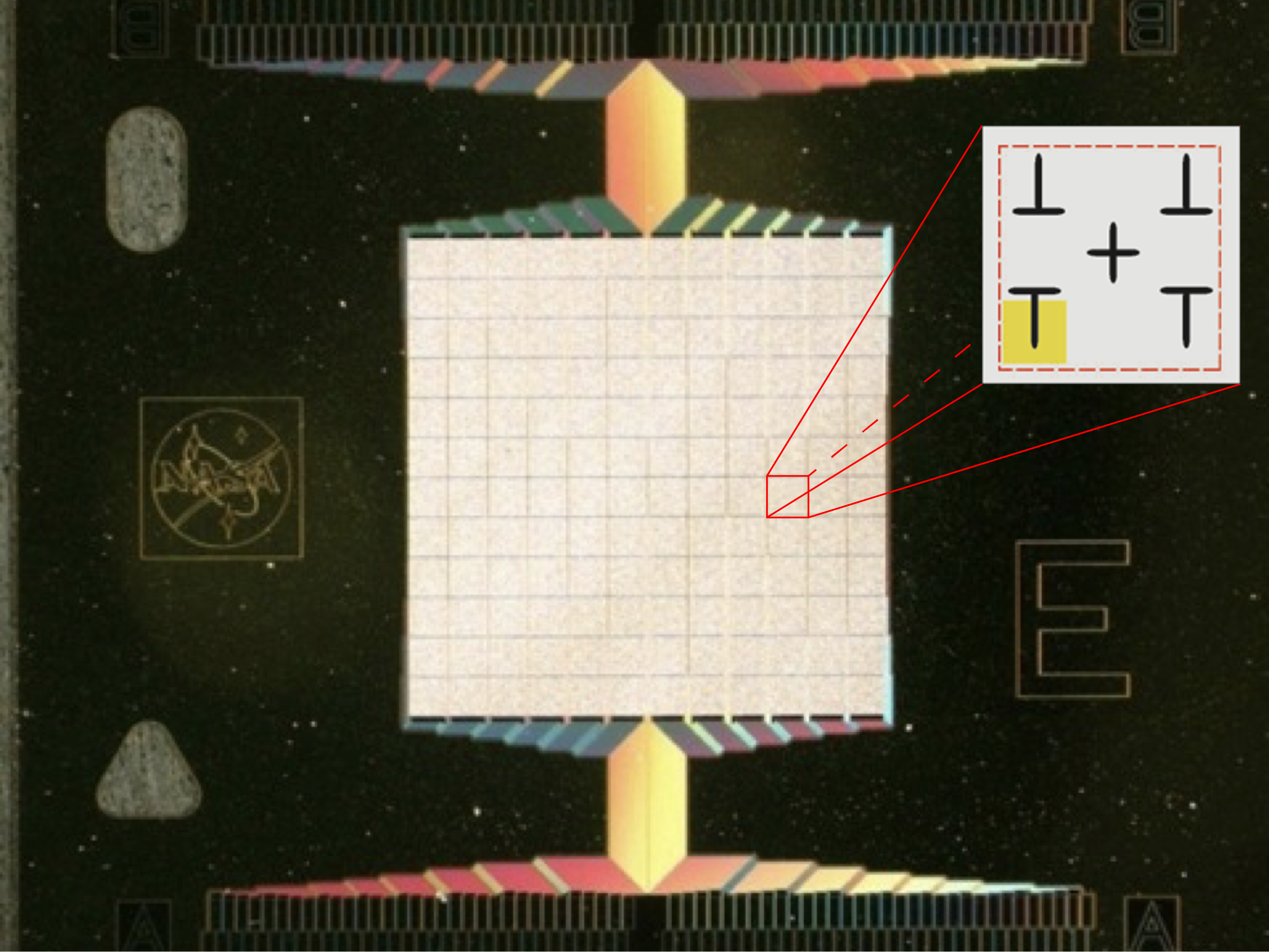}
    \includegraphics[height=1.8in]{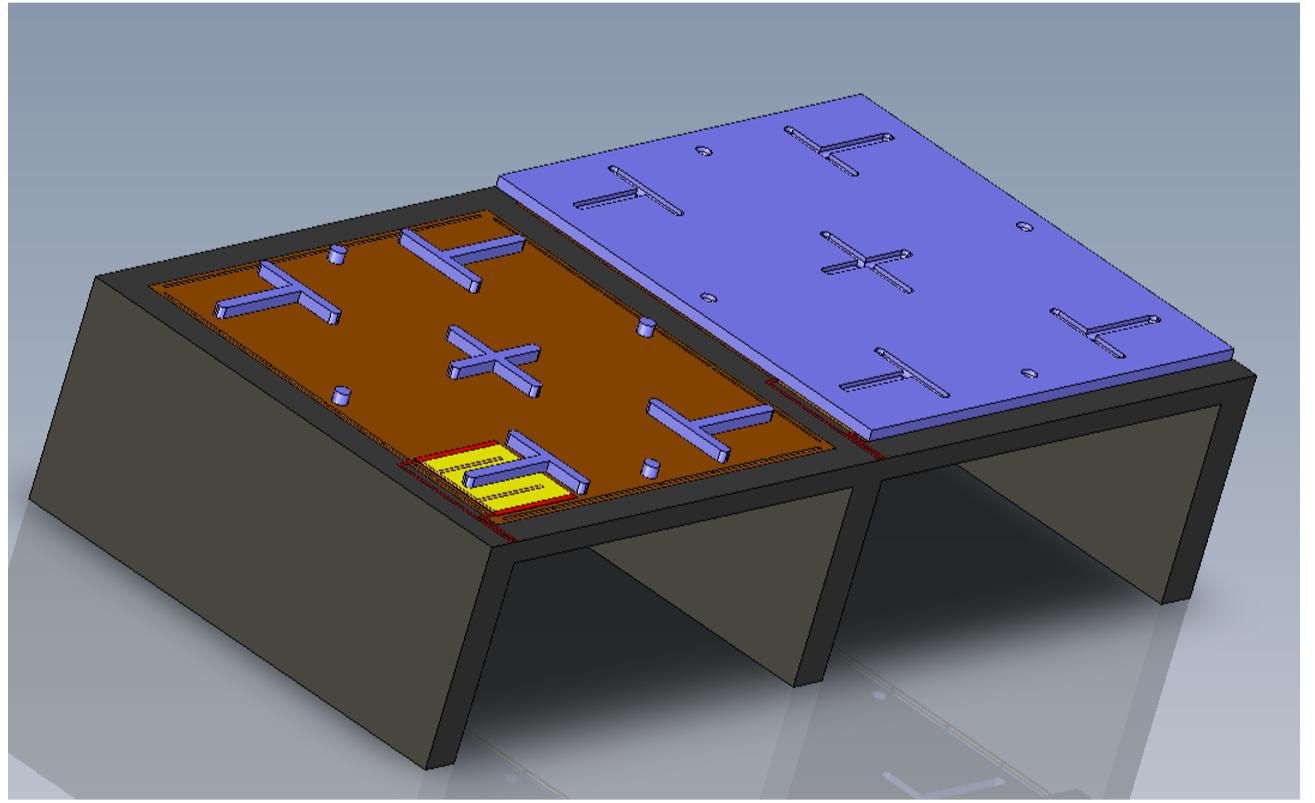}
\end{center}
    \caption{The Micro-X array (left) includes 128~active pixels in a $12 \times 12$ array, with 4 unwired pixels in each corner. The inset shows a schematic of one pixel without the absorber. The SiN membrane is shown in grey and the TES in yellow. The black crosses are the stems, and the red dashes are perforations in the membrane. A drawing of two pixels (right) shows one with the absorber (purple) removed to expose the underlying SiN membrane (brown), stems (purple), and TES (yellow). The bias leads (red) and two gold stripes are visible on the TES. Both pixels sip atop the silicon grid (black). }
    \label{fig:array}
\end{figure}

The core of the instrument is the 128-pixel TES microcalorimeter array~\citep{Eckart:2009, Eckart:2013}. The pixels are arranged in a 12$\times$12 grid, shown in Figure~\ref{fig:array} (left), with the 16 corner pixels left unwired to match the 128-pixel readout. Each pixel holds a TES on top of a 1~$\upmu$m-thick silicon nitride (SiN) membrane, which serves as the thermal link to the underlying 300~$\upmu$m-thick silicon grid, as shown in Figure~\ref{fig:array} (right). The absorber is mechanically supported by nine stems and overhangs the rest of the pixel. One stem connects to the TES, and the other eight support the absorber over the membrane and the TES wiring.  

Each Micro-X TES is made from a 250~nm-thick molybdenum/gold proximity-effect bilayer. The proximity effect modifies the TES transition temperature by adjusting the thickness of the superconductor (Mo) and normal metal (Au) components. Three $\sim$350~nm gold stripes are deposited on the TES to improve the noise performance~\citep{Ullom2004}. Each absorber is made of 3.4~$\upmu$m-thick bismuth electroplated onto 0.6~$\upmu$m-thick gold (\cite{Rutherford:2013}).
% \rcom{in Megan's 2013 paper it says The absorber is 0.5 $\mu$m Au and 3 $\mu$m Bi (See image in paper as well) for one of the designs considered for micro-x but I don't see these specific numbers listed - are these conflicting or am I missing something in the manufacturing process? Otherwise the numbers quoted here are in John's, Sarah's, and David's thesis.} \dcom{changed reference to John's thesis.}
The high thermal conductivity of the gold layer provides fast thermalization across the large absorber, while the bismuth layer increases structural stability and stopping power without significantly increasing the heat capacity. 

The energy from an incident X-ray is quickly thermalized in the absorber, the stems, and the TES. The pixel as a whole then cools back to its original operating temperature. A layer of copper on the back side of the silicon grid increases conductivity to reduce crosstalk between pixels \cite{Iyomoto:2008c}. Perforations in the SiN are designed to limit the thermal conductivity to the cold bath by controlling the length of the boundary between the pixel and the silicon grid. The pixels were specifically designed for Micro-X to produce pulses with a 1.6~$\pm$~0.3~ms decay time. 

\bigskip

\subsubsection{Front End Assembly and SQUIDs}
\label{sec:FEA}

TESs are conventionally read out with Superconducting QUantum Interference Devices (SQUIDs), which serve as superconducting, low-noise current amplifiers. When magnetic flux is coupled into a SQUID, it produces a quasi-sinusoidal response in current or voltage (depending on the bias). This response is referred to as the SQUID's characteristic V-$\Phi$ relation, and it is periodic with the magnetic flux quantum, $\Phi_0$ ~\cite{{VanDuzerTurner}}. When an X-ray hits the TES absorber and changes the resistance of the voltage-biased TES, the current through the input circuit of the SQUID changes. Inductively coupling the TES to the SQUID converts this changing current into magnetic flux, which modulates the signal across the SQUID. A second inductor is coupled to an active feedback circuit. The flux-locked feedback loop actively nulls changes in flux from the SQUID input coil and returns the SQUID to its initial signal level or ``lockpoint" (in feedback parlance). This feedback linearizes the response of the SQUID amplifier chain, and the response of the feedback loop is therefore proportional to the TES current. 

\begin{figure}[htb]
\begin{center}
    \includegraphics[width=0.45\textwidth]{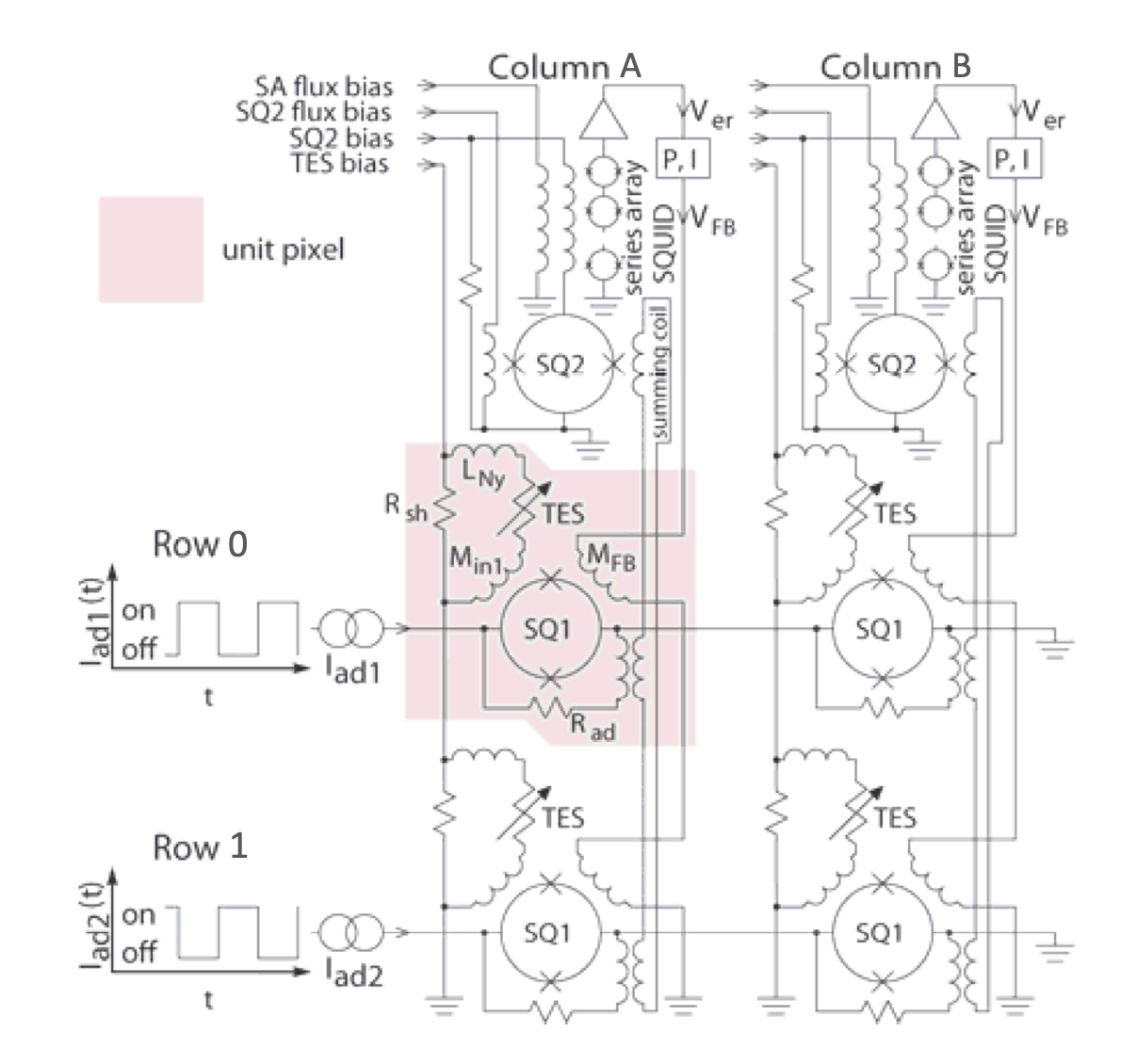}
\end{center}
    \caption{The Micro-X readout is a three-stage TDM SQUID system. Each SQ1 reads out a single TES, and the SQ1 currents are cycled sequentially to read out one TES at a time for each column. The output from the SQ1 is amplified by the SQ2 and SA stages. Only two rows and two columns are shown. Figure from~\cite{Doriese:2015}.}
    \label{SQUIDs}
\end{figure}

\begin{figure}[htb]
\begin{center}
    \includegraphics[height=3in]{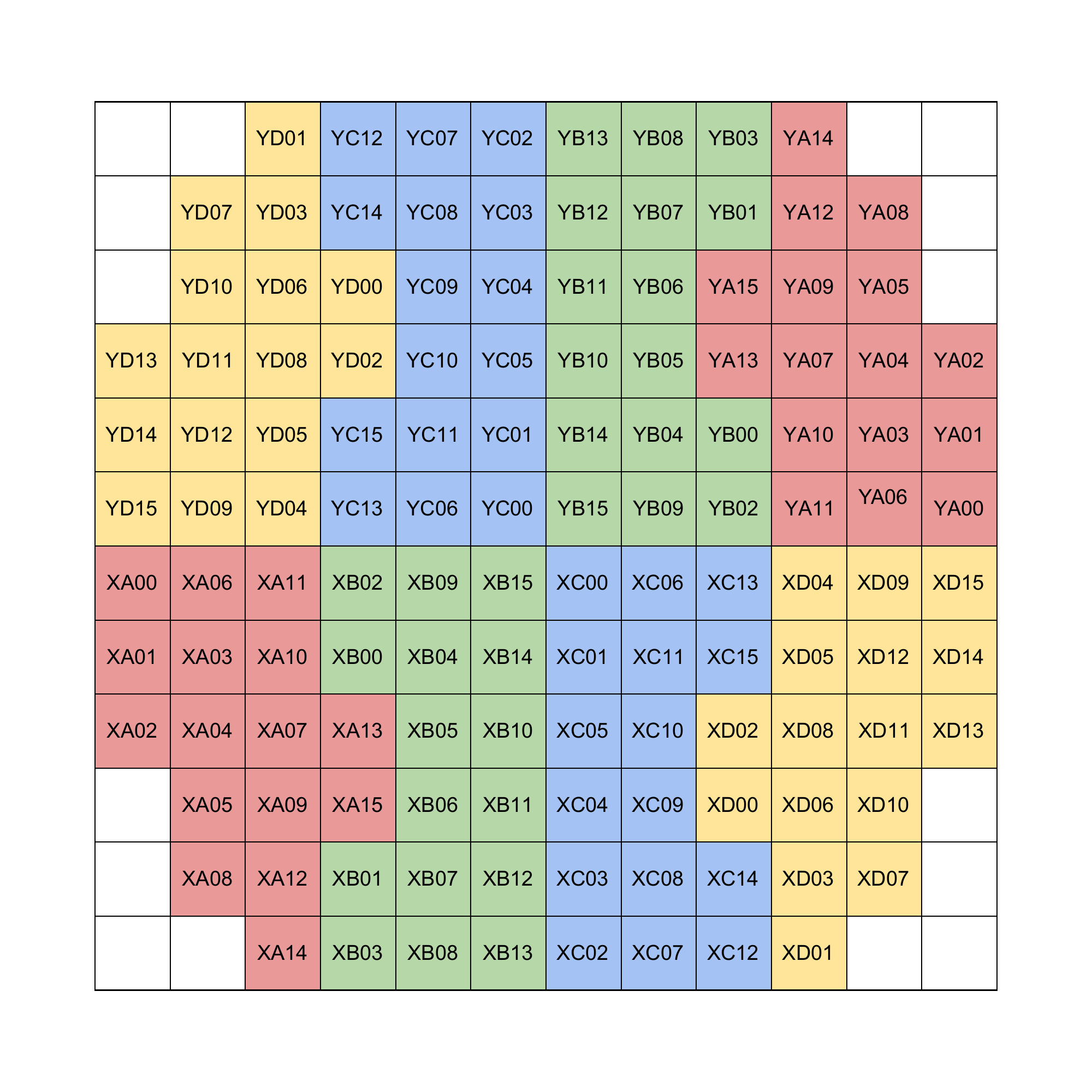}
    \includegraphics[height=3in]{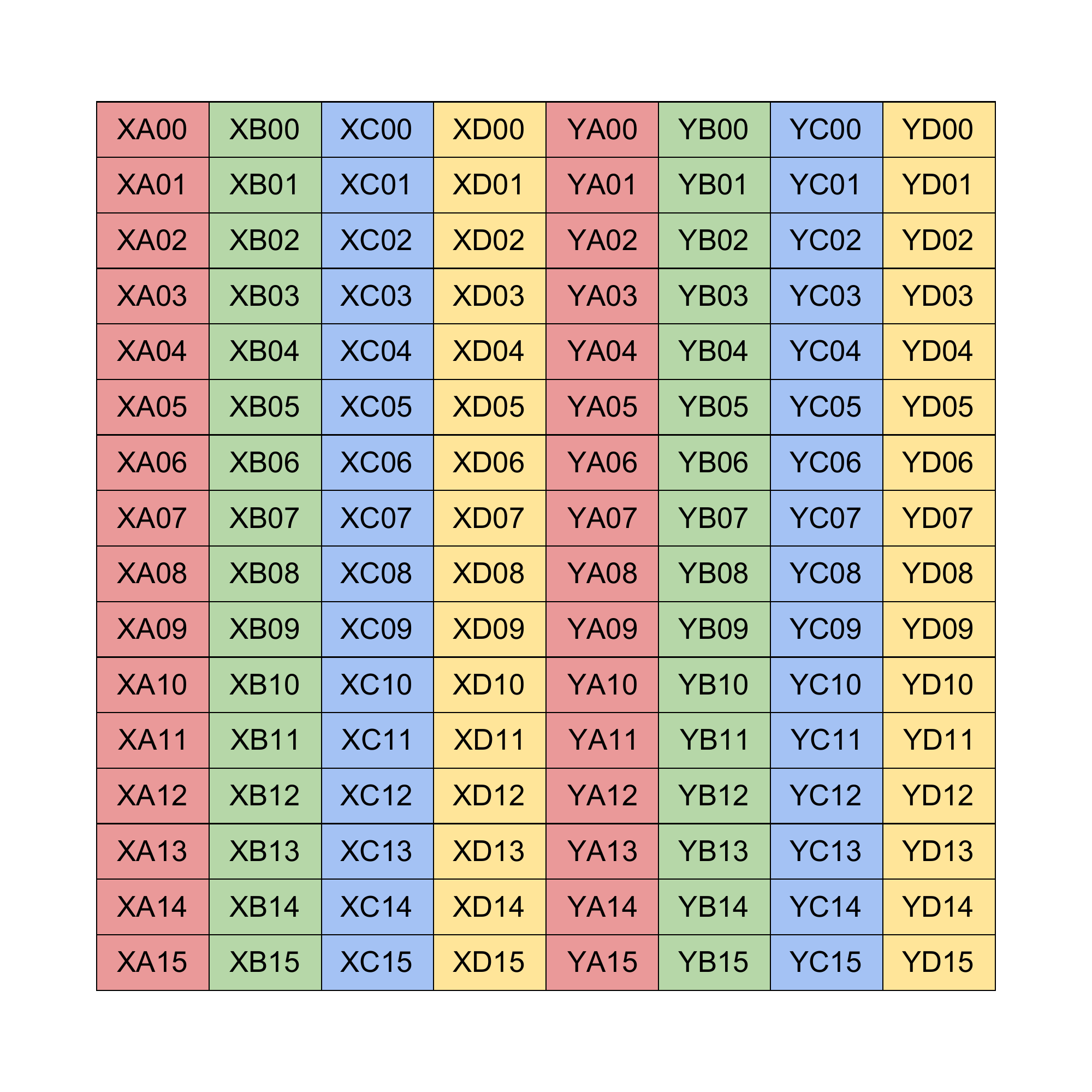}
\end{center}
    \caption{The nomenclature used to identify each pixel based on the science chain side (X or Y), column (A--D), and row (0--15). The physical layout of the array (left), where each pixel is identified on the 12x12 array, is directly comparable to Figure~\ref{fig:array} (left). The TDM scheme (right), where each pixel is arranged in its TDM column and row, is directly comparable to Figures~\ref{fig:NEP_readout_config} and \ref{fig:NEP_readout_flight}.}
    \label{fig:array_naming}
\end{figure}

Micro-X uses a three-stage SQUID multiplexing (MUX) readout scheme~\citep{Reintsema:2009}. The MUX readout decreases the number of wires to the detector stage, reducing the heat load and increasing the number of detectors that can be instrumented with a modest increase of complexity in the SQUID readout circuitry. As shown in Figure~\ref{SQUIDs}, there is an independent 1st stage SQUID (SQ1) for each TES sensor in the array. Sensors are arranged in columns and rows. A discrete multiplexer circuit element (NIST MUX06A design~\cite{Reintsema:2009}) instruments a single column of sensors. The design employs a superconducting transformer summing coil to couple the current signals from the SQ1s to a common 2nd stage SQUID amplifier (SQ2). In Time-Division Multiplexing (TDM), the SQ1s are activated sequentially, coupling the signal for a single TES in the column to the SQ2. SQ1 activation signals (address currents) are shared in series across the columns, activating an entire row of SQ1s in the array simultaneously. The SQ2s are voltage biased at a fixed setpoint in series with the input coil of a 3rd stage cryogenic amplifier. The 3rd stage amplifier is a SQUID series array (SSA)~\cite{Welty:1991} that resides on the 2~K stage (due to power dissipation) and amplifies the switched dynamic signal from the multiplexer to a level suitable for the room temperature electronics. There is a single SSA per column of the array (i.e., per multiplexer). The readout is divided into two independent, 64-channel DC SQUID TDM systems~\cite{TDM, de-Korte:2003}, identified as ``X" and ``Y." The 64 pixels on each science chain are divided into four TDM ``columns," (labeled A--D) with 16 ``rows" (labeled 0--15) each. The pixel nomenclature is visualized in Figure~\ref{fig:array_naming}.

The TDM timing must be compatible with the TES pulse parameters. The Micro-X readout uses a 50~MHz master clock and activates each row address signal for 48~samples (960~ns). A 42-sample settling time is applied to allow for the switching transient to decay. This is followed by the accumulation of four ADC samples of the settled signal from the readout channel. The last two samples (40~ns) are not read out to conservatively reject any possible switching noise. It takes 16.96~ns to sequence through the 16 rows in the array, resulting in a frame rate of 65.1~kHz. Downsampling is implemented in the room temperature electronics (\S \ref{sec:Science Chain Electronics}) to limit the data rate. It only keeps every third sample from each pixel, so while a single pixel is read out at 65.1~kHz, the downsampling produces an effective sampling frequency of 21.7~kHz. This timing is fast enough to capture 20~samples across the leading edge of the pulse, avoiding pulse height dependence on subsample arrival time~\cite{Witthoeft:2021}. 

\begin{figure}[htb]
\centering
    \includegraphics[width=\textwidth]{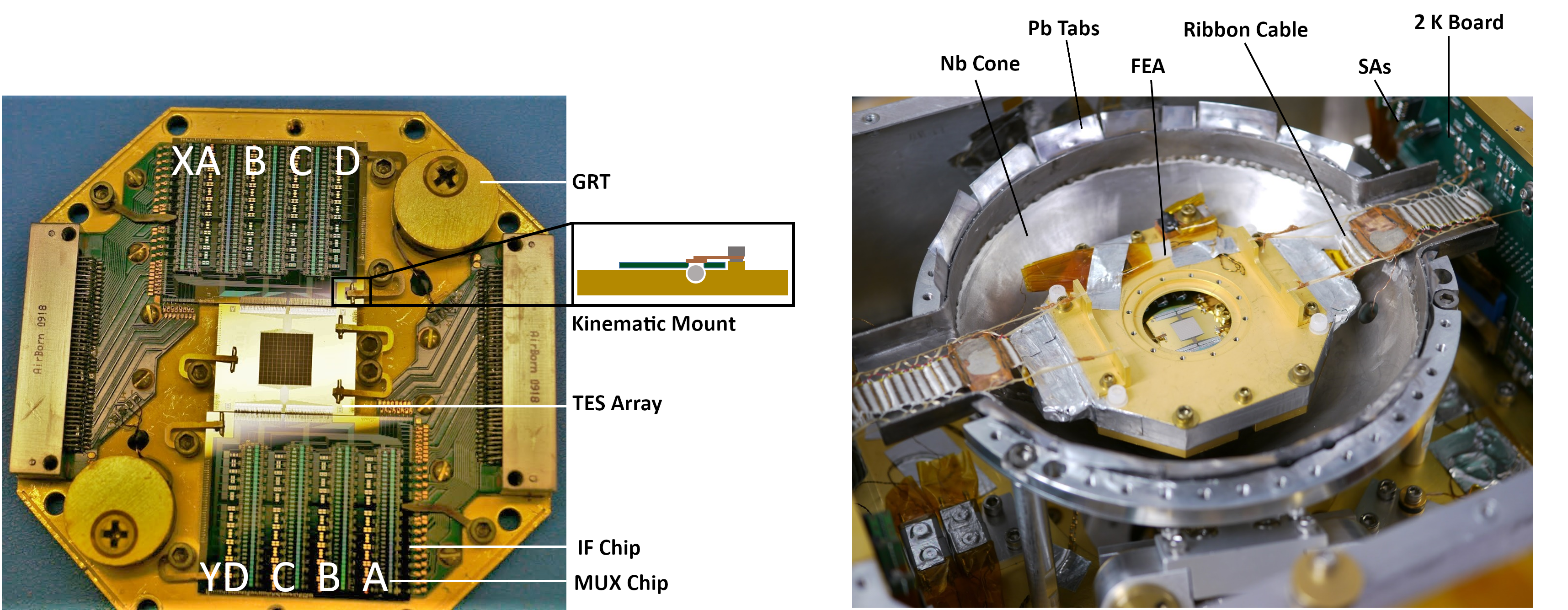}
    \caption{The Front End Assembly (FEA) includes the TES array and first two SQUID readout stages (left). The MUX chips are identified by science chain side (X or Y) and column (A to D). The inset shows a schematic of the kinematic mounts that hold the TES and IF chips. The Germanium Resistance Thermometers (GRTs) are used for ADR temperature control (\S \ref{sec:ADR}). The FEA bolts to the ADR cold stage and is surrounded by a superconducting Nb magnetic shield (right, \S \ref{sec:Shielding}), which is open in this image. The Pb tabs used to create a superconducting joint between the lid and base of the Nb shield can be seen along the edge of the Nb can. Woven ribbon cables run from the FEA to the 2~K boards, visible in the upper right corner, that hold the third SQUID stage. The TES array is visible through the filter aperture on the FEA lid.}
    \label{fig:fea}
\end{figure}

The TES array and the SQUID multiplexer (SQ1 and SQ2) are housed on the Front End Assembly (FEA), shown in Figure~\ref{fig:fea} (left). This lightweight magnesium box is thermally connected to the ADR salt pill, which keeps the FEA at a stable 75~mK (\S \ref{sec:Cryostat}). The SQUID MUX chips are glued to an interface (IF) chip, designed specifically for Micro-X, that holds bandwidth-limiting 1~$\upmu$H Nyquist inductors (L$_{Ny}$ in Figure~\ref{SQUIDs}) and 0.5~m$\Omega$ TES shunt resistors (R$_{sh}$ in Figure~\ref{SQUIDs}). The TES array and IF chips are kinematically mounted to the FEA, protecting them from any stress due to the different thermal expansion coefficients of the silicon chip and magnesium plate \cite{mccammon2002}. Each chip rests on three tungsten carbide balls. Beryllium-copper clips ensure that the chips maintain contact with the balls, as shown in the inset of Figure~\ref{fig:fea}. The entire focal plane is covered by a magnesium lid, which has an aperture above the TES array for an optical/infrared blocking filter (\S \ref{sec:Shielding}). 

Communication with the FEA is done separately for each science chain. A superconducting woven ribbon cable with a niobium-titanium core and copper-nickel cladding mates to a Nanonics connector on the FEA through a light-tight interface. The cable runs to a printed circuit board mounted at the 2~K stage that holds the SAs, which can be seen in Figure~\ref{fig:fea} (right). Each SA board holds four SAs. They are mounted on four PCBs and enclosed in a Nb box, open on one end for the harnessing to the main SA board. A separate cable from the 2~K board, heat sunk with copper sheets to the 50~K stage, runs to the lid of the cryostat at 300~K to interface with the room temperature electronics (\S \ref{sec:Science Chain Electronics}). This cable is primarily manganin, aside from the SA bias signals, which are carried on alloy30 wire. 

\subsubsection{Science Chain Electronics}
\label{sec:Science Chain Electronics}

The room-temperature science chain electronics run the SQUID feedback loops and deliver the raw science data. The two independent readout chains are each responsible for communication with half of the array, and they interface individually to the FEA and to telemetry. This arrangement mitigates risk in the case of a failure of one of the chains. Science data passes out of the cryostat and into a Low-Noise Amplifier (LNA) stack that mounts to the cryostat lid. The LNA stack provides the SA bias and amplifies the output science signals by a factor of 100. The LNA stack also serves as the star ground for the system. All signal return lines, the analog ground, and the digital ground are tied to chassis here.

\begin{figure}[htb]
    \includegraphics[width=0.7\textwidth]{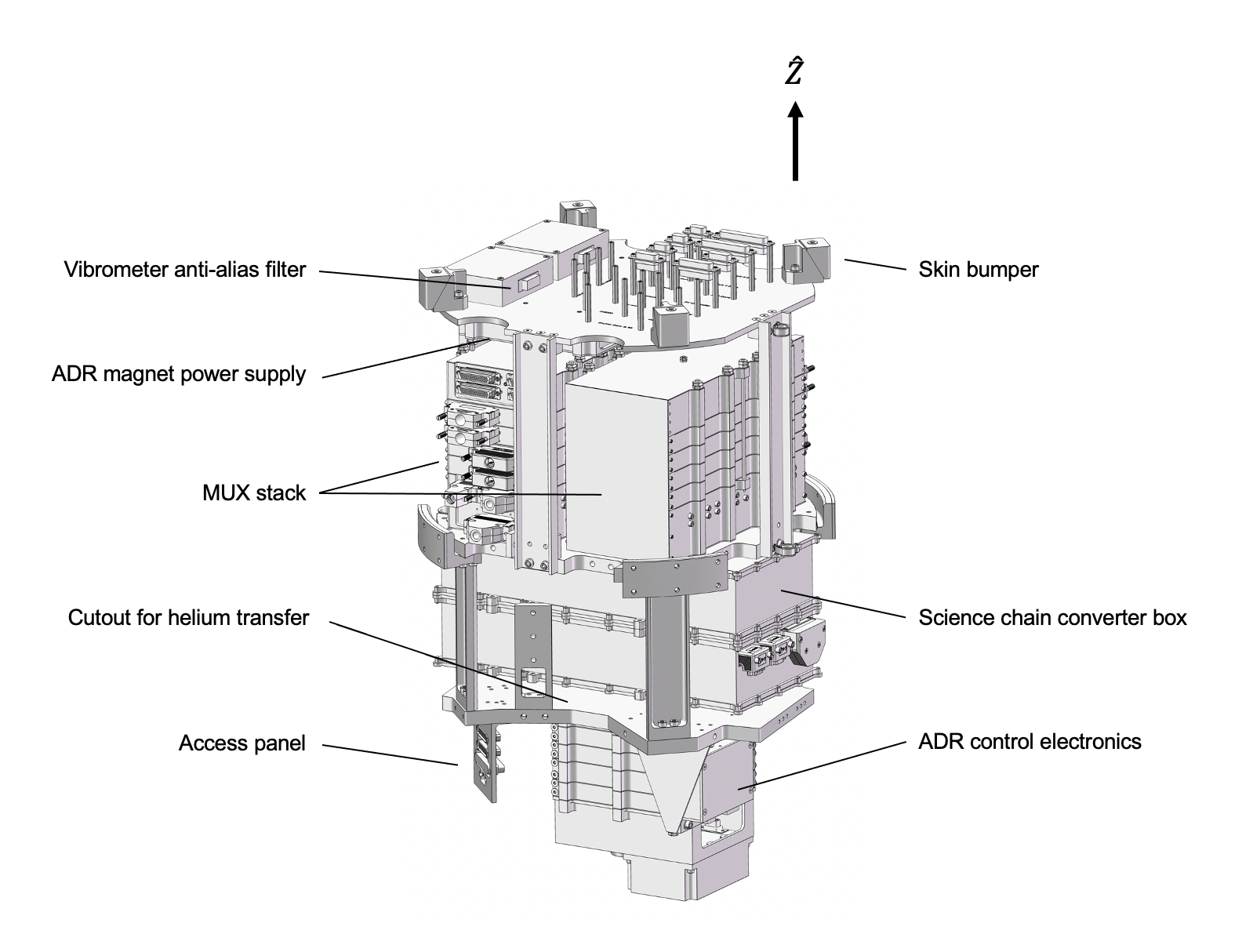}
    \caption{The electronics assembly holds the room temperature readout and controls for the science chains and the ADR. The science data comes out of the cryostat, passes through the LNA stacks shown in Figure~\ref{fig:cryostat}, and goes into the digital feedback circuits of the MUX stacks. The converter box circuits convert power from telemetry into clean power rails for the science chains. The ADR electronics are described in \S~\ref{sec:ADR Electronics}.}
    \label{AvionicsAssembly}
\end{figure}

The digital electronics, housed in the ``MUX stack," control the feedback loops for the detector chain and digitize the science data before transmitting it to telemetry. The MUX stack contains separate boards for each function. The Row Addressing board switches the SQ1 bias signals, while the Quiet DC board provides low-pass-filtered DC current for the TES bias, SQ2 bias, and SA feedback. These boards are described in~\cite{Reintsema:2009}. The Digital Feedback boards apply an anti-alias filter to the analog signals from the LNA stack, digitize the filtered input signal, compute and apply the multiplexed feedback signal to the SQ1, apply a multiplexed quasi-static signal to the SQ2 feedback coils (for optimum tuning of the amplifier chain), and stream both the error and feedback signals to the Master Controller Board for storage and transmission. The transmitted feedback is set independently for each pixel, and the 50~MHz master clock in the Master Control Board synchronizes the digital feedback output with the Row Addressing board so that each pixel receives its own feedback. 

The Master Control Board controls the other MUX stack boards, and it is responsible for timing, commanding, and data control. It downsamples the raw science data stream output to meet bandwidth requirements (three-fold decimation and an anti-alias filter with 8.4~kHz rolloff) and writes the downsampled 66 Mbit/s data stream to its flash memory. It also sends the data from 54 pre-selected pixels to telemetry for real-time transmission in case the data cannot be recovered from the flash memory after flight.

The Master Control Board also receives telemetry commands and the system clocks generated in the MV encoder. Each science chain receives 28~VDC from its dedicated power bus in telemetry. A set of power conversion and filter boards in the instrument electronics steps down the voltage to multiple power rails without introducing a significant amount of noise to the system. These boards, developed for Micro-X, run the incident power through a set of DC-DC converters before filtering them across multiple cards in a dedicated, fully enclosed aluminum box. Noise tests demonstrate that this system contributes $<$1~mV$_{\mathrm{RMS}}$ noise in a 100~MHz bandwidth, which is quieter than a standard commercial power supply. 

The interface between the electronics assembly and the exterior of the rocket is designed to minimize thermal coupling that could heat up the system. The assembly mounts to the rocket skin with four stainless steel brackets; four shims coated in Room-Temperature-Vulcanizing (RTV) silicone sit between the brackets and the skin for thermal isolation. For additional mechanical support, four delrin bumpers press against the rocket skin while minimizing thermal conduction.

%%% CRYOSTAT %%%
% !TEX root = ../instrument_paper_revB.tex
\subsection{Cryostat}
\label{sec:Cryostat}

\begin{figure}[htb]
	\includegraphics[width=0.8\textwidth]{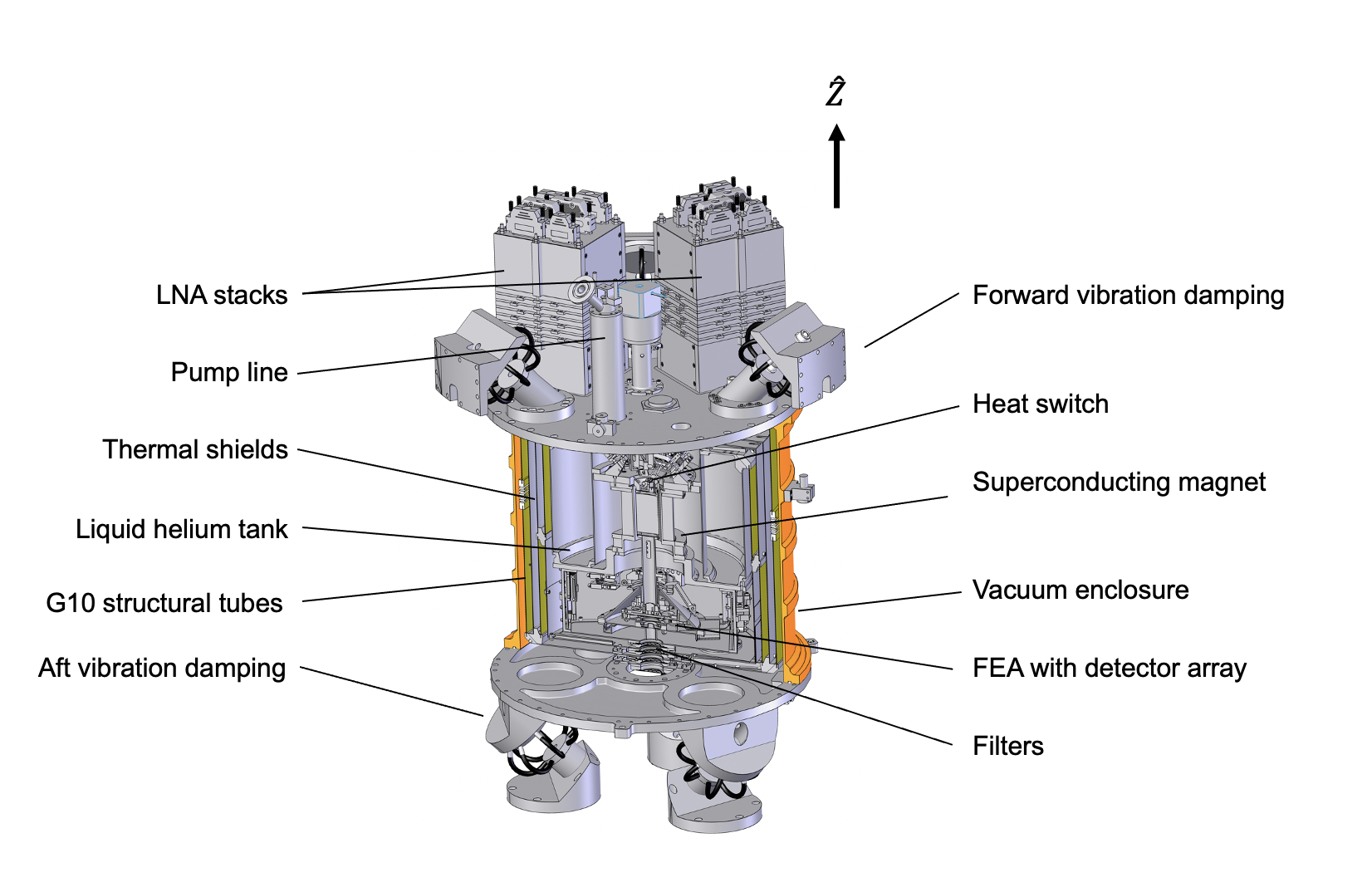}
	\caption{The cryostat provides a vibrationally-isolated 75~mK environment for the detector array. Vapor-cooled thermal stages are mounted to G10 structural tubes at the 50~K and 150~K stages. A pumped LHe ADR system provides temperature control. An external pumpline allows the vacuum of space to pump on the LHe tank in flight.}
	\label{fig:cryostat}
\end{figure}

\begin{figure}[htb]
	\centering
	\includegraphics[width=\textwidth]{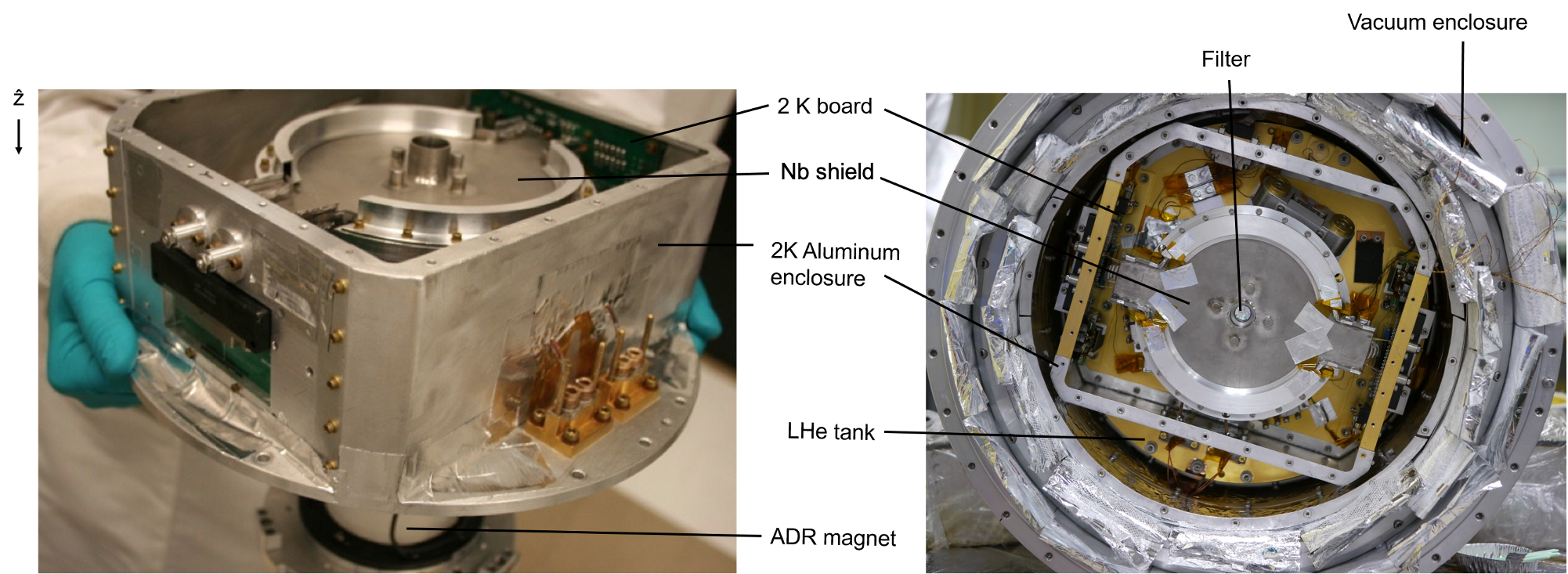}
	\caption{The insert (left) holds the 2~K and 75~mK elements of the instrument, including the ADR and the detector array. The insert mounts onto the LHe tank. The cryostat (right) is shown facing aft, with the magnetic shielding enclosing the FEA, and the FEA filter visible in the center aperture.}
	\label{fig:insert}
\end{figure}

The cryostat, shown in Figure~\ref{fig:cryostat}, is an aluminum vacuum enclosure that houses the FEA (\S \ref{sec:FEA}) and the cryogenic refrigeration system (\S \ref{sec:ADR}). It is modified from the design developed for the XQC sounding rocket~\citep{mccammon2002}. The cryostat houses an outer vacuum layer and four temperature stages: 150~K, 50~K, 2~K, and 75~mK. Thermal radiation shields at the 150~K and 50~K stages are passively cooled by vapor passing through the LHe pumping line. Each shield uses a G10 cylinder to provide mechanical support and Multi-Layer Insulation (MLI) to provide thermal insulation, as seen in Figure~\ref{fig:insert} (right). 

The 2K stage includes all components that are thermally connected to the LHe bath, including the LHe tank and portions of a removable assembly known as the ``insert." The LHe tank is an annular cylinder, and the insert is bolted to it, extending into the central bore of the LHe tank. The insert, shown in Figure~\ref{fig:insert} (left), holds both 2~K and 75~mK elements, including an aluminum enclosure on the aft side and the ADR magnet (\S \ref{sec:ADR}) on the forward side. The aluminum enclosure holds the FEA, magnetic shielding, and the two SA boards. The 75~mK elements are described in \S \ref{sec:ADR}.

The detectors sit within the cryostat vacuum space, protected from the outside atmosphere by the aperture valve that opens to the optics in flight. A 50~kft pressure switch keeps the valve from actuating if the optics pressure is above 116~$\pm$~24~mbar. An onboard microcontroller controls the valve and transmits monitoring information to telemetry for real-time transmission to the ground. 

\subsubsection{Cryogenic System}
\label{sec:ADR}

The cryogenic refrigeration system uses a bath of pumped LHe and an Adiabatic Demagnetization Refrigerator (ADR) to keep the detector stage at a stable 75~mK. The 6.5~L LHe bath is pumped down to a few mbar to lower its temperature from 4.2~K to 2.1~K, measured by a GaAs thermometer bolted to the aft side of the LHe tank. The pump for the LHe tank is external to the payload and connects through the pumping valve on the rocket skin. The valve closes just before launch, and it stays closed until the payload is above an altitude of 85~km. It then opens to allow the vacuum of space to pump on the bath. The valve is controlled by an identical microcontroller to the one used for the aperture valve. To contain the LHe while in the microgravity environment of space, a sintered stainless steel disk in the pump line keeps liquid in while allowing gas to pass through. The LHe can thus be pumped on without losing LHe to space. 

\begin{figure}[htb]
	\centering
	\includegraphics[width=0.8\textwidth]{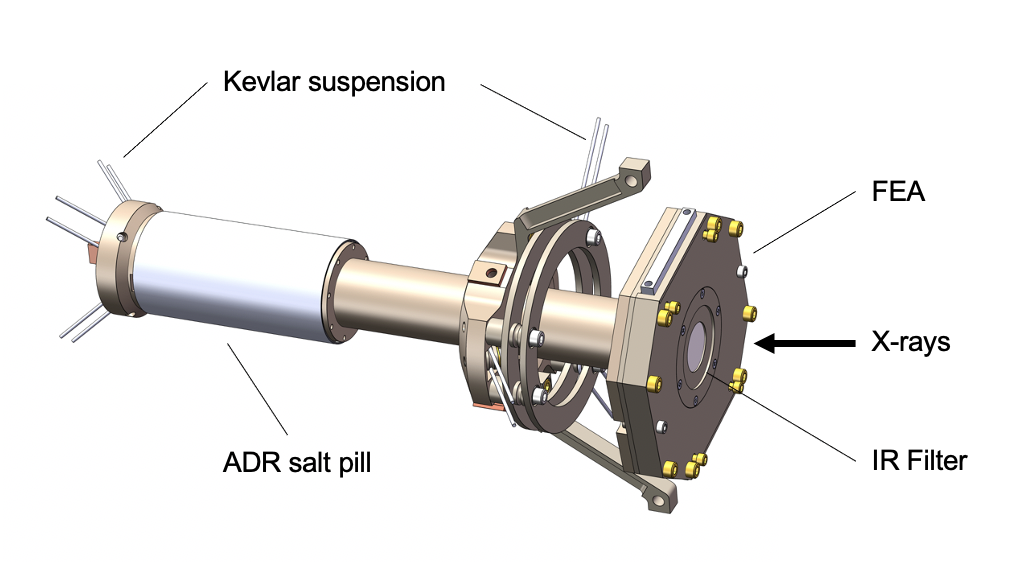}
	\caption{The cold stage, including the FEA, is kept at 75~mK. Kevlar suspension provides thermal and vibrational isolation, and filters provide IR and optical light blocking (\S \ref{sec:Shielding}). The ADR salt pill and the FEA are connected mechanically via two thin-walled Mg tubes attached to the aft Kevlar suspension mount (shown center-right). Inside these tubes is a solid Mg rod (not visible) that provides the thermal connection from the FEA to the salt pill. The ADR magnet and the 2~K components of the insert are not shown.}
	\label{fig:ColdStage}
\end{figure}

The ADR cools the FEA from 2~K to 75~mK. The ADR magnet cycle takes 80~minutes, so it is done before flight (\S \ref{sec:Operations}). The ADR has the cooling capacity to stay at 75~mK~$\pm$~10~$\upmu$K for $>$9 hours with a measured parasitic heat load of 1.7~$\upmu$W in laboratory conditions. The magnet is a custom, multi-filament NbTi/Cu double solenoid (Cryomagnetics, Inc). % that is wound around the salt pill. 
It reaches its full 4~T field with 9.4~A of magnet current, and it has an inductance of 22~H. The salt pill is 75~g of paramagnetic Ferric Ammonium Alum (FAA), grown around a mesh of 200 gold wires and kept inside a hermetically-sealed stainless steel can~\citep{Heine:2014}. The gold wires are brazed to a copper thermal bus, which in turn is bolted to a cold finger machined from high-purity magnesium, providing the thermal link to the FEA. This magnesium rod is enclosed inside two 0.83 inch diameter thin-walled, higher strength magnesium tubes, attached to the aft Kevlar mount to provide mechanical support, as shown in Figure~\ref{fig:ColdStage}. The entire assembly is supported by a Kevlar suspension system to provide thermal isolation from the 2~K stage and minimize coupling to any input vibration from the outer stages. The suspension scheme uses 8520~denier Kevlar string loops under tension. Four loops are attached to the forward side of the salt pill, while two attach to a stiffening ring partway down the magnesium rod. An additional Kevlar loop attached to the stiffening ring connects to a spring that provides $\sim$400~N of tension to the suspended system, which is half of the breaking strength of a single loop ~\citep{Heine:2014}. 

A mechanical heat switch controls the thermal isolation of the 75~mK stage. The 75~mK stage is thermally connected to the 2~K stage while the ADR magnet is ramping up and magnetizing the salt, and it is isolated during the adiabatic phase, when the salt is demagnetized to cool the cold stage to its operating temperature~\cite{PobellADR}. The heat switch uses two clamps, actuated by a Kevlar string, that are mounted to a copper plate in thermal contact with the LHe bath. To thermally connect the cold stage to the 2~K stage, the string is pulled, and the clamps grab the forward end of the salt pill to establish good thermal contact. To thermally isolate the cold stage, the string is relaxed, opening the clamps and breaking thermal contact. 
%The Kevlar string is controlled by a mechanical actuator that sits on the forward side of the cryostat and connects to the vacuum enclosure through a hermetic feedthrough. The actuator uses a spring to pull the kevlar string and close the heat switch. A stepper motor on the forward end of the spring moves to compress the spring and open the heat switch. A potentiometer mounted to the stepper motor reports the position of the spring to determine the state of the heat switch~\citep{Goldfinger:2020}. 
The heat switch is only actuated on the ground, and it is kept open in flight.

\subsubsection{ADR Control Electronics}
\label{sec:ADR Electronics}

The ADR control electronics include a precision readout for the cold stage GRT, a Proportional-Integral-Derivative (PID) controller that compares the temperature with the 75~mK setpoint, and a power supply that controls the superconducting magnet terminal voltage using the output from the PID. A second (``witness") GRT monitors the temperature on the opposite side of the FEA. The electronics control the current drawn by the ADR magnet, and if commanded to ramp the magnet up or down, the voltage across the magnet is changed to produce the desired change in current~\citep{mccammon2002}.  These electronics are a duplicate of the unit used for the XQC instrument~\cite{mccammon2002}, with a redesigned housekeeping telemetry interface.

The ADR system receives 28~VDC power from the same telemetry power buses that supply the science chains. 
%A diode-or circuit is included so that the ADR can draw from either bus if one were to unexpectedly fail in flight. 
The busses are redundantly connected so that either bus could power the entire instrument if the other were to fail. Power is supplied to the Magnet Power Supply, developed for the XQC payload, which uses DC-DC converters to step down the incident voltage to multiple power rails. The Magnet Power Supply is enclosed in a dedicated box with filtered connectors to minimize system noise, and the magnet current lines pass through an additional in-line common-mode choke filter box before entering the cryostat. 

The ADR controller stack includes three XQC-supplied boards for temperature control and two additional boards to create the telemetered data stream. Output data from the ADR controller is used to monitor the state and health of the system (``housekeeping data"). The ADR controller housekeeping boards read in the analog housekeeping signals, multiplex them, and convert them into a single serial asynchronous RS422 output stream. This 82~kbit/s data stream is transmitted in flight and recorded by the ground support equipment. The ADR controller also receives commands in-flight from on-board timers, notably to start active temperature regulation once the payload is above 51~km altitude (\S \ref{sec:Operations}). 
%%%%%%%%%%%%%%%%%%%%%%%%%%%%%%%%%%%%%%%%
\bigskip 

\subsubsection{Shielding}
\label{sec:Shielding}

To achieve a stable response from the TESs in flight, the detectors must be kept in a magnetically and thermally stable environment, and multiple forms of shielding are required to achieve this. Shielding from magnetism, vibrations, optical/infrared light, and high-energy radiation were considered, as described below: \\

\begin{figure}[htb]
	\includegraphics[width=0.9\textwidth]{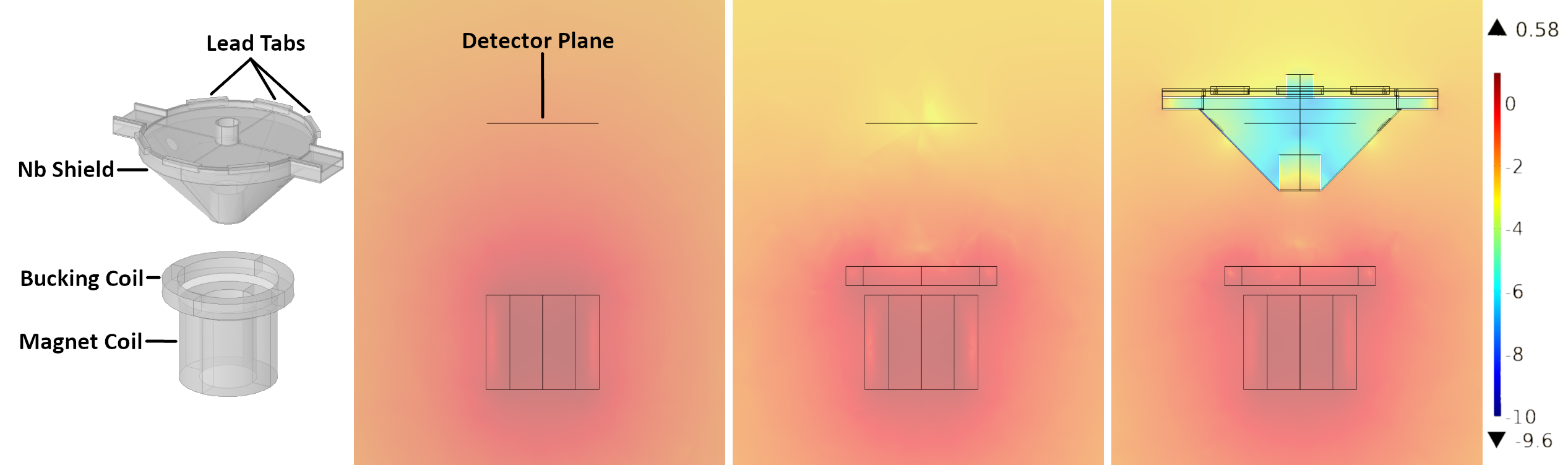}
	\caption{3D COMSOL Multiphysics simulations of two of the fundamental magnetic shielding stages used to protect the magnetically-sensitive detectors from the ADR magnet. The color-bar expresses the simulated field magnitude in a vertical slice through the center of the magnet and Nb can, and is in $\log_{10}$(Tesla). The location of the detector plane is indicated by a horizontal grey line in each simulation. All simulations model the magnetic coil as energized to its full field of 4~T. The left subfigure shows the simulation geometry. The simulations predict the magnetic field from the ADR magnet under the following conditions, from left to right: unshielded; bucking coil only; bucking coil and Nb shield. The Pb tabs are modeled as superconducting regions connecting the lid and base of the Nb shield. The third stage of magnetic shielding, the field coil, is modeled as off in these simluations. The shielding is designed to keep the ADR field below 500~nT at the detector array. }
	\label{Nb-Shield-with-Magnet}
\end{figure}

\bigskip

\paragraph{Magnetism}
\label{sec:magnetism}
TESs and SQUIDs are highly sensitive to magnetic fields, and a 3-stage system shields the detector array from both the Earth's magnetic field and the field generated by the ADR magnet. Magnetic fields can lower the transition temperature of the TES detectors, degrading their energy resolution. The SQUIDs are magnetometers, and a change in the magnetic field environment therefore impacts their outputs. Changes in the magnetic environment can manifest as changes in the DC baseline level of the SQUID response. Small changes in the SQUID response can adversely affect the signal to noise ratio and dynamic range, while large changes move the SQUID response far enough from its stable lockpoint that it is outside the stable range of the flux-locked loop, as discussed in \S\ref{sec:flight_susceptibility}. 

The three-stage magnetic shielding system is designed to keep the field from the ADR magnet below 500~nT at the array, with a field uniformity across the array of better than 100~nT. It is designed to keep time-varying magnetic fields below 3 pT/$\sqrt{Hz}$~\citep{Rutherford:2013}. There is no passive shielding on the magnet to minimize weight. Sensitivity to magnetic fields - and the resulting shielding requirements - is one of the fundamental differences between the Micro-X and XQC instruments. 
%Figure~\ref{Nb-Shield-with-Magnet} shows three COMSOL Multiphysics simulations of the Micro-X ADR system side by side, modeling the ADR magnet alone, with the bucking coil, and with both the bucking coil plus Nb shield (left to right). All three cases are full 3D models with the magnetic coil energized to its full field of 4 Tesla. 

The first stage of magnetic shielding is a bucking coil. It is in series with the primary ADR magnet coil, but wound in the opposite direction to generate an opposing magnetic field. It sits between the primary magnet coil and the FEA, with a slightly larger radius than the primary magnet coil. Its cross section and location are optimized so that the ADR magnet field is partially cancelled at the detector array without significantly affecting the field at the ADR salt pill. The impact of the bucking coil on the magnetic field at the detector array is shown in the third panel of Figure~\ref{Nb-Shield-with-Magnet}. In this configuration, there is a low field region around the detector plane with a field magnitude of 1.2~mT at its center, compared with 47~mT without the bucking coil. 

The second stage of magnetic shielding is a superconducting Nb shield that is thermally connected to the LHe can and surrounds the FEA to reject external magnetic fields through the Meissner-Ochsenfeld effect~\cite{Meissner:1933}. The top (aft cover) of the Nb shield is visible in Figure~\ref{fig:insert}. The combined impact of the bucking coil and the Nb shield is shown in the right panel of Figure~\ref{Nb-Shield-with-Magnet}. The Nb shield has a welded cone on the forward side, shaped to accommodate the Kevlar suspension and minimize sharp corners that could concentrate the magnetic field and potentially exceed the Nb superconducting critical field. A Nb lid on the aft side completes the shield. The two pieces are connected with a Pb tape ``zipper" to make a superconducting connection between the two pieces~\cite{Garfield:1997}. The Pb tape zipper is made of a set of tabs, attached to the Nb shield with a layered indium and Rose's alloy solder. The efficacy of the Pb tab connection was found to be variable with each installation, complicating the input assumptions for the simulations. A 2~mm stainless steel braided wire was added at the interface between the Nb cone and lid to prevent a thermal touch between the 2~K Nb and the FEA. This wire is inside the Pb zipper, and its impact on the effectiveness of the shield is unknown. The wire was replaced with a superconducting Pb wire for the second flight to provide a more robust shield in the case that there are any undetected gaps in the zipper (\S \ref{sec:Reflight}). 
There are six penetrations in the superconducting shield. The lid on the aft end has a $\varnothing$16~mm hole with raised walls 16~mm deep to allow X-rays to reach the detectors. The forward side of the Nb cone has a $\varnothing$25~mm hole with depth 23~mm for the rod between the FEA and the ADR salt pill. Two rectangular channels (30~mm $\times$ 5~mm with length 26~mm) provide a path for the science chain cables from the FEA, and two small ($\varnothing$4~mm) holes accommodate these cables' heat sinking wires. As described in \S\ref{sec:FEA}, the SAs are located outside the Nb shield and are enclosed in their own Nb boxes. 

Finally, a field coil inside the Nb shield can be used to counteract any remaining trapped static magnetic field. This 500-turn copper-clad Nb-Ti wire coil is wrapped around a 23~mm copper disk inside the Nb shield, forward of the detectors. It can be biased to produce up to 10~$\upmu$T across the detector plane. The field coil was not used in the first flight due to a broken wire that developed inside the cryostat during integration.

In addition to this shielding system, a high-magnetic-permeability Metglas blanket~\cite{Metglas} is wrapped around the cryostat each time it is cooled from room temperature. The blanket minimizes the amount of Earth's field that is frozen into the Nb shield as it goes superconducting. A blanket is used instead of a mu-metal bucket due to the mechanical constraints of the rocket configuration. The blanket is removed before flight. 

%\hl{PW: rework the remainder of this section. Needs tightening up.} 
Due to the complicated geometry of the shielding system, its shielding factor (SF) to external magnetic fields cannot be derived analytically and is solved numerically. 
%above is JA/PW proposed replacement for "from outside the shield to the inside depends on the direction of the applied field, the test location inside the shield, and the relevant direction of the field at the test location"
It is represented by a position-dependent $3\times 3$ matrix, with each component of the matrix defined as: 
\begin{equation}
    SF^{\hat{i}}_{\hat{j}}\rvert_{Location} = \frac{\vec{B}_{ext} \cdot \hat{i}}{\vec{B}_{int} \cdot \hat{j}}\Big\rvert_{Location}
\label{equation:sf}
\end{equation}
where $\hat{i}$ and $\hat{j}$ are unit vectors in the rocket frame and $Location$ is where the SF is evaluated inside the magnetic shield. A higher SF is therefore better at expelling external magnetic fields. 
When one or both of the components in Equation~\ref{equation:sf} are the total magnetic field magnitude, they are designated as $tot$ instead of a unit vector.
%The magnetic field magnitude is designated as $tot$ in Equation~\ref{equation:sf}. 
With the Nb shield in place, %the pre-flight 
COMSOL simulations predict a field magnitude of 370~nT at the center of the detector plane, corresponding to a SF of $SF^{tot}_{tot}\rvert_{TES}=3200$ compared to the field at the detectors without the Nb shield. The combined efficacy of the bucking coil and Nb shield to protect the detectors from the ADR magnet is $SF^{tot}_{tot}\rvert_{TES}=127000$. As the shield’s primary design goal is to protect the detectors from the field of the ADR magnet, this SF passed detector shielding requirements. 

While the sensitivity of the SQUIDs to applied fields is a known risk, the shielding efficacy at the SQUIDs was not specifically investigated because it was projected to be a subdominant effect due to the nature of the fields and shielding in this system. The primary magnetic sensitivity of the SQUIDs is due to the topology of the summing coil, which leads to a relatively large SQ2 effective area (468.5~\micron{}$^2$~\cite{Stiehl_2011}) through which external flux can couple to the device. This area vector is in the z-direction, so any external fields in the xy-plane were not expected to couple significantly to the SQUIDs. The largest source of magnetic fields inside the cryostat is the ADR magnet, which primarily produces fields in the z-direction at the detector plane, thus this direction is particularly well-shielded. Additionally, static field that does couple through the summing coil can be accounted for by adjusting the SQ2 flux bias. As expected, the slowly varying magnetic field of the ADR magnet during TES operation is not observed to significantly affect the SQUIDs in laboratory operation. However, as will be discussed in \S\ref{sec:flight_susceptibility_interpretation}, the changing orientation of the rocket with respect to Earth's magnetic field during flight caused a time-varying magnetic field through the SQ2 summing coil.
%While the sensitivity of the SQUIDs to applied fields is a known risk, the shielding efficacy at the SQUIDs was not specifically investigated because it was projected to be a subdominant effect for several reasons. First, the primary source of magnetic fields inside the cryostat is the ADR magnet, which primarily produces fields in the z-direction that the Nb shield expels quite effectively. Second, the major magnetic sensitivity of the SQUIDs is due to the topology of the summing coil, which leads to a relatively large SQ2 effective area (468.5~\micron{}$^2$~\cite{Stiehl_2011}) through which external flux 
%(parasitic as opposed to signal) \hl{JA: remove this parenthetical} 
%can couple to the device. However, this area vector is in the well-shielded z-direction, so any external fields in the xy-plane were not expected to couple significantly to the SQUIDs. Thirdly, any static field that does couple through the summing coil can be accounted for by adjusting the SQ2 flux bias. The slowly varying magnetic field of the ADR magnet during TES operation was not observed to significantly affect the SQUIDs in laboratory operation. However, as will be discussed in \S\ref{sec:flight_susceptibility_interpretation}, the changing orientation of the rocket with respect to Earth's magnetic field during flight caused a time-varying magnetic field through the SQ2 summing coil.

\bigskip

\paragraph{Vibration}
%\begin{wrapfigure}{r}{0.6\textwidth}
%   \centering\vspace{-10pt}
%	\includegraphics[width=0.6\textwidth]{figs/launch_specs.png}
%	\caption{Vibration isolation during powered flight, with ignition marked in red. The detector array experienced no discernible heat load during powered flight, despite significant vibrational loads on the rocket.}
%	\label{fig:vibe}
%\end{wrapfigure}
Successfully isolating the FEA from external launch vibrations was a significant engineering challenge. 
%The largest challenge for the instrument is maintaining a stable, cryogenic temperature at the detector array shortly after launch. 
The sounding rocket flight affords minimal time to return to operating temperature if mechanical energy from the launch heats the cold stage. A three-stage isolation system staggers the resonant frequencies of the structural components and keeps mechanical energy from being transmitted to the detectors; it is designed so that the increasingly higher resonant frequencies on the inner stages match poorly to the outer stages. The system includes (\cite{Danowski:2016}): 
\begin{itemize} \setlength \itemsep{-0.25em}
	\item Wire rope isolators that mount the cryostat to the rocket (44~Hz resonant frequency)
	\item G10 plastic tubes that support the helium can inside the cryostat (96~Hz resonant frequency)
	\item Kevlar suspension that hangs the detectors from the helium can (325~Hz resonant frequency)
\end{itemize}

The wire rope isolators (Isotech, Inc.) are made of stainless steel ropes that are glued to mounting plates. Three isolators on the forward end of the cryostat mount the cryostat to the rocket skin, and three at the aft end mount it to a bulkhead that connects directly to the rocket skin. These isolators are labeled ``vibration damping" in Figure~\ref{fig:cryostat}. The isolators have a much lower spring constant under shear than under compression, so they are placed at a 45\degree angle to offset their shear planes from one other. The isolators have a well-defined equilibrium point to which they reliably return after displacement. To produce the desired resonant frequency, the G10 tube at the 150~K stage is split in half, with an annular metal spring placed around its circumference. This reduces longitudinal stiffness and produces a resonant frequency of 96~Hz, which is within the desired frequency range for the vibration isolation scheme ~\citep{Danowski:2016}. The Kevlar suspension geometry achieves its desired resonant frequency with a stabilizing ring and optimized orientation and thickness of the Kevlar strings~\citep{Heine:2014}. 

%The instrument successfully met its vibration isolation requirements during flight. The payload saw a vibration load of 7.5~g$_\mathrm{rms}$ for the first 44~s of flight (\S\ref{NSROC Systems}). At the time of landing after the flight the ADR system had more than one hour of hold time left, indicating a large margin for heat due to launch vibrations. \S \ref{sec:Flight Performance} provides more details of the instrument response during the launch vibration period in flight. 

%\begin{figure}[htb]
%   \centering
%	\includegraphics[width=\textwidth]{figs/launch_specs.png}
%	\caption{The instrument's vibration isolation successfully isolated the cold stage from launch vibrations during powered flight, with ignition marked in red. The detector array experienced no discernible heat load during powered flight (top), despite significant vibrational loads on the rocket (bottom). The uncertainty of the temperature is driven by the readout precision of the GRT thermometer at this temperature (\S \ref{Cryostat and ADR}). This plot spans powered flightl the full flight analysis is provided in \S~\ref{sec:Flight Performance}). \acom{Remove - redundant}}
%	\label{fig:vibe}
%\end{figure}

\bigskip

\begin{figure}[htb]
   \centering
                \includegraphics[width=0.7\textwidth]{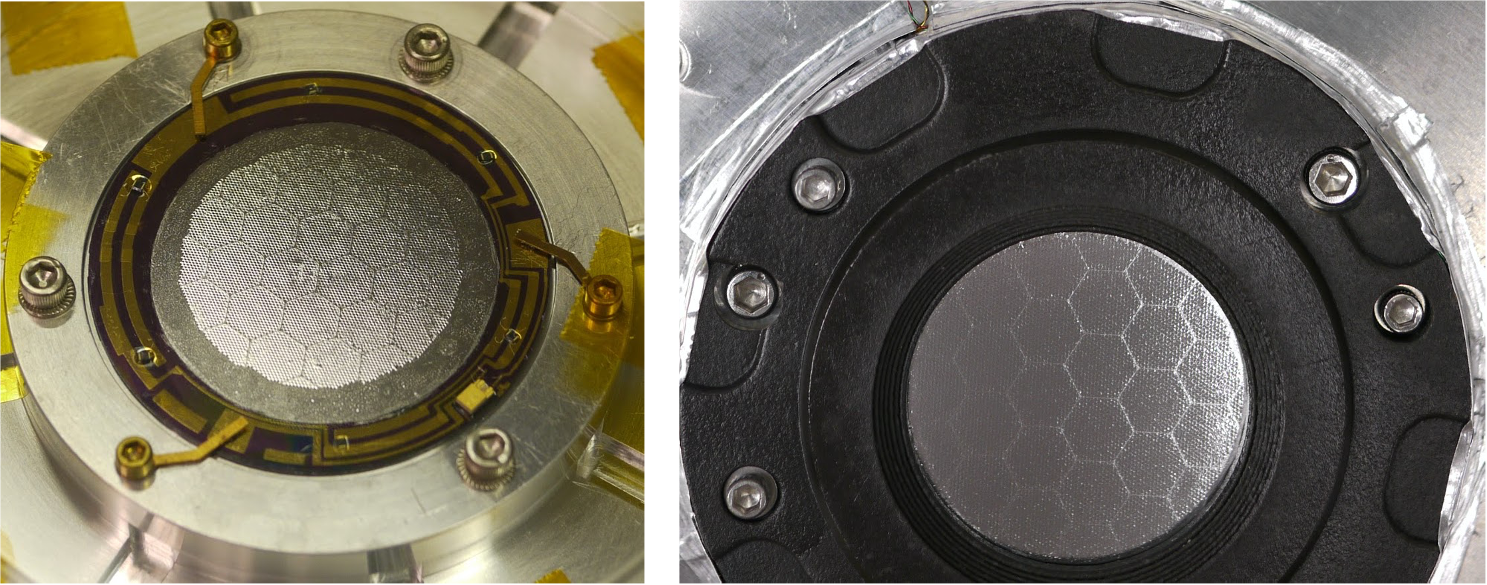}
                \caption{The mesh filters are kinematically mounted to aluminum interfaces (left) and surrounded by baffles (right) that provide a path for air while removing a direct path for IR radiation or water. }
                \label{fig:mesh_filter}
\end{figure}

\paragraph{Optical/infrared light}
\label{sec:filters}
Six thin-film filters sit between the detectors and the aft opening of the cryostat to reduce the transmission of blackbody radiation from the warmer stages of the instrument, which can heat the detectors and increase shot noise. Each filter film is made by Luxel Corp. and composed of 200--250~\angstrom\ of aluminum deposited on 500~\angstrom\ of polyimide. Two $\varnothing$14~mm filters are located on the magnesium FEA lid separated by a 0.5 mm spacer. A $\varnothing$28~mm filter is located at each thermal stage (2~K, 50~K, 150~K, 300~K) between the FEA and the aperture valve. The FEA-lid filters have 200~\angstrom\  aluminum, while the outer four filters have 250~\angstrom\ aluminum. The $\varnothing$28~mm filter films are mounted on micromachined silicon support meshes that have fine and coarse hexagonal meshes with pitches of 330~\micron{} and 5.28~mm~\cite{Crowder:2012}. Figure~\ref{fig:mesh_filter} shows a mesh filter kinematically mounted to its interface. Each filter is placed at a different angle from the others to prevent reflected photons from being trapped between them and ultimately transmitted. A single silicon mesh filter has an open fraction of 0.943 (considering only the silicon mesh), and an average transmission of 0.92 at 1 keV taking into account the mesh and thin film, calculated using literature values of mass absorption coefficients~\cite{Henke:93,CXRO_opt_const} and the specified film thicknesses. The full set of six filters provides an average transmission of 0.67 at 1 keV.
 
A combination of heaters and baffles keep the filters free of water and any stray IR radiation. Surface-mount resistors act as filter heaters, wired to the filter frames with gold traces on the silicon filter mount. They keep the filters warmer than the surrounding surfaces to discourage ice accumulation during cooldown, and they sublimate any ice that may build up.
A set of Ecosorb baffles, one at each mesh-mounted filter, remove any direct path for IR radiation or water transport. Water passing through the baffle will freeze onto the baffle wall rather than reaching the next inner filter. Figure~\ref{fig:mesh_filter} shows a mesh filter before (left) and after (right) baffle installation.
 
\paragraph{Radiation}
The short sounding rocket flight means that trace radioactive elements in the instrument materials do not produce a measurable background. Cosmic rays deposit minimum ionizing radiation in the detectors and are modeled as a background in the astrophysics analysis. There is no requirement to use radiopure materials. 

\subsubsection{Calibration source}
\label{sec:Calibration Source}
\begin{figure}[htb]
	\includegraphics[width=0.8\textwidth]{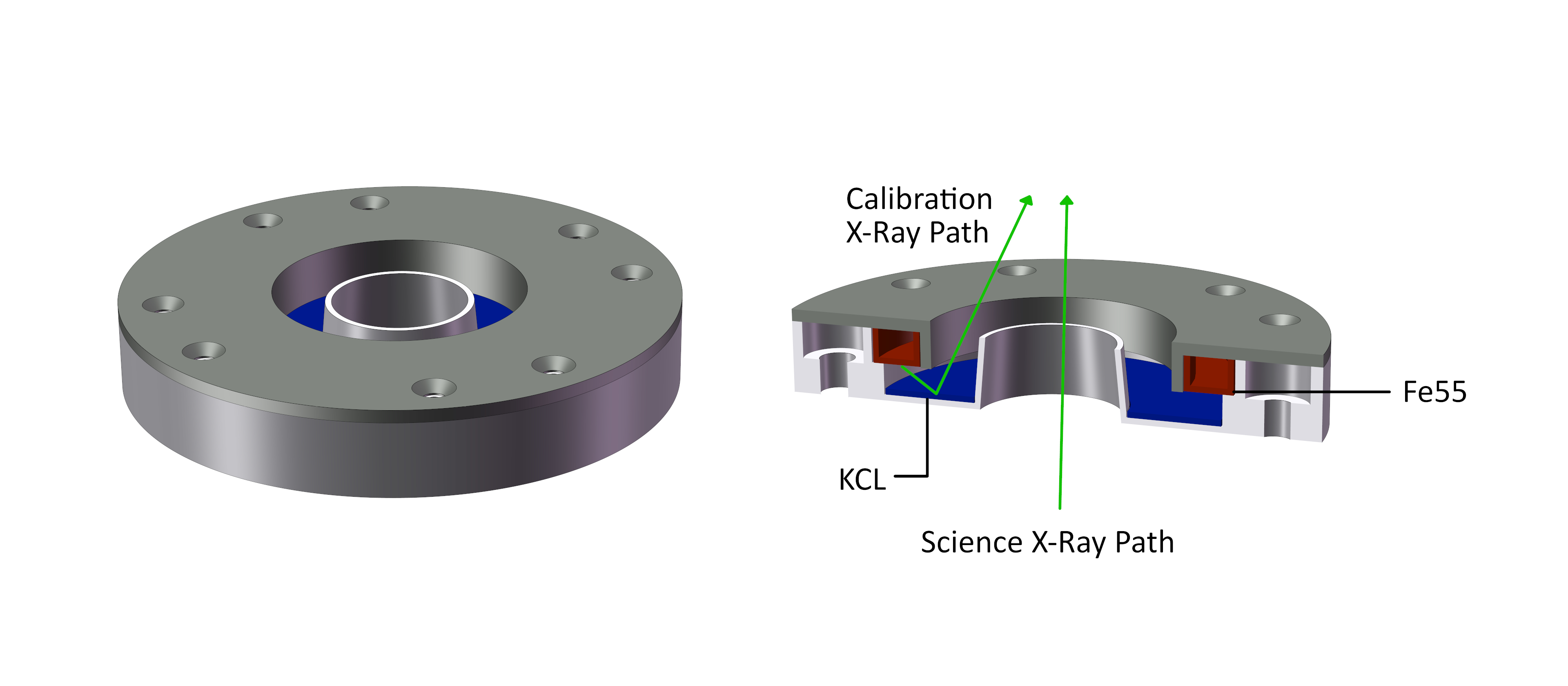} 
	\caption{The calibration source uses a ring of $^{55}$Fe to illuminate a KCl target, providing a constant rate of characteristic X-rays at energies just above the science bandpass. These X-rays are used to track the detector response throughout the flight. }
	\label{fig:cal_source_lines}
\end{figure}

A calibration source inside the cryostat provides a continuous source of X-rays just outside the science band. The source is used to track detector response in flight, allowing any observed changes to be calibrated out in post-processing. It is mounted on the inside of the Nb shield lid, roughly 1~cm from the detector array. The calibration source uses a 15~mCi $^{55}$Fe source to illuminate a KCl target, shown in blue in Figure~\ref{fig:cal_source_lines}. The KCl produces fluorescent lines at 2.62~keV (Cl~K-$\alpha$), 2.81~keV (Cl~K-$\beta$), 3.31~keV (K~K-$\alpha$), and 3.58~keV (K~K-$\beta$), along with a few backscattered Mn~K X-rays at 5.89~keV from the $^{55}$Fe source. Pb walls shield the $^{55}$Fe from the detectors so that only emission from KCl and a small fraction of backscattered 5.89~keV events reach the array. The source rate in flight was measured to be $0.7$~counts/s/pixel, with a K-K$\upalpha$ rate of 0.23~$\pm$~0.06~counts/s/pixel. 

%%% OPTICS %%%
% !TEX root = ../instrument_paper_revB.tex
\subsection{Optics}
\label{sec:Optics}

\begin{figure}[htb]
	\begin{center}
		\includegraphics[width=\textwidth]{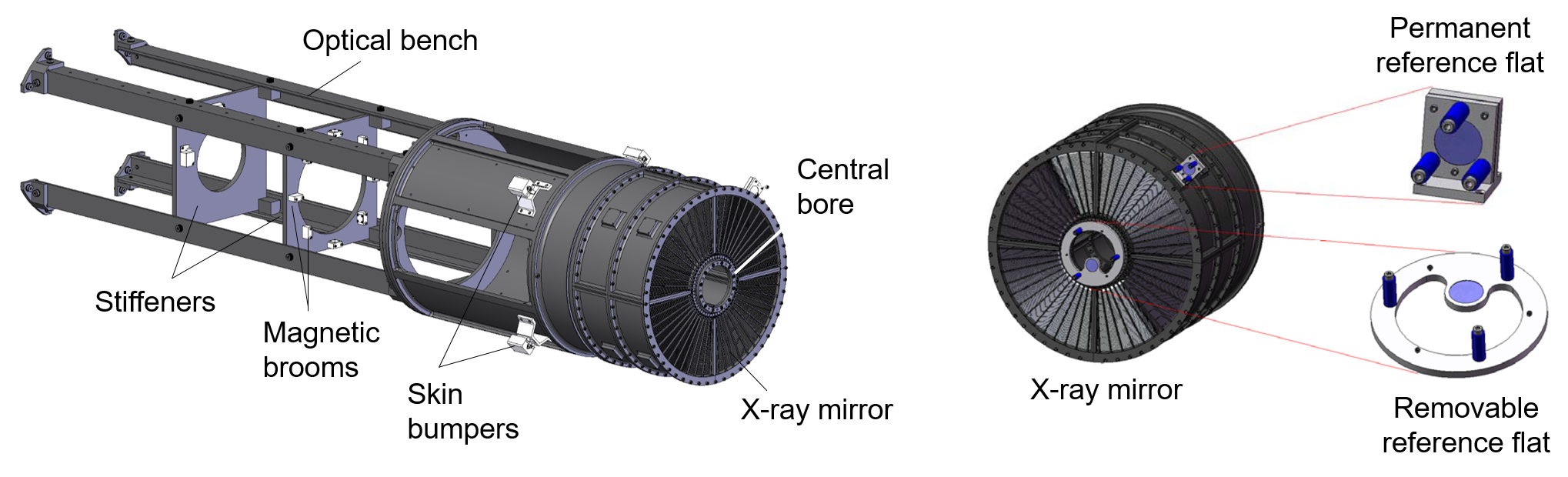}
	\end{center}
	\caption{The optics assembly (left) keeps the X-ray mirror mounted to the rest of the instrument without touching the rocket skin. The optics alignment to the celestial star tracker uses two reference flats that are not used in flight (right).} 
	\label{OpticalBench}
\end{figure}

\begin{figure}[htb]
	\includegraphics[width=0.6\textwidth]{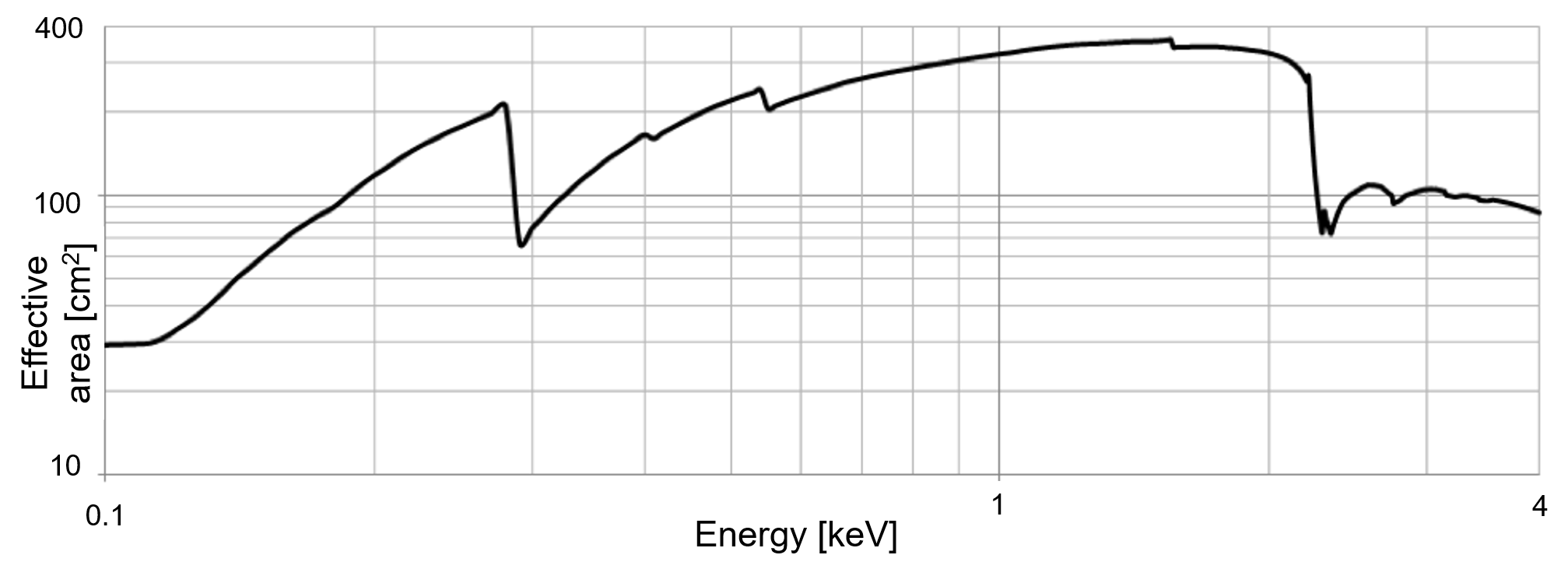}
	\caption{Total effective area of the mirror and filters combined. The X-ray mirror provides a broad energy acceptance, with more than 300~cm$^2$ effective area at 1.5~keV. The low energy effective area is reduced by the aluminized polyimide in the optical/IR filters (\S~\ref{sec:filters}), while the decrease above 2.5~keV is due to the gold coating on the mirror foils.}
	\label{fig:optics_eff_area}
\end{figure}

The X-ray optics assembly, shown in Figure~\ref{OpticalBench}, includes an X-ray mirror, an optical bench, and a star tracker. The system provides a broad energy response, exceeding 300~cm$^2$ of effective area at 1.5~keV, as shown in Figure~\ref{fig:optics_eff_area}. The mirror provides a 2.4~arcmin angular resolution (Half-Power Diameter) and a 2.1~m focal length, which sets the 11.8~arcminute field of view (FOV) for the 7.2$\times$7.2~mm$^2$ array. The mirror is composed of 68 nested, reflecting foils ranging from 80 to 200~mm in diameter. The foils, spares from the Astro-E2 mission, are 0.152~mm-thick aluminum substrates with gold plating~\citep{Kelley:1999}. They are configured into quadrants and held in place by precisely-engineered combs within the mirror housing, which was flown on the Supernova X-Ray Spectrometer (SXS) mission~\citep{Marshall:1987, Serlemitsos:1988}. The X-ray mirror was subsequently refurbished and calibrated for Micro-X. 

The optical bench is also from SXS and was shortened to accommodate the Micro-X instrument. It is mounted to a hermetic bulkhead at its forward end, and bumpers brace its aft end against the skin to prevent the assembly from rocking. Magnetic brooms, made from permanent magnets, are mounted on the optical bench to funnel charged particles away from the detector array~\citep{Wikus:2010a}. 

A ST5000 celestial star tracker~\citep{Percival:2008} for the ACS is accommodated in the central bore of the mirror, mounted with an aluminum adapter. The star tracker compares the stars in its field of view with a pre-loaded star map, and the ACS adjusts the pointing of the payload until the star tracker is centered on the science target. The X-ray mirror and star tracker are co-aligned before flight using the reference mirrors in Figure~\ref{OpticalBench}~(right). A permanent flat reference mirror is used for the star tracker alignment, and a removable flat reference mirror verifies the alignment of the permanent flat to the X-ray mirror optical axis. The alignment of the X-ray mirror to the detector array is centered by shimming the base of the optical bench such that the optical axis of the X-ray mirror hits the center of the aperture valve lid. A hermetic alignment tool is then mounted to the aperture valve housing to measure how well the detector array, held at vacuum, is centered on the aperture valve.  

\begin{figure}
		\begin{center}
		\includegraphics[width=0.9\textwidth]{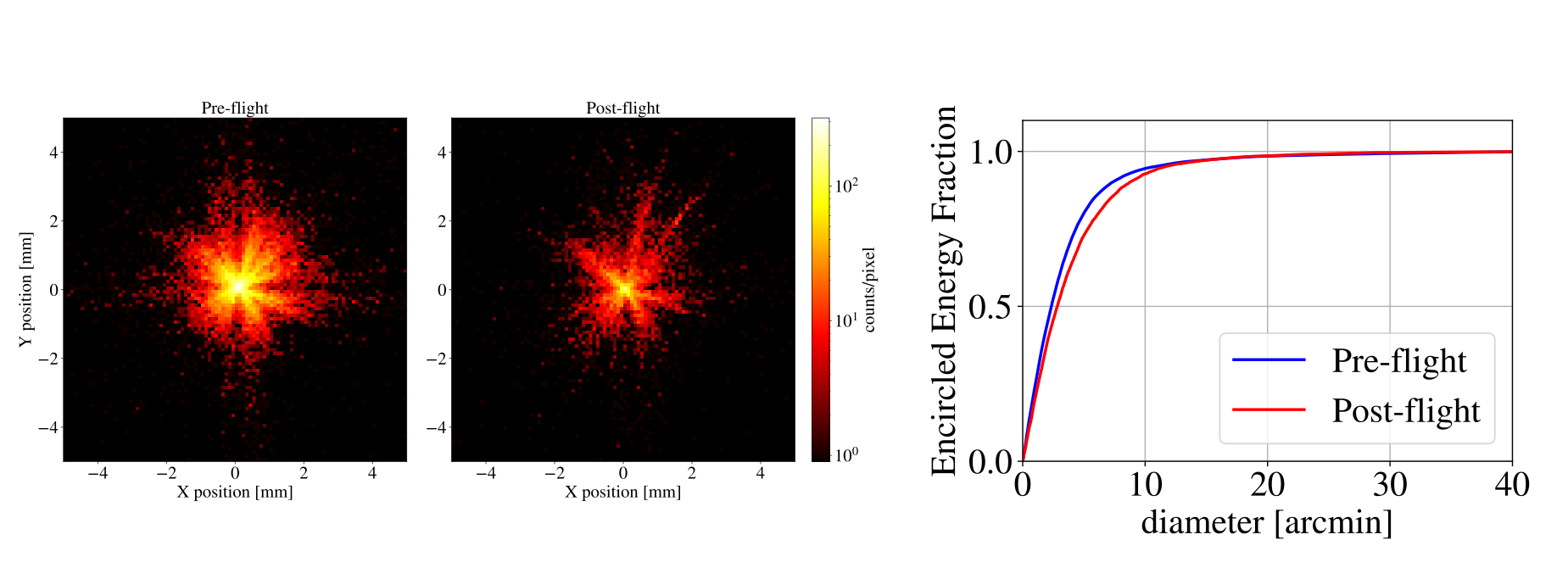}

		\caption{X-ray mirror angular resolution before (left) and after (center) flight. The heat map images show the photon distribution at the focus of the mirror during beam tests. The right image converts these distributions to an Encircled Energy Fraction (EEF), which is the fraction of photons that fall within a circle of a given radius. The EEF shows that the Half Power Diameter of the optics was not significantly affected by the flight. }
		\label{fig:optics_alignment}
		\end{center}
	\end{figure}

The angular resolution of the X-ray mirror was measured before flight, using a 1.5~keV X-ray beam. It is expressed as the Half-Power Diameter (HPD), which encloses 50\% of all focused X-rays. After correcting for beam divergence, the angular resolution of the X-ray mirror was measured at the 2.144~m focus of the mirror to be 140~arcseconds HPD, as shown in Figure~\ref{fig:optics_alignment} (left). Pre-flight vibration testing was used to verify the alignment stability of the X-ray mirror to the center of the aperture valve under flight vibration conditions. The alignment of the mirror to the aperture valve shifted by 1~arcminute, and the alignment of the star tracker to the flat reference mirror shifted by 0.5~arcminutes under a launch vibration load, meeting the alignment stability requirement of 2.5~arcminutes. Post-flight analysis of the X-ray mirror, shown in Figure~\ref{fig:optics_alignment} (right), indicated a 30~arcsecond degradation in the angular resolution, likely due to the force of impact. This degradation is consistent with expectation and is small enough that the mirror was reused for the second flight. 

%\input{sections/flight-optics.tex}
%%%%%%%%%%%%%%%%%%%%%%%%%%%%%%%%%
\clearpage
% !TEX root = ../instrument_paper_revB.tex
\section{First Flight Results}
\label{sec:Flight Performance}

\subsection{First Flight Overview}
\label{sec:flight-overview}
\begin{figure}
	\begin{center}
		\includegraphics[width=0.95\textwidth]{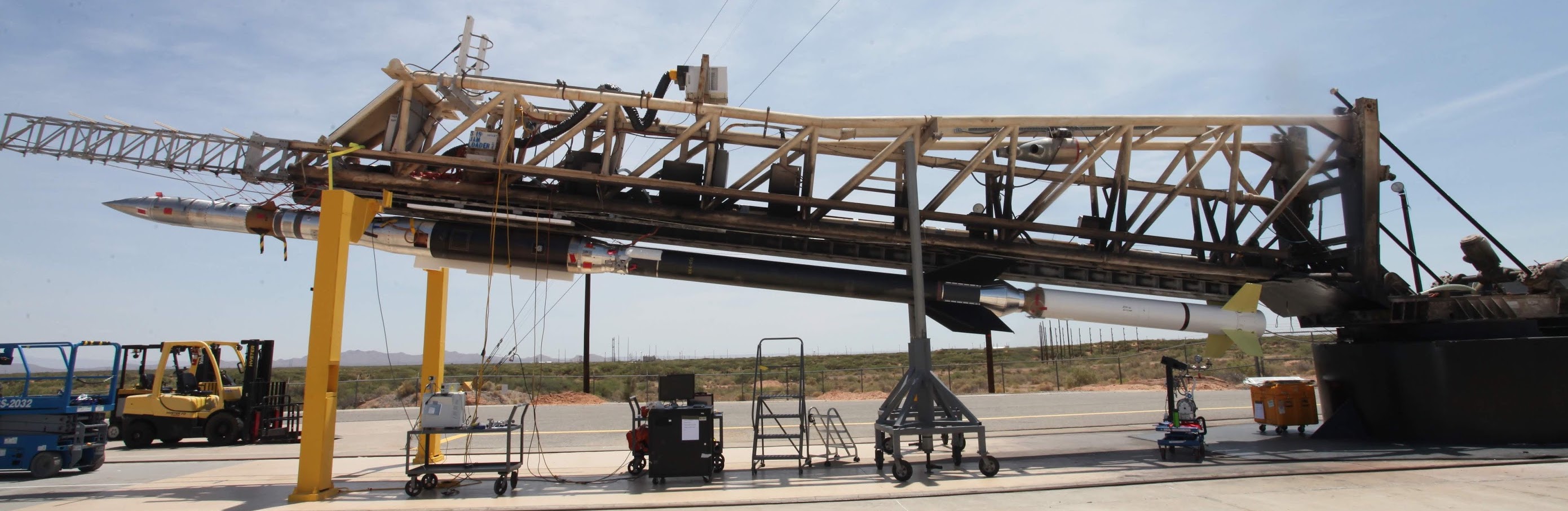}
	\end{center}
	\caption{Micro-X on the launch rail for pre-flight operations.  The two rocket motors can be seen aft of the payload. The rail was raised to 85$^\circ$ from horizontal for launch. }
	\label{fig:rail}
\end{figure}

Micro-X launched from the White Sands Missile Range (WSMR) in New Mexico at midnight on July 22, 2018. Figure~\ref{fig:rail} shows the rocket on the launch rail before flight. The target was the Cassiopeia~A SNR. An Attitude Control System (ACS) failure~\cite{GLNMAC} caused the payload to slowly spin for the duration of the flight, with no time on target. Despite this, data from the on-board calibration source was used to analyze the instrument performance in flight. The payload was recovered in good condition, with minimal changes required for the second flight. 

The optics performance could not be evaluated without time on target, but the discussion of pre- and post-flight optics measurements is provided in \S\ref{sec:Optics}. The first flight experience motivated the changes that were successfully implemented for the second flight (\S\ref{sec:Reflight}). 

%%%%%%%%%%%%%%%%%%%%%%%%%%%%%%%%%
% !TEX root = ../instrument_paper_revA.tex
\subsubsection{Operations}
\label{sec:Operations}

Operating the cryogenic Micro-X detectors under the conditions of a sounding rocket flight requires a precise combination of pre-flight and flight events, shown in Table~\ref{table:countdown}. Pre-flight operations began by preparing the cryogenic system. The LHe tank in the cryostat, initially cooled down from room temperature days before launch, was fully refilled with LHe nine hours before launch and then pumped on continuously. Once pumping began, the launch rail was evacuated for rocket arming by NSROC, after which point all operations were performed remotely from the ground station. Communication between the ground station and the launch rail passed through cables that run through the 1000~ft long underground trench between the sites. These cables mated to a set of fly-away connectors at the rocket, facilitating both commanding and direct data acquisition while on the ground. Once the telemetry antennae were turned on for flight, the flight data streams (\S\ref{Electrical Health}) were transmitted to the ground support equipment, which received, recorded, and displayed the data in real-time. This transmission continued without interruption throughout the flight. %Figure~\ref{fig:rail} shows Micro-X on the rail before it was raised for launch. 

\begin{table}
\begin{center}
	\begin{tabular}{c|l|c|l}
	\hline
		{\bf Task} & {\bf Time} & {\bf Altitude} & {\bf Event} \\
		& & {\bf [km]} & \\
		\hline
		Prepare cryogenics & T - 9:00~hr & 1.2 & LHe transfer \\ 
		  & T - 8:25~hr & 1.2 & Pump down LHe tank \\ 
		\hline
		\multicolumn{4}{c}{\textit{NSROC operations}} \\
		%Arming & T - 4:15~hr & 1.2 & \textit{Payload power off for arming}\\ 
		\hline
		%& T - 180~min & 1.2 & ADR ramp up \\ 
		%ADR & T - 125~min & 1.2 & Open heat switch \\
		%magnet cycle & T - 120~min & 1.2 & ADR ramp down \\ 
		%& T - 105~min & 1.2 & ADR regulate to 75~mK \\
		%\hline
		& T - 180~min & 1.2 & ADR magnet cycle \\
		& T - 100~min & 1.2 & Tune detectors \\
		Prepare instrument & T - 10~min & 1.2 & Ramp up to 300~mK \\
		& T - 6~min & 1.2 & Close LHe pumping valve \\ 	
		%& T - 2~min & 1.2 & Check go-no-go criteria for launch \\
		\hline
		\textbf{Launch} & \textbf{T = 0} & 1.2 & \textbf{Ignition}\\
		%\multicolumn{4}{c}{\textbf{Ignition (T = 0)}} \\
		\hline
		Powered flight & T + 0.1~s & 1.2 & Start writing to on-board flash memory\\
		& T + 43.5~s & 37.8 &\textit{Rocket motor burnout} \\
		\hline
		& T + 51.0~s & 52.4 & Begin ADR temperature regulation\\
		Prepare for & T + 62.0~s & 72.7 & \textit{Rocket de-spin}\\
		observation & T + 64.0~s & 76.3 & \textit{ACS align to target*} \\
	    & T + 69.0~s & 86.8 & Open LHe valve (pump on LHe)\\
		& T + 72.0~s & 90.2 & Open shutter door\\
		\hline
		& T + 111.3~s & 149.7 & Open aperture valve (open detectors to space)\\
		Science observation & \textbf{T + 256.4~s} & 245.7& \textbf{Apogee} \\
		& T + 437.8~s & 106.7 & Close aperture valve\\
		\hline
		Prepare for & T + 433.0~s & 103.1 & Close LHe pumping valve \\
		re-entry & T + 438.0~s & 94.9 & Close shutter door\\
		\hline
		 & T + 609.0~s & 4.6 & \textit{Deploy parachute}* \\
		%\cline{2-2} \cline{3-3}
		Descent* & T + 835.0~s & 1.7 & ADR ramp down*\\
		& T + 870.0~s & 1.2 & Experiment power off*\\
		\hline
		\textbf{Payload impact*} & \textbf{T + 902.0~s} & 1.2 & \\
		\hline
		%\clineB{1-3}{5}
		\end{tabular}
\caption{A combination of pre-flight and flight operations prepared the instrument for the science observation while keeping it in a protected configuration during launch and impact. Flight trajectory milestones are shown in bold and NSROC events in italics. Starred events significantly deviated from the planned timeline, as described in the text.}
\label{table:countdown}
\end{center}
\end{table}

After rocket arming, the ADR and detectors were prepared for flight. The ADR magnet cycle cooled the detectors to 75~mK at 100~minutes before launch. The SQUIDs then underwent three ``de-Gaussing cycles," in which the ADR magnet was ramped up to 300~mK and back down to 75~mK. Pre-flight testing had demonstrated that the SQUIDs exhibited a hysteretic response to the ADR magnet being ramped up and down. This effect decreased with each deGaussing cycle, with three de-Gaussing cycles rendering this hysteresis negligible. Afte de-Gaussing, the SQUIDs were then tuned. Tuning is the manual process by which the biases shown in Figure~\ref{SQUIDs} are set and the phase relationships between the detectors and SQUID stages are established. These operational settings were maintained from this point through flight. The settings were stored on the Master Control Board (MCB) in the unexpected case of a power glitch. After tuning, the ADR ramped up to its 324~mK launch temperature to maximize the available cooling capacity. At six minutes before launch, the pumping valve closed off the LHe tank. The launch criteria were assessed, including the number of operational pixels ($\ge$80), the available ADR magnet current ($>$480~mA), the parasitic heat load on the cold stage ($<$150\% nominal heat load), the LHe level ($>$10\%), the pressure in the optics section ($\le$100~mbar), and the system health monitors (predefined acceptable ranges; see \S\ref{Electrical Health}). 

After ignition, the rocket left the ground, and lanyards that were tied to the ground pulled out pins that started the clock for the on-board timers. No commanding was done from the ground during flight; all operations were pre-programmed into timers. The rocket motors burned for roughly 40~s~\cite{SRHB}, accelerating the rocket upward as it spun at $\sim$4~Hz about the thrust axis for stability. During this time (``powered flight"), the vibrational and thermal loads to the outside of the rocket were significant. All instrument valves remained closed during this period, and the temperature of the detector array was not controlled. 

Immediately after rocket motor burnout, the rocket de-spin system engaged to stop the payload from spinning, and the ACS attempted to align the instrument with the science target. The shutter door then opened, exposing the optics to space. The LHe pumping valve opened to allow the vacuum of space to pump on the bath, and the ADR started actively regulating to 75~mK. The ADR control system maintained this temperature for the rest of the flight. At an altitude of 160~km, the aperture valve opened, allowing any X-rays from outside the instrument to reach the detector array; this occurred 67.8~seconds after rocket burnout. X-rays from the on-board calibration source continuously illuminated the array for the duration of the flight. During the science observation, the ADR is designed to regulate its temperature to within $\pm$10~$\upmu$K, and the ACS is required to maintain a pointing precision on the target of $<$0.25~arcminutes. 

Under a normal flight trajectory, the science observation continues until the payload falls below 160~km. Below this altitude, the aperture valve, the pumping valve, and the shutter door to the optics are closed for atmospheric re-entry. At 15,200~ft (4.6~km), the parachute deploys to slow the descent. Directly before impact, the ADR magnet is ramped down and the experiment is powered off. In this flight, the descent was anomalously fast due to the ACS failure, and impact occurred before the scheduled ADR ramp down and instrument power down. 
%%%%%%%%%%%%%%%%%%%%%%%%%%%%%%%%%
% !TEX root = ../instrument_paper_revA.tex
\subsubsection{Flight Environment}
\label{NSROC Systems}

\begin{figure}
	\begin{center}
		\includegraphics[width=\textwidth]{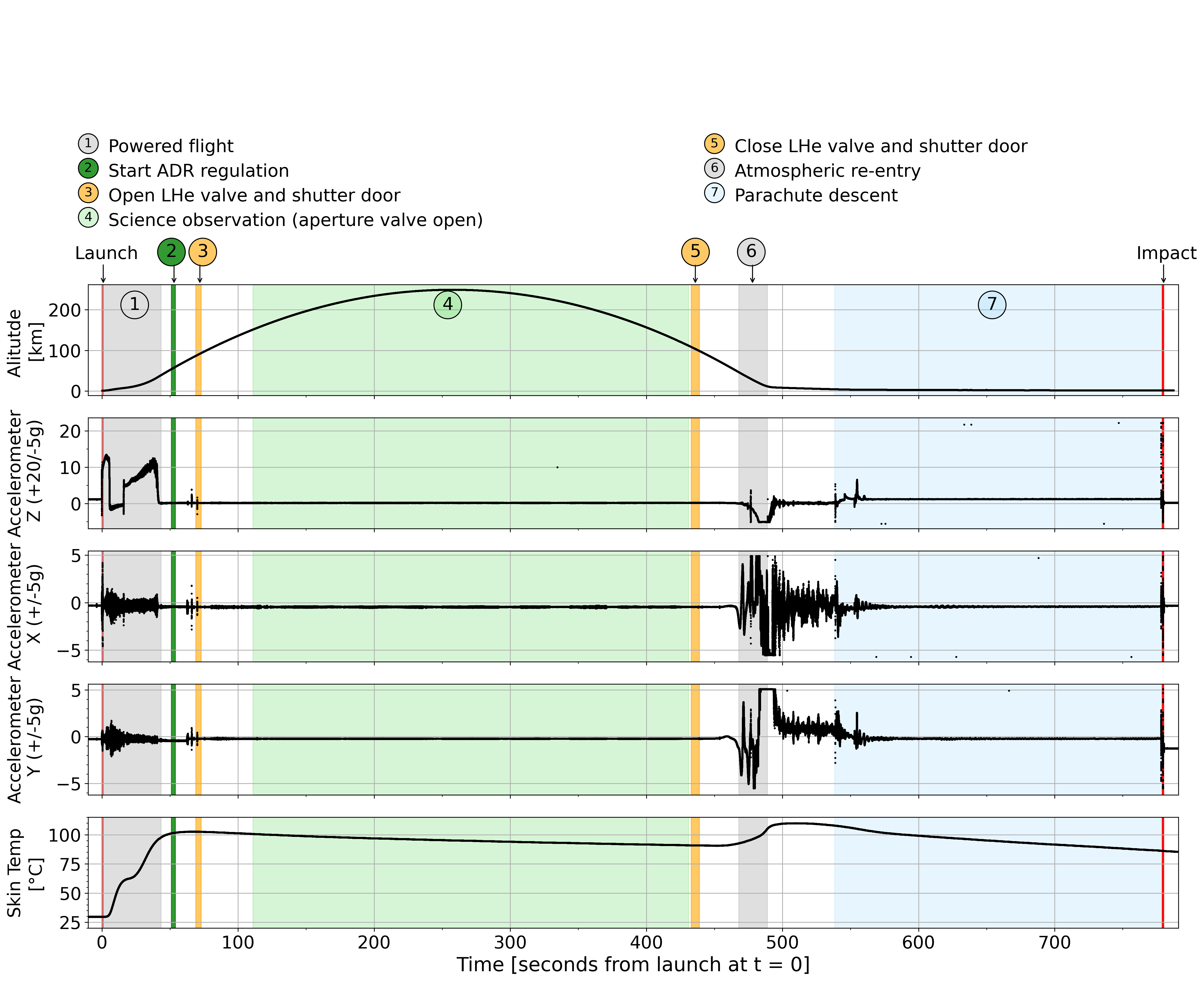}
	\end{center}
    \caption{The rocket trajectory was nominal through the end of the science observation. The payload was above the required 160~km altitude for 279.9~s, reaching a maximum altitude of 249.2~km (top). It reached its maximum vertical velocity of 2003.2~m/s at the end of powered flight. The ACS failure created an anomalously fast descent, but the instrument was recovered with no significant damage. Powered flight, atmospheric re-entry, and impact were the three most significant contributions to energy coupling into the instrument, as measured by accelerometers (from top: z-direction (thrust), x-direction, y-direction) and the rocket skin temperature (bottom). Powered flight induced 13.4$g$ peak acceleration in the thrust axis. After the science observation, re-entry saturated all of the 5$g$ accelerometers, while impact saturated all accelerometers, including the 20$g$ sensor.}
	\label{Radar-general}
\end{figure}

The flight lasted for 780~seconds from launch to impact, and the mechanical and thermal environments were monitored for the entire flight. The trajectory, with the significant exception of the pointing direction, was nominal through the end of the science observation. The payload reached a maximum altitude of 249.2~km and a horizontal range of 73.3~km. Figure~\ref{Radar-general} shows that it was above the required 160~km altitude for 279.9~s. It experienced a maximum vertical velocity of 2003.2~m/s on the upleg and 1925.0~m/s on the downleg. The ACS failure created an anomalously fast descent after the trajectory check at T+440~s. The parachute deployment, actuated by an altitude switch, occurred at the correct altitude and 70.6~s early. Impact occurred 122.2~s early, but the instrument was not damaged. 

The thermal isolation scheme successfully minimized thermal coupling into the electronics and optics sections during flight. As shown in Figure~\ref{Radar-general}, thermometer on the rocket skin in the telemetry section measured the skin to be 29$^\circ$C (85$^\circ$F) at launch and up to 104$^\circ$C (216$^\circ$F) during powered flight. It cooled to 91$^\circ$C (195$^\circ$F) during the science observation, and jumped up to 110$^\circ$C (230$^\circ$F) during atmospheric re-entry. No thermal damage to the electronics or the optics components was observed. 

The vibration isolation system successfully isolated the cold stage from vibrational loads. Three accelerometers in the telemetry section monitored the vibrational load on the payload, with coordinates as shown in Figure~\ref{InstrumentImage}. The readout range was [-5$g$, +5$g$] in the x- and y-directions, and [-5$g$, +20$g$] in the thrust z-direction. The two rocket motors were ignited sequentially during powered flight. As shown in Figure~\ref{Radar-general}, the payload saw a vibration load of 7.4~g$_{rms}$ during powered flight, and the first and second stage motors produced a maximum acceleration of 13.4$g$ and 12.5$g$, respectively. The science observation was comparatively calm, with a maximum acceleration of 0.25$g$ in the thrust axis and 0.28$g$ in the radial direction. Re-entry was more violent than launch in the radial direction, saturating both 5$g$ accelerometers. Impact saturated all accelerometers, notably exceeding 20$g$ in the positive thrust axis. Mechanical damage was observed on three components after flight: the crush bumper, which compressed as intended to absorb the force of impact; all bolts at the joint of two rocket skins, which were sheared when the payload laid down during impact; and three of the wire rope isolators that mount the cryostat to the rocket skin, which were disconnected, as described in \S \ref{Cryostat and ADR}. No other mechanical damage was found on the instrument.
%%%%%%%%%%%%%%%%%%%%%%%%%%%%%%%%%
% !TEX root = ../instrument_paper_revB.tex
\subsection{Operations and Instrument Electronics Flight Performance}
\label{Electrical Health}

The instrument electronics successfully received timer commands from telemetry, maintained instrument operating conditions, and read out data. All data (both transmitted and recorded on board) was recovered. Data continuity, measured with embedded checksums and timing information, was 99.86\% for the on board flash data and 99.44\% for the data telemetered after T=0. On-board recording is programmed to stay on until telemetry power is turned off or the 16~GB memory is filled (equivalent to $\sim$32~minutes for the full data stream). The telemetry command for on-board data recording dropped out to one science chain for a short ($<$1~$\upmu$s) period between the science observation and the parachute deployment. The flash memory recording is designed to handle interruptions, and it successfully recovered to record the remainder of the flight. The risk of this interruption was mitigated for the second flight (\S \ref{sec:Reflight}). 

Housekeeping data is incorporated into the readout streams of the science chains and the ADR controller. It includes over 100~values to monitor the health of the instrument and the electronics themselves. These include command monitors, voltage monitors, temperature monitors, sync words, and checksums, among others. Separate housekeeping streams track the power draws, pressures, and limit switches at the LHe pumping valve and aperture valve. All housekeeping data is transmitted in-flight, and this data was used for the systems analysis of the flight presented here. The housekeeping data indicated no damage to the instrument electronics in flight. Those electrical components considered susceptible to overheating were instrumented with temperature sensors, and none of them exceeded 33.2$^\circ$C, which is well within their design specifications. The power rails were nominal throughout flight; each regulated power rail supplied by the instrument power boxes was stable to within 0.5\%. As expected, power bus X (Y) battery voltage steadily decreased throughout the flight, from 24.920~$\pm$~0.002~V (29.426~$\pm$~0.005~V) at ignition to 24.474~$\pm$~0.002~V (28.277~$\pm$~0.003~V) at impact. Current draws in flight were consistent with pre-flight testing and were well within the available margin. 

The ACS failure drove the flight trajectory into an anomalously fast descent, and impact occurred before the timer commands to ramp the ADR magnet down and turn off the power to the experiment were programmed to actuate. At impact, there was 113~$\pm$~2~mA of current on the ADR magnet, and the electronics were still on. A switch in the telemetry power system tripped upon impact, shutting off power to the experiment immediately. Post-flight testing of the instrument showed no damage to the electronics, the ADR, or the detectors from this sudden loss of power. Any telemetered data received after impact is meaningless since the instrument was powered off, although the antennae continued to transmit. 
%%%%%%%%%%%%%%%%%%%%%%%%%%%%%%%%%
% !TEX root = ../instrument_paper_revB.tex
\subsection{Thermal Flight Performance}
\label{Cryostat and ADR}
\begin{figure}
	\begin{center}
	\includegraphics[width=0.9\textwidth]{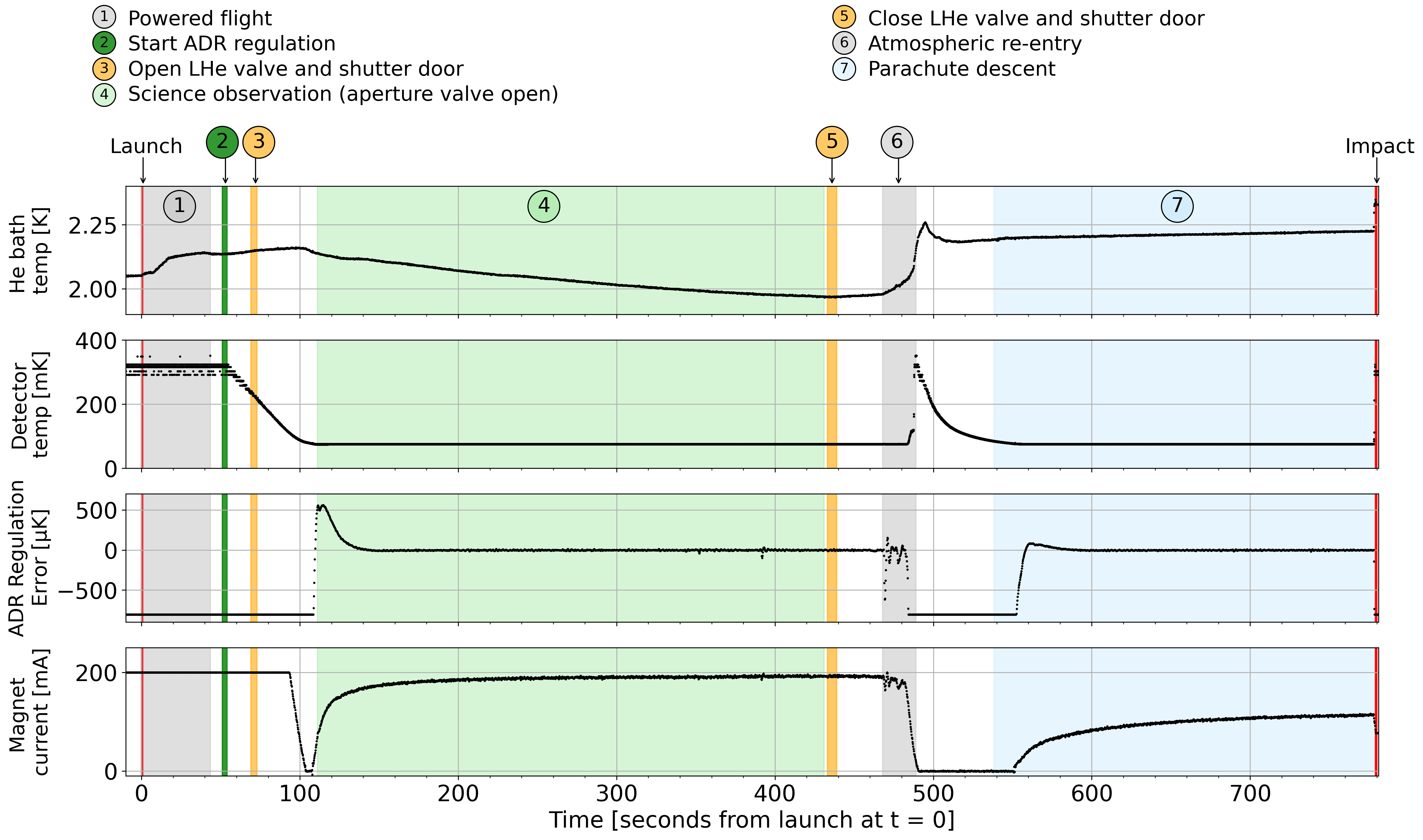}
 %{figs/flight_therm_colors.png}
	\includegraphics[width=0.9\textwidth]{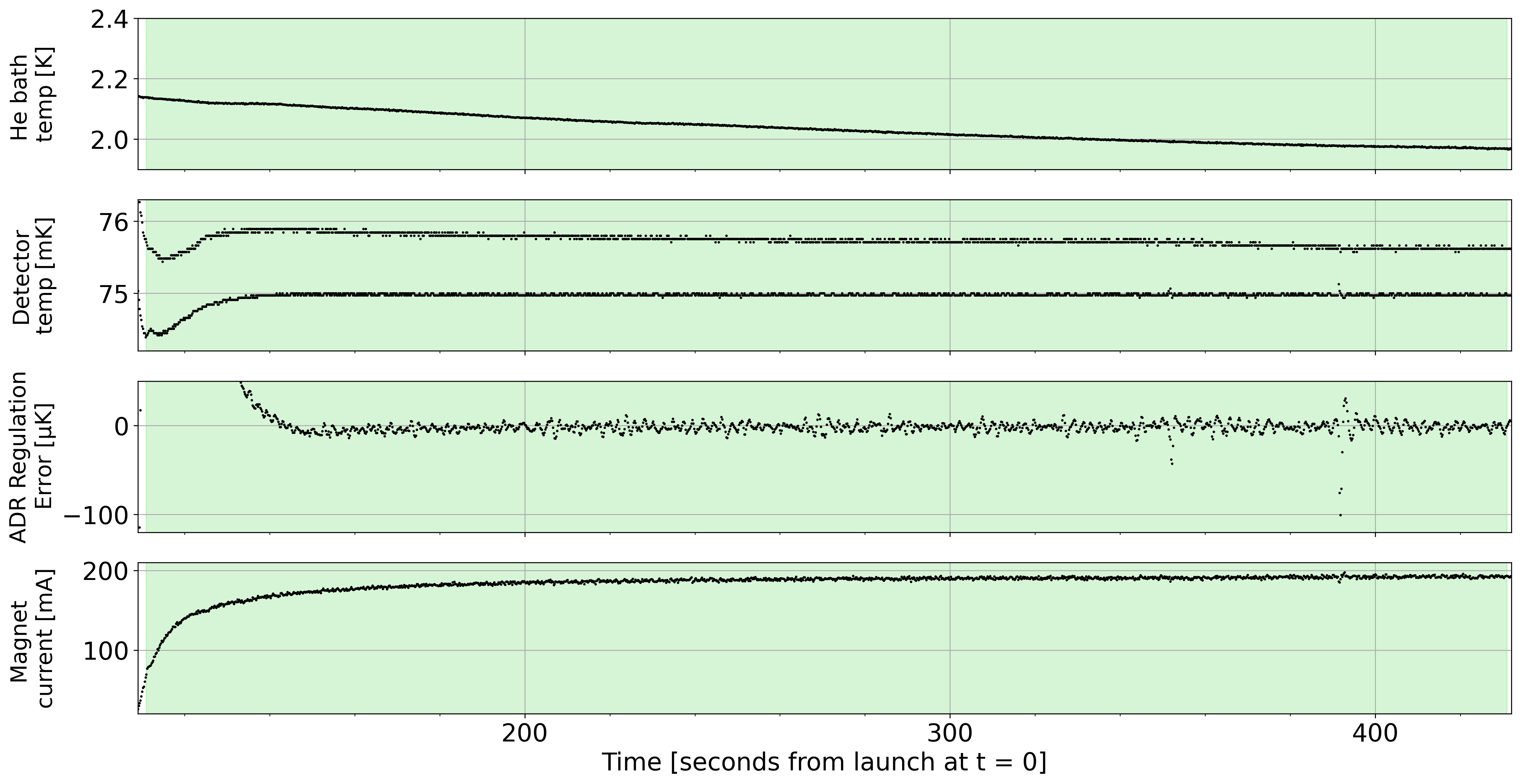}
	\end{center}
	\caption{Thermal performance of the instrument through flight (top) and during the science obervation (bottom), as measured by the (from top): LHe bath temperature, detector housing (FEA) temperature, difference from FEA temperature to its regulation set point (ADR error when in regulation), and ADR magnet current. As expected, the LHe bath temperature increased while the pumping valve was closed, and decreased while it was open to the vacuum of space. The ADR held the detector stage at 324~mK during powered flight; the apparent bit-noise of the FEA temperature signal during this phase is driven by the limited readout sensitivity of the GRT thermometers at the higher launch temperatures. The ADR then began regulating. Once stable temperature regulation was reached at T+150~s, the ADR temperature control maintained the control thermometer at 74.971~mK with $\pm$5~~$\upmu$K  (1 $\sigma$ rms) precision. The monitor GRT, also shown in the detector temperature plots, continued to cool during this period; this difference in behavior is possibly due to thermalization temperature gradients across the FEA. The ADR regulation error and magnet current signals are optimized to be high resolution during active ADR regulation periods; they saturate outside of a narrow readout range, as can be seen before active regulation and during re-entry. Their values in these periods reflect this dynamic range of the monitoring signal and are not valid. }
	\label{ADR_Temp_TM1}
\end{figure}

The vibration isolation system, described in \S\ref{sec:Shielding}, minimized heating of the detector stage in flight, as shown in Table~\ref{table:flight_thermals} and Figure~\ref{ADR_Temp_TM1}. The ADR held the detectors at a comparatively high temperature during launch (324~mK). At this higher temperature the ADR has more cooling capacity and can absorb a larger heat load from launch. It was imperative that the detector array not warm up substantially during powered flight, as there were only 60~s between the start of ADR regulation and the start of the science observation. Figure~\ref{ADR_Temp_TM1} shows that powered flight increased the temperature of the LHe stage by $\sim$100~mK, but there was no discernible change in the temperature of the detector array. The temperature of the detectors in powered flight is consistent with the temperature directly before launch to within the measurement precision of $\pm$21~mK at 324~mK of the control GRT. 

Once the payload was through powered flight and above 85~km altitude, timer commands prepared the cryogenic system for the science observation. The LHe pumping valve opened to the vacuum of space, and the temperature of the LHe tank continuously decreased, improving cryogenic performance and indicating successful pumping. The ADR took 99~s to reach a stable operating temperature, in large part due to the selected higher launch temperature. This meant a $\sim$10\% loss of science data at the beginning of observation, but the trade-off was deemed acceptable due to the increase in cooling capacity during launch. This evaluation was re-optimized for the second flight. 

From the start of stable regulation at T+150~s, the ADR maintained a stable operating temperature of 74.971~mK~$\pm$~5~$\upmu$K for the duration of the science observation, according to the control thermometer. Over this period, as shown in Figure~\ref{ADR_Temp_TM1}, the monitor thermometer on the other side of the FEA cooled from 75.892 to 75.574~mK; this difference in readings is possibly due to thermalization temperature gradients across the FEA. Two discernible temperature excursions (at T+350~s and T+390~s) during the science observation were successfully mitigated by the ADR. These events are likely to be from cosmic rays depositing minimum ionizing radiation directly in the control thermometer, as the monitor thermometer did not see these excursions. 
\setlength{\tabcolsep}{8pt}
\begin{table}
	\begin{centering}
	\begin{tabular}{lllll}
		\hline
		\bf{Time} & \bf{Stage} & \bf{He bath} & \bf{Detector Housing} & \bf{Magnet current}\\
		\hline
		%T = 0 & Launch & 2.054 $\pm$ 0.001~K & 324 $\pm$ 21 mK & 1.191 $\pm$ 0.005 A \\
		%& & & 317 $\pm$ 33 mK \\
		T = 0 & Launch & 2.054 $\pm$ 0.001~K & 324 $\pm$ 21 mK & 1.191 $\pm$ 0.005 A \\
		T + 0.1 -- 44~s & Powered flight & 2.054 -- 2.143~K %$\pm$ 1~mK 
		%& 324 $\pm$ 28 mK & 1.191 $\pm$ 0.005 A\\
		%& & & 317 $\pm$ 33 mK \\
		& 324 $\pm$ 28 mK & 1.191 $\pm$ 0.005 A\\
		%T + 111 -- 431~s & Science observation & 2.139 -- 1.967~K $\pm$ 2~mK &  74.418-- 75.129~mK $\pm$ \hl{x}~$\upmu$K & 76.6602 -- 197.852 $\pm$ \hl{x}~mA\\
		%& & & 75.440-- 75.892~mK $\pm$ \hl{x}~$\upmu$K & \\
		T + 150 -- 431~s & Stable science observation &  1.967 -- 2.109~K %$\pm$ 2~mK 
		& 74.971~mK $\pm$ 5~$\upmu$K & 171.973 -- 197.852~mA \\%$\pm$ \hl{x}~mA\\
		%%%%%%%%& & & 75.892 -- 75.574~mK %\pm$ 24~$\upmu$K & \\
		%T + 484 -- 485~s & Re-entry & 2.041 -- 2.260~K $\pm$ \hl{x}~mK & 80.284 -- 351.450~mK $\pm$ \hl{x}~$\upmu$K & 131.445 -- 0 $\pm$ \hl{x}~mA\\ 
		%& & & 83.009 -- 349.440~mK $\pm$ \hl{x}~$\upmu$K & \\
		%T + 538 -- 780~s & Parachute descent & 2.189 -- 2.242~K $\pm$ 1~mK & 74.877 -- 84.059~mK $\pm$ \hl{x}~$\upmu$K & 0 -- 116.113 $\pm$ \hl{x}~mA \\
		%& & & 75.709 -- 85.690~mK $\pm$ \hl{x}~$\upmu$K & \\
		%T + 596 -- 780~s & Stable parachute descent & 2.202 -- 2.227~K%2.20214 -- 2.22723~K $\pm$ 29~$\upmu$K
		%& 74.997~mK $\pm$ 4~$\upmu$K & 78.906 -- 116.113 \\%$\pm$ \hl{x}~mA \\
		%& & & 75.801~mK $\pm$ 4~$\upmu$K & \\
		T + 596 -- 780~s & Stable parachute descent & 2.202 -- 2.227~K%2.20214 -- 2.22723~K $\pm$ 29~$\upmu$K
		& 75.399~mK $\pm$ 3~$\upmu$K & 78.906 -- 116.113~mA \\
		\hline
	\end{tabular}
\end{centering}
\caption{Successful thermal and vibrational isolation kept excessive heat from coupling into the cold stages of the instrument. The ADR was not in stable regulation at the start of the science observation or the start of the parachute descent, so this data is not used; the thermal performance after stable regulation was achieved in each of these flight phases is shown in the table. The small difference in control temperature between science observation and parachute descent may be due to residual thermalization of the system from the heat input incurred during atmospheric reentry.}
\label{table:flight_thermals}
\end{table}

After the science observation, a significant thermal load was experienced when the payload re-entered the atmosphere; atmospheric re-entry is more similar to a mechanical impact than a gradual evolution. Re-entry exceeded the 5$g$ range of the accelerometers mounted to the telemetry section and increased the temperature of both the LHe bath and the FEA by an apparent 290~mK. The three forward-side wire rope isolators that mount the cryostat directly to the rocket skin failed in flight, likely due to a thermal failure of their glue during atmospheric re-entry. Thus the heating and acceleration incurred by the cryostat during re-entry is likely higher than if this failure had not taken place.
Post-flight testing demonstrated that the isolators begin to fail at temperatures above 60$^\circ$C, and the skin was significantly hotter than this temperature during re-entry ($\sim$95$^\circ$C, see Fig. \ref{Radar-general}). The three aft isolators remained intact, and no damage (other than the isolators themselves) was done to the instrument from this unexpected failure. The isolators have been replaced with a newer version made with glue that has been tested up to 170$^\circ$C (\S \ref{sec:Reflight}).

The ADR had enough cooling power to regain temperature control less than a minute after the heat input at atmospheric re-entry.
Once the ADR reached a stable temperature at T+596~s, it regulated to a temperature of 75.399~mK~$\pm$~3~$\upmu$K until impact. 
The ADR magnet had 113~$\pm$~2~mA of current remaining when power was shut off upon impact, which corresponds to a significant margin of over 2~hr of additional available hold time at 75~mK. 

%
%\bigskip
%
%\begin{figure} [htp]
%	\begin{center}
%		\vcenteredhbox{\includegraphics[width=\textwidth]{figs/tdas-skin.png}}
%	\end{center}
%	\caption{Skin temperature (top) and pressure measured inside the He bath (bottom). Bath pressure is still not understood; it does not match the He bath thermal behavior, which behaved as expected. }
%	\label{Skin_bath}
%\end{figure}
%
%\clearpage
%%%%%%%%%%%%%%%%%%%%%%%%%%%%%%%%%
\subsection{TES and SQUID Flight Performance}
\label{sec:flight_detectors}

The detectors observed X-rays from the onboard calibration source in flight. The count rate and energy of the observed events are consistent with the internal source, with no additional photons from astrophysical sources. Due to the ACS failure, the payload tumbled during flight at an average rate of 1.6$^\circ$/s, and did not scan through any substantially strong X-ray sources. Less than 1~photon is expected across the array from astrophysical point sources during the roughly 5~minute science observation. Up to 7~photons are expected from the Cosmic X-ray Background \cite{Hickox_2006}. The average observed count rate was 0.7~counts/pixel/s, corresponding to a K-K$\upalpha$ rate of 0.23$\pm$0.06~counts/pixel/s and matching the pre-flight rate observed from the calibration source. For the subsequent analysis, all photons are assumed to be from the calibration source.

\subsubsection{Impact of Magnetic Environment: Flight Data}
\label{sec:flight_susceptibility}

Sensitivity of the TESs and SQUIDs to magnetic fields is a known flight risk and a key difference between TES and Si thermistor systems (\S\ref{sec:Detectors}). The magnetic field environment in flight is a combination of the stray field generated by the ADR magnet and the Earth's magnetic field (\S\ref{sec:Shielding}). A three-axis magnetometer in the telemetry section monitors the Earth's field as the rocket flies. A correlation between the science chain output and the magnetometer data was observed during flight. 

The coupling of external magnetic fields to the SQUIDs is dominated by the large SQ2 effective area caused by the superconducting summing coil, as discussed in \S~\ref{sec:magnetism}. The MUX06a is a 2006 NIST TDM SQUID design, and newer designs have mitigated this parasitic coupling (see \cite{Stiehl_2011} and \S\ref{sec:Reflight}). A change in the magnetic flux through the SQ2 summing coil changes the relative phase of the V-$\Phi$ response between the three SQUID stages~\cite{Stiehl_2011,Fuhrman:2021}. For small flux changes this effect is observed as a change in the baseline of the feedback signal while the SQUIDs are locked in their flux-locked loops.  However, larger flux changes can move the phase offsets between the SQUID stages far enough from their tuned lock point that no stable lock point exists for the system. In this state, the system is considered ``unlocked," and it is unable to read out the TES signal.

The sensitivity of the TESs is difficult to disentangle from the SQUID effects when the TESs are in their transition. However, the ADR was at 324~mK through powered flight (from T+0 to T+43.5~s, see Fig.~\ref{ADR_Temp_TM1}). At 324~mK, the TESs are in their normal-state resistance and thus produce no signal; any observed magnetic response in this period is then solely due to the SQUIDs. While the magnetic field of the ADR and the detector temperature start changing upon the start of regulation at T+51~s, the ADR fields are well shielded and the detectors did not enter the superconducting transition until after the rocket de-spin at T+62~s. Therefore the primary driver of the baseline response from T+0 to T+62~s is believed to be the Earth's magnetic field coupling to the SQUIDs. 
 
\begin{figure}
    \begin{center}
    \includegraphics[width=\textwidth]{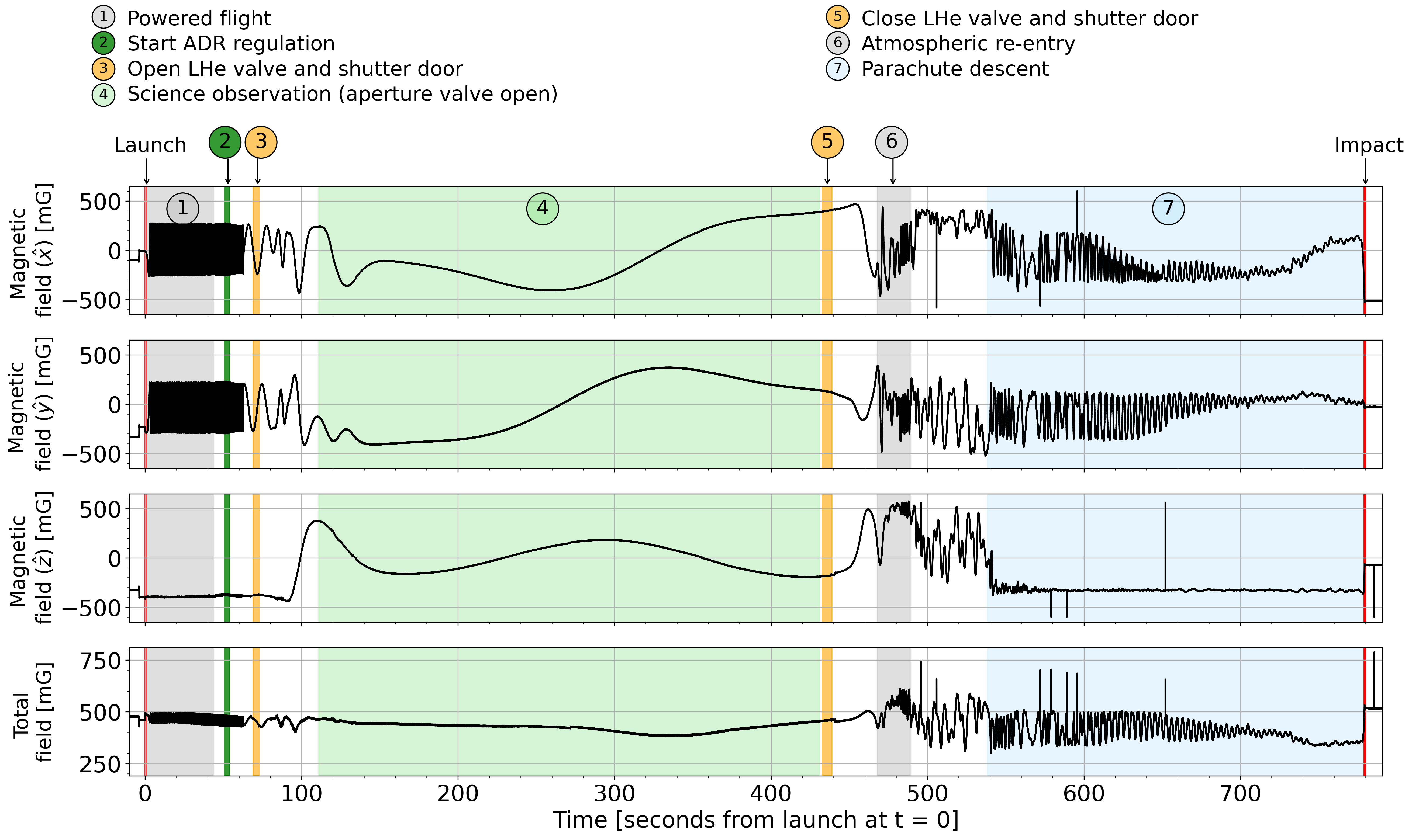}
    \end{center}
    \caption{Three-axis magnetometer data showed the rocket spinning in the Earth's field during powered flight (region 1). The payload slowly tumbled for the duration of the science observation (region 4) due to the ACS failure. The large changes in magnetic field during the rocket descent are characteristic of sounding rocket descents; their source has not been confirmed, but it is not dominated by the Earth's field.}
    \label{fig:Magnetometers}
\end{figure}

\begin{figure}
    \begin{center}
    \includegraphics[width=0.7\textwidth]{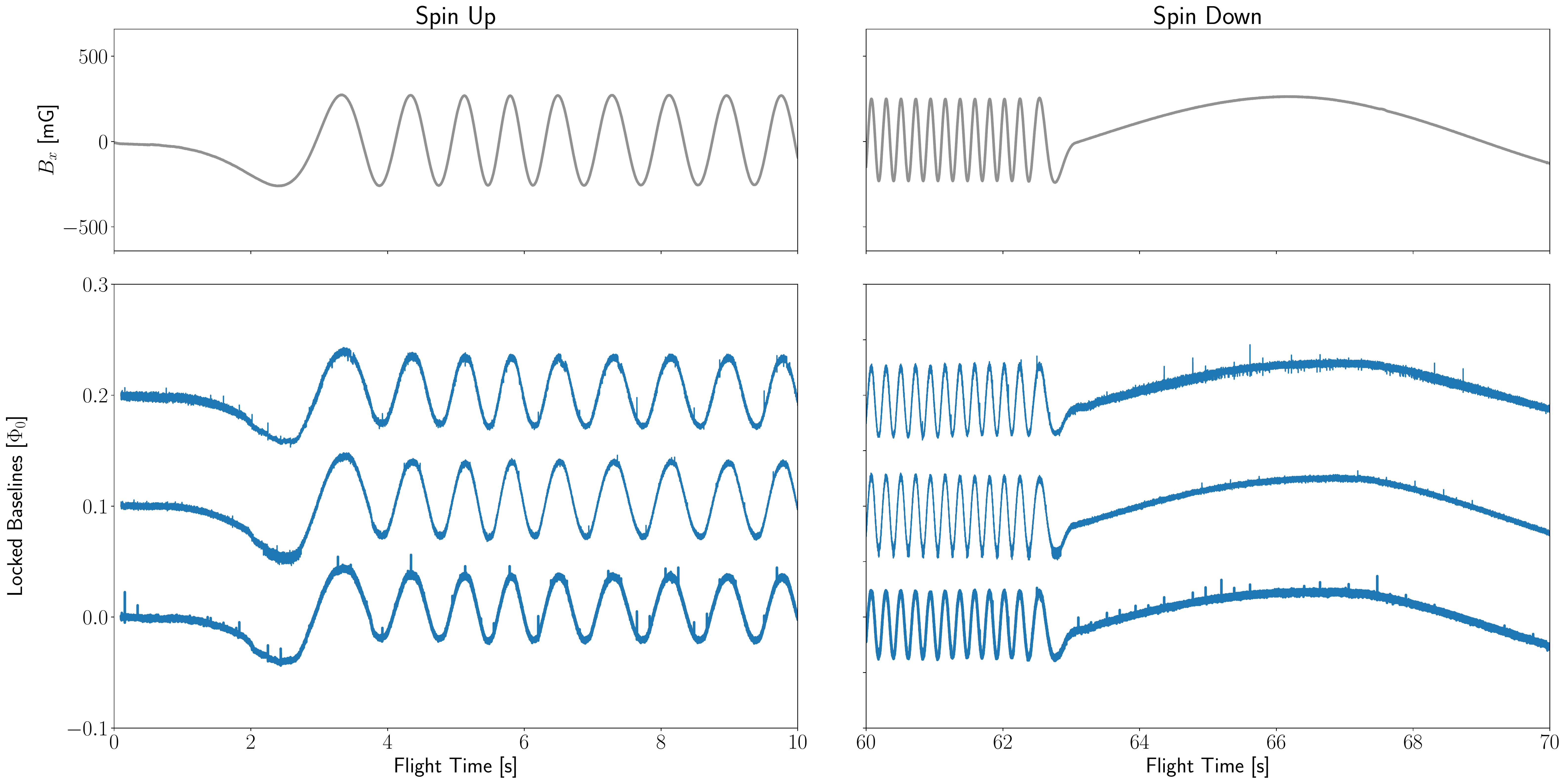}
    \end{center}
    \caption{The baselines of the side~X locked SQUID feedback signals during powered flight through de-spin (bottom panels) were strongly correlated with the Earth's magnetic field in the x-direction (top panels). The SQUID response remains correlated with the external field from the initial spin-up period at T+2~s (left) through its de-spin mechanism at roughly T+63~s (right). The feedback signals from 3~pixels are shown, with their feedback signals set to 0 at T=0 and a $0.1\Phi_0$ per-pixel offset for clarity. The magnetometer data is a subset of the data shown in Figure~\ref{fig:Magnetometers}. The TESs are in the normal state during this period. All 3~pixels are from FEA side~X; side~Y was largely unlocked during this period.}
    \label{fig:powered_flight}
\end{figure}

 During powered flight, the rocket spins about the thrust axis, accelerating from zero rotations per second at the beginning of flight to up to 4.5 rotations per second at the end of powered flight. Measured from the rocket coordinate frame, the Earth's magnetic field rotates around the thrust axis at the spin rate. This can be inferred in the x- and y-axis magnetometer data shown in Figure~\ref{fig:Magnetometers}, and directly seen in the x-axis magnetometer data shown in Figure~\ref{fig:powered_flight}, which focuses on the initial spin-up and final de-spin of the rocket. Figure~\ref{fig:Magnetometers} shows that, in powered flight, the locked SQUIDs exhibited changes in the baseline of their locked feedback signals with a strong positive correlation to the x-axis magnetic field. The locked feedback baselines tracked the external x-axis magnetic field at a rate of approximately 1.1e-4 $\Phi_{0}$ per mG. 

The field inside the superconducting shield is estimated from this data and knowledge of the effective area of the SQ2 loop (\S\ref{sec:FEA}). The SQUID MUX chips are located on the detector plane, with their loop area vectors oriented in the the rocket z-axis. Modeling of the flown NIST MUX06a SQUIDs predicts the locked feedback baseline should respond to a z-axis field inside the superconducting shield with a shift of 1.4e-2 $\Phi_{0}$ for a change of 1~mG through the SQ2 summing loop~\cite{Fuhrman:2021}. The flight data is used to calculate the ratio of the external x-axis magnetic field to the implied z-axis field inside the superconducting shield threading the SQ2 summing loop. The shielding factor (SF) is found to be approximately $SF^{\hat{x}}_{\hat{z}}\rvert_{SQ2}=120$, which is much smaller than expected, as described below in \S\ref{sec:flight_susceptibility_interpretation}. From this calculation, the baseline oscillations during powered flight can be roughly converted to a magnetic field inside the FEA with a z-component on the order of a few mG.

\begin{figure}
    \begin{center}
    \includegraphics[width=0.7\textwidth]{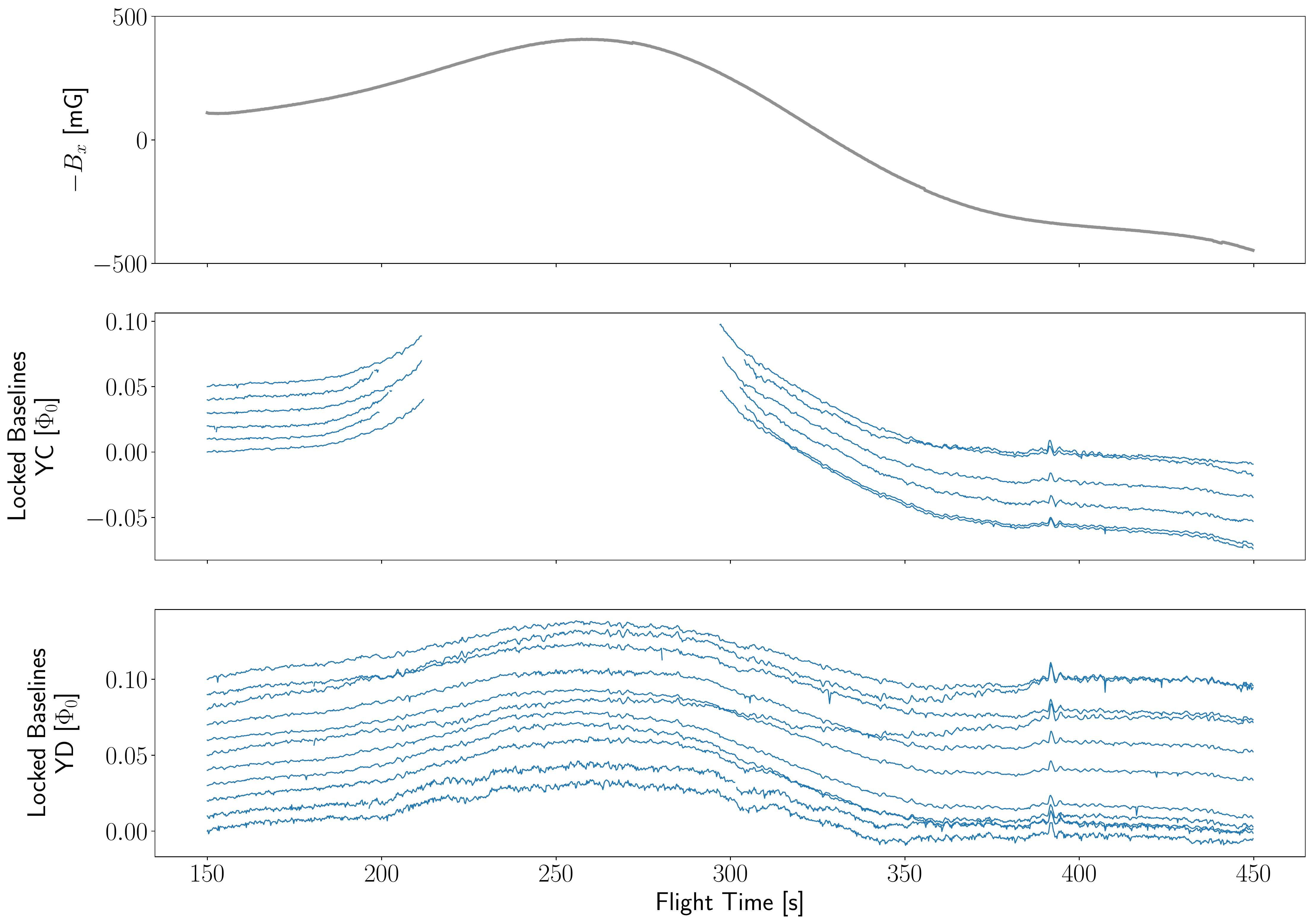}
    \end{center}
    \caption{The locked SQUID feedback signals on side Y was negatively correlated with the external magnetic field in the x-direction during the science observation (top panel). As described in the text, the SQ2s experienced varying levels of impact from the magnetic field. The pixels on column YC (middle panel), which share a SQ2, were impacted to the point of becoming unlocked, while the pixels on YD (bottom panel) remained locked but saw changes in their locked baseline levels. Note that the magnetic field sign has been switched to better show the correlation, and the SQUID feedback signals were set to 0 at $t=150s$ plus a $0.01\Phi_0$ per-pixel offset for clarity. Those portions where the baseline signal disappears in the middle panel are due to the pixels becoming unlocked, as described in the text. }
    \label{fig:squids_science_obervation}
\end{figure}

A consequence of the ACS failure during launch was the inability to point the payload during the science observation. The payload orientation drifted during this time, with the telescope pointing direction changing with an average rate of 1.6$^\circ$/s, and a maximum rate of 4.7$^\circ$/s. During this period, the locked feedback signal baselines of the side~Y pixels were observed to strongly negatively correlate with the external magnetic field in the x-direction, as shown in Figure~\ref{fig:squids_science_obervation}. Pixels on side~X were largely unlocked during the science observation, but the few that had data were observed to be positively correlated with the x-axis magnetometer, which we discuss further in the next section. Reliable shielding factors cannot be deduced from this data, however, because the locked baseline response is the sum of SQUID and TES effects, and the rocket orientation is changing in roll, pitch and yaw simultaneously. Further details of the SQUID response to magnetic fields are discussed in~\cite{Fuhrman:2021}.

\begin{figure}
    \begin{center}
    \includegraphics[width=.7\textwidth]{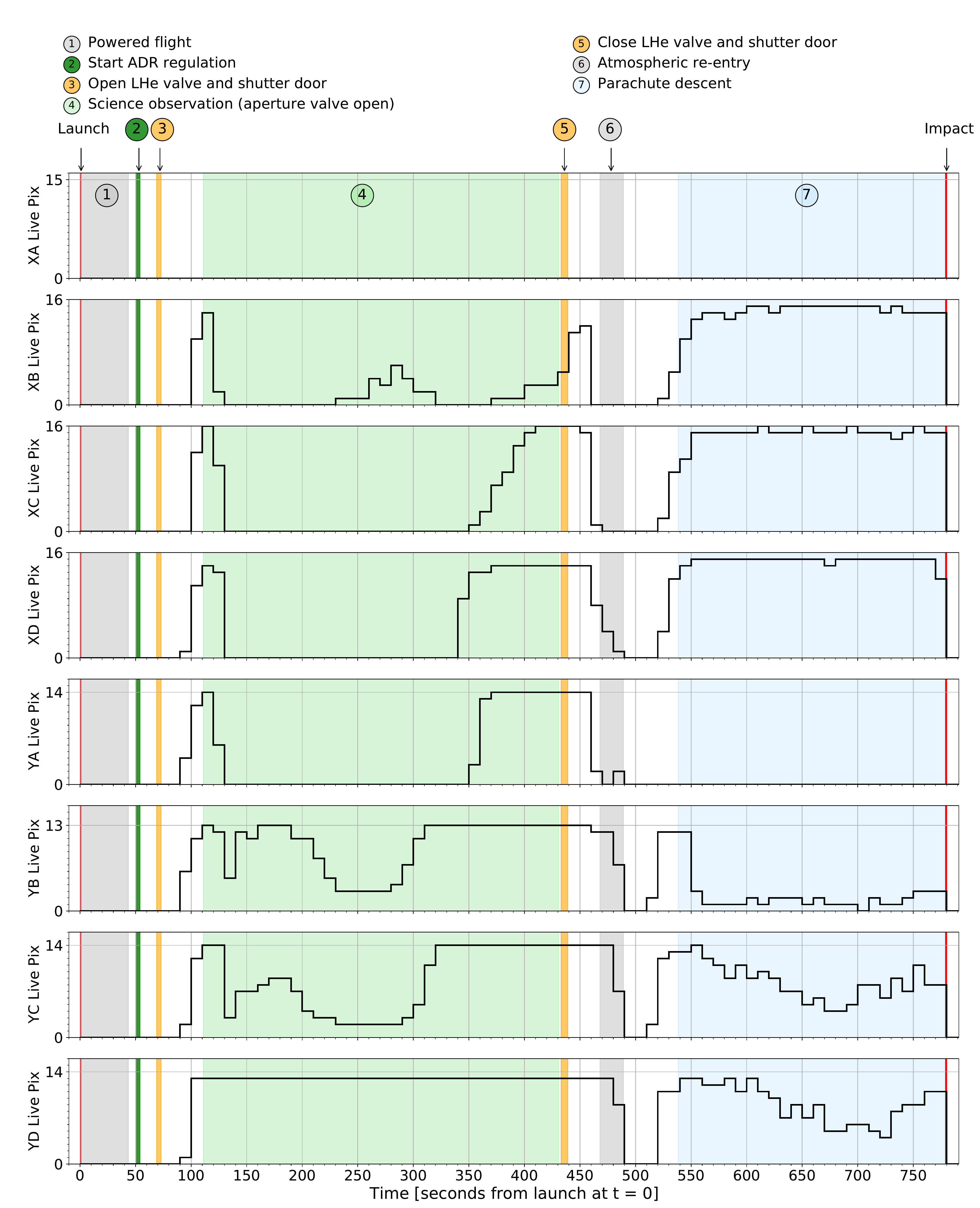}
    \end{center}
    \caption{Number of live pixels, grouped by column, throughout the flight. Pixels are considered live within a 10~s time bin if at least one X-ray is observed during that period.
    X-rays are only expected in regions 4 and 7 when the detectors are cold. Some pixels were known to be non-functional prior to flight, so the number of functional pixels per column is indicated with a horizontal gray line. 
    }
    \label{fig:Pulse_times}
\end{figure}

Many SQUIDs were observed to unlock throughout flight as the payload tumbled through Earth's magnetic field. Pixels tended to unlock and re-lock as a column when the locked baseline values were above a threshold inside the FEA. As shown in Figure~\ref{fig:squids_science_obervation}, column-to-column variations are attributed to the spatial layout of the SQUIDs inside the superconducting magnetic shield, which will be discussed more in the next section. Pixel-to-pixel variations are attributed to variation in the bias values selected while tuning the SQUIDS and variations in the individual detector response. This unlocking behavior significantly limited livetime, as shown in Figure~\ref{fig:Pulse_times}. Pixels are considered live during a 10~s time bin if they records at least one X-ray in that period. This metric requires that the TESs are at or near their operating temperature and that the full readout chain, including the SQUIDs, is functional. Given the X-ray rate of the internal calibration source,
there is a 99.9\% chance that there will be an X-ray emitted within a given 10~s window.
% a live pixel only has a 0.1\% chance to not record any X-rays during a 10~s window. \dcom{stated this way, it implies a trigger eff that we haven't calculated}
During an ideal flight, all working pixels should be live during the observation and parachute periods (regions 4 and 7 in Figure~\ref{fig:Pulse_times}). 

\subsubsection{Impact of Magnetic Environment: Post-flight Studies}
\label{sec:flight_susceptibility_interpretation}

Post-flight Helmholtz coil testing and magnetic field modeling yielded results which are consistent with the susceptibility behavior observed in flight. Helmholtz coil testing applied static fields in the three magnetometer axes individually. These tests corroborated that the SQ2 summing loop effective area is the dominant coupling between external magnetic fields and the SQUID/TES system, in agreement with~\cite{Stiehl_2011}. As observed in flight, changes to the locked baseline response were observed to be much stronger from fields applied in the xy-plane rather than in the z-direction. The baseline response during Helmholtz measurements was negatively correlated between sides X and Y, likely due to the field bending inside the magnetic shield. This agrees with the flight observations of the locked feedback baseline of pixels on side~X (side~Y) being positively (negatively) correlated with the x-direction magnetic field. 

To understand the observed behavior, a three-dimensional finite-element model of the superconducting Nb magnetic shield (\S\ref{sec:Shielding}) was made using COMSOL Multiphysics software. The magnetic field was taken from the flight magnetometer data. The lid and cone of the Nb shield were modeled individually and connected with thin, superconducting connections to model the Pb zipper tabs, described in \S \ref{sec:Shielding}. The Nb was modeled as a magnetic insulation boundary~\cite{Bergen2016}. Modeling the field inside the superconducting shield during flight is non-trivial due to the complex geometry of the Nb can, the changing direction of the external magnetic field, the off-axis location of the SQUIDs in the FEA, the possibility of trapped flux in the Nb, and the possibility that the Pb zipper tabs did not form a continuous superconducting connection between the lid and base of the Nb shield. However, the model was instrumental in understanding how the Pb tab configuration, the direction of applied external field, and the location of a SQUID within the FEA impact the magnetic field seen by that SQUID.

\begin{figure}
    \begin{center}
    \includegraphics[width=0.8\textwidth]{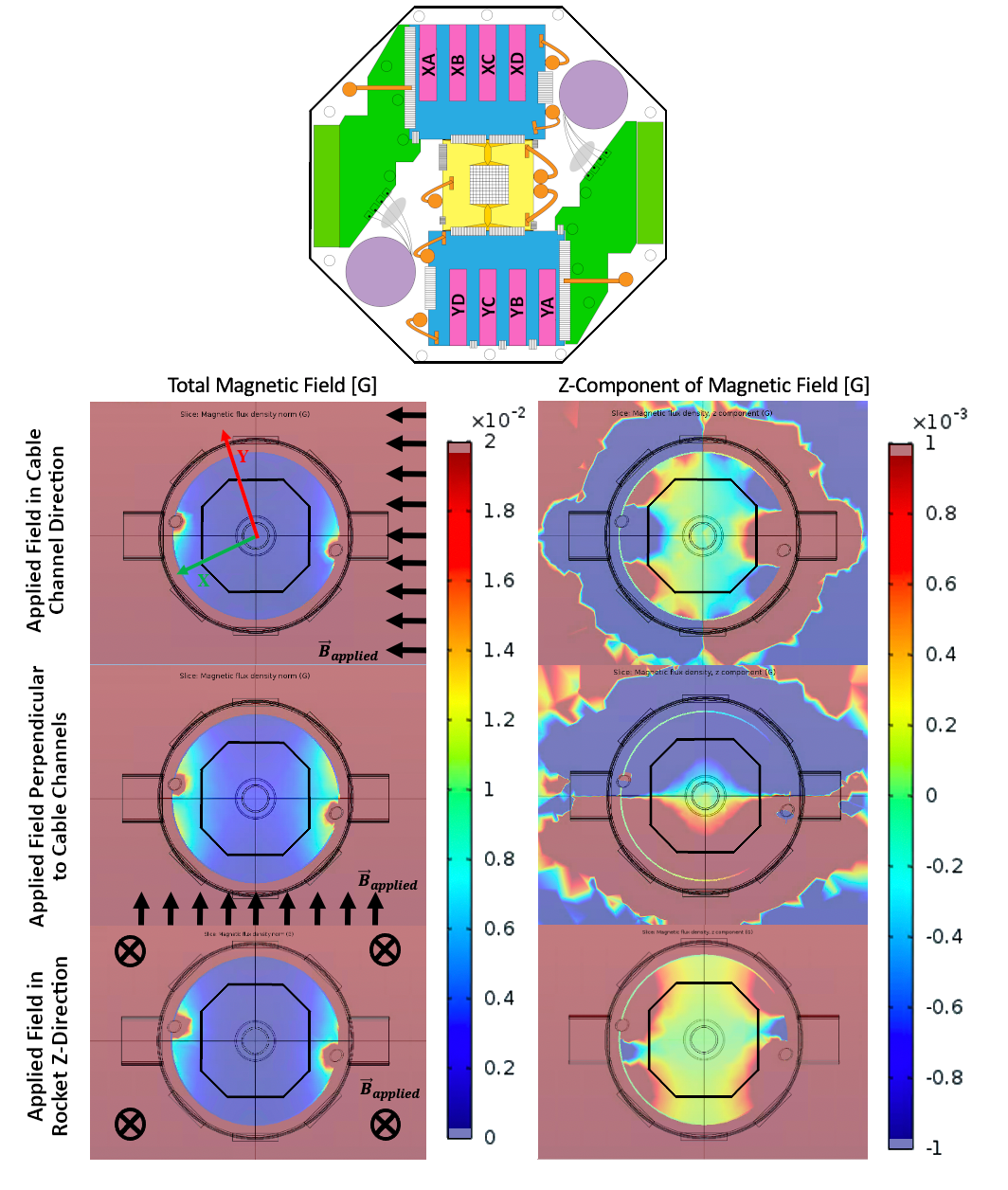}
    \end{center}
    \caption{COMSOL models of the field inside the Nb shield for an external magnetic field of 0.5~Gauss applied along the science chain cable channels (top), perpendicular to the channels (center), and in the thrust axis (bottom). These axes are rotated from the rocket x- and y-axes (used by the magnetometer and Helmholtz measurements), shown in the top left figure, by 22.5$^\circ$. The remnant fields shown are at the detector plane, which is 21 mm from the Nb can lid inside the Nb cone. The octagonal outline of the FEA is shown in the center of the Nb shield for each model and the layout of the SQUID MUX chips inside the FEA is shown in the schematic above the models (compare with Figure~\ref{fig:fea}). Results are shown for the total magnetic field (left) and the thrust axis (right). Note the different color scales between the two images.
    }
    \label{fig:COMSOLmodel}
\end{figure}

Figure~\ref{fig:COMSOLmodel} shows a COMSOL rendering of the magnetic field inside the Nb can and across the FEA for a 0.5~Gauss external field applied in the principal axes of the shield geometry. Six pieces of superconductor connect the top lid to the aft cone to model the Pb tabs. The model predicts that an external field applied in certain directions produces a residual B$_z$ with different polarity at SQUIDs on side~X and side~Y, as observed in flight and Helmholtz coil testing. Many similar models were made with different regions of superconducting connection between the lid and base of the Nb can, varying the number of such regions, their locations, and the symmetry between the two readout sides. These variables significantly change the phase between an applied field rotating about the z-axis and the predicted locked baseline responses of both sides. The particular correlation with B$_x$ observed in flight (as opposed to correlating with some other direction in the xy-plane) is believed to be due to the Pb tab shielding efficacy.

A model with six symmetric Pb tabs predicts a high shielding efficacy at the TES array for external fields originating in the rocket z-direction, $SF^{z}_{tot}\rvert_{TES}=9000$, in good agreement with the predicted shielding of the ADR magnet (\S\ref{sec:Shielding}). The model also predicts minimal $B_Z$ at the FEA; all simulated homogeneous external fields produced $SF^{\hat{i}}_{\hat{z}}\rvert_{TES}>7000$ for any $\hat{i}$. However, the model also revealed several flaws in the practical shielding of the SQUIDs. While the TES array is in the center of the Nb shield, the SQUIDs are positioned radially outward where the shielding factors are generally lower. External fields in the xy-plane especially produce larger fields at the SQUID locations, including a substantial z-direction component which can thread the summing coil loop area. The quantity and location of Pb tabs included in the model significantly varied the magnitudes of $SF^{\hat{x}}_{\hat{z}}\rvert_{SQ2}$ and $SF^{\hat{y}}_{\hat{z}}\rvert_{SQ2}$, as well as their ratio. The SQ2 shielding factors varied from $SF^{\hat{x}}_{\hat{z}}\rvert_{SQ2}<100$ in a model with no Pb tabs, to $SF^{\hat{x}}_{\hat{z}}\rvert_{SQ2}>1800$ in a model with 6 Pb tabs.

The discrepancy in shielding performance between the COMSOL model to the in flight performance is attributed to several factors. First, the COMSOL model failed to model the lateral efficacy of the Pb tabs with enough detail. The shielding factor predicted by modeling a continuous superconducting joint between the Nb can and its lid is too large, while that predicted by modeling a gap between them is too low. The exact coverage of the Pb tabs is unknown, varying with each installation of the Nb shield lid, and changing this parameter in the model changes the predicted shielding factor between the two extremes. This parameter can be adjusted to agree with observation. Second, in the lab environment, Earth's magnetic field is approximately static, so any remnant magnetic fields at the SQ2 are simply accounted for by adjusting the SQ2 flux bias. Lastly, the ACS failure highlighted this issue as the rocket swept through orientations in Earth's magnetic field where the SQUIDs came unlocked. Had the instrument been on-target during observation any SQUIDs that remained locked would have appeared to be operating normally. It is unknown how many SQUIDs, if any, would have come unlocked at the target orientation.

The magnetic sensitivity of the TESs is difficult to disentangle from the changing SQUID response. Those pixels that were locked during the science observation exhibited minimal variation in pulse shape with changing magnetic field. The largest observed response was on pixel YD08, which had a pulse height variation of 1.8e-4 $\upmu$A per mG in the x-direction. The corresponding change in pulse fall time was -2.0$\times$10$^{-4}$~ms per mG in the x-direction. Comparing the relatively small and linear change in pulse shape with the results found in~\cite{Smith:2013} for similar Mo/Au TESs suggests that the effective shielding is better for the TES array than the SQUIDs. This is consistent with the COMSOL model, which predicts a smaller residual magnetic field closer to the central z-axis. 
%%%%%%%%%%%%%%%%%%%%%%%%%%%%%%%%%
\subsubsection{Spectral Resolution}
\label{sec:flight_resolution_methods}

To understand the impact that integrating and flying the instrument had on spectral resolution, three data sets were analyzed and compared: 

\paragraph{Laboratory data:} This data was taken at Northwestern University in April 2018. Only one of the science chains was powered and read out at a time (this will be referred to as ``one-sided readout"). The instrument electronics were run on a table, physically separated from each other.

\paragraph{Integration data:} This data was taken at WSMR, 11 and 12 days before flight in July 2018. Both science chains were powered and read out at the same time (this will be referred to as ``two-sided readout"), which means that two power supplies were on, two master clocks were running, and twice as many cables were installed compared to the laboratory configuration. The instrument electronics were mounted in their flight configuration above the cryostat. In keeping with standard sounding rocket practice, electrical connectors were potted and encased in plastic backshells. Metal sheaths surrounded the power cables, with nylon sheaths on the other cables. All cables were tightly bound and laced for routing.
%The TES bias was 6500~DAC units. \com{JA: remove TES bias lines}

For the comparison in Fig.~\ref{fig:NEP_readout_config}, an extra dataset is plotted that is also from integration at WSMR. It was taken 10 days prior to flight so the instrument was in the same integrated configuration, but the science chains were powered and read out one at a time for a more direct comparison to the Laboratory data.
%The TES bias was 7500~DAC units.

\paragraph{Flight data:} This data was taken in flight in July 2018. Both science chains were powered and running, and the instrument electronics were mounted in their flight configuration. The reported data is the full data stream recovered from on-board flash memory.

All three data sets are from different operating runs of the instrument, between which the cryostat was allowed to warm up. Some amount of performance variation is expected due to the amount and distribution of trapped flux in the superconducting magnetic shield as it is re-cooled to LHe temperatures (4.2~K). The performance of each pixel is very stable between ADR cooling cycles (from 4.2~K to 75~mK) as long as the shield remains below $T_c=7$~K. Therefore, when possible, comparisons will be made between data within the same operating run. 

All data was recorded without triggering, and then played back for event selection. A derivative-based edge trigger was used to select pulses, and an inverted level trigger was used to collect noise traces from the data between the pulses. For each trigger, a trace of 4096~samples was saved, including 512~pre-trigger samples. The digitizer samples at a rate of 46.08~$\upmu$s/sample. This triggered data was then loaded into IGOR Pro (WaveMetrics) for analysis, including: data selection, Optimal Filtering (OF)~\cite{McCammon2005}, drift correction, and energy calibration. Quality cuts were used to identify traces with single pulses and noise with a stable baseline. Traces with multiple pulses or unstable baselines were rejected.

The OF uses the average pulse and the average noise record for each pixel to construct an OF template to estimate the energy of each event. The average pulse is derived from the K-K$\upalpha$ pulses, selected by pulse height. The $\upchi^2$ of each pulse to the average pulse template is calculated, and events at the edge of the distribution are removed, representing approximately 10\% of the total number of events. The same process was used to reject outlying noise samples. An additional noise cut was used for the flight data, since the effects from the SQUID unlocking required that noise traces be selected only from those times when the SQUIDs were locked. Comparatively stable periods in which the feedback signal baseline was not rapidly changing due to system unlocking were selected by a combination of the feedback signal baseline level and the baseline slope. The template was then convolved with each pulse, returning the uncalibrated energy estimate for that event. %A rough energy calibration using the pulse height and the known energy of the template pulse was used to scale the data.

The average pulse and noise are also used to predict the energy resolution of each pixel by using the noise-equivalent power (NEP). The NEP is a measure of the system noise at each frequency compared to a reference response from a known energy input. The estimate of the baseline resolution is given by: 
\begin{equation}\label{resolution_definition}
\Delta E_{NEP} = 2 \sqrt{2 \ln 2} \left( \int_{0}^{\inf} \frac{4~ df}{{NEP}^2(f)} \right) ^{-1/2}
\end{equation}

where the NEP is defined as the noise divided by the square of the responsivity of the system~\cite{moseley1984}. The NEP is: 
\begin{equation}\label{nep_definition}
{NEP}^2(f) = \frac{J_n(f)~E_{cal}^2}{{\Delta t}^2~S_p^2(f)}
\end{equation}
where $J_n(f)$ is the average power spectral density of noise traces, $E_{cal}$ is the energy of the calibration line used to create the average pulse, $\Delta t$ is the digitization time step, and $S_p$ is the Fourier transform of the average time pulse.
This estimate, measured in eV, applies a linear energy scale from zero to the known energy input, which was the K-K$\upalpha$ line at 3.3~keV. 

\begin{figure}[htb]
\centering
    \includegraphics[width=0.9\textwidth]{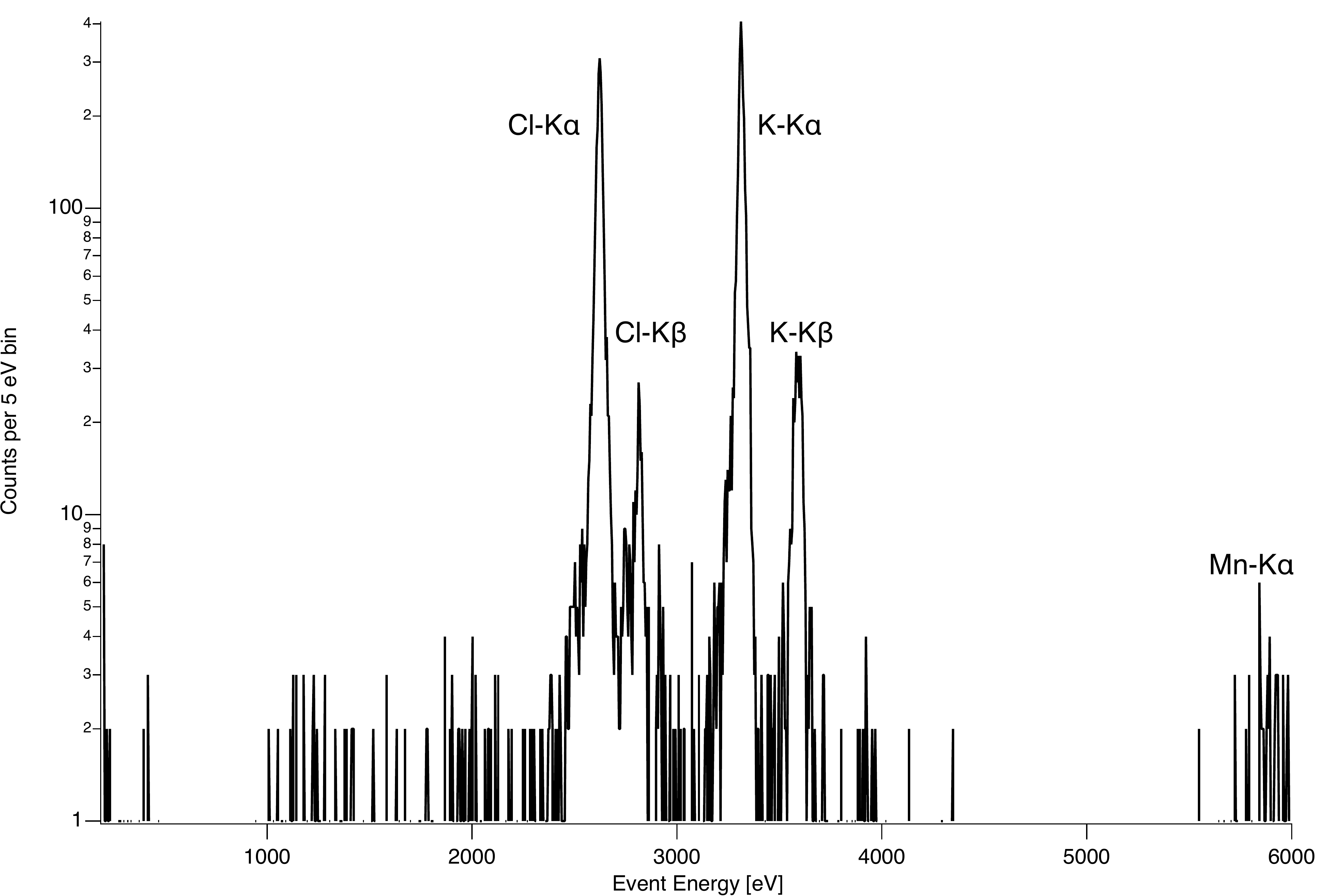}
    \caption{Energy spectrum from the science observation window of flight with all operational pixels co-added. All labeled peaks are from the on-board calibration source.}
    \label{fig:flight_spectrum}
\end{figure}

The baseline resolution of each pixel is a measurement of the intrinsic detector system noise. It is measured by applying the optimal filter to gain-calibrated noise traces and calculating their full-width at half-maximum (FWHM) around 0~keV.
The estimated baseline resolution using the NEP ($\Delta E_{NEP}$) agreed with the FWHM measurement of the baseline noise peak to within 10\% in all three data sets. When low statistics preclude a full spectral analysis, the $\Delta E_{NEP}$ is used as a proxy for energy resolution. Eleven pixels from the flight data had a much higher level of broadband noise and more pronounced lines that resulted in $\Delta E_{NEP}$ greater than 30 eV. The noise from these pixels was deemed pathological and they were excluded from the co-added spectrum in Figure~\ref{fig:flight_spectrum} and the scatter plots in Figures~\ref{fig:scatter_lab_int} and \ref{fig:scatter_int_flight}.

When there are enough statistics to do a spectral resolution analysis, IGOR Pro is used to implement baseline corrections, lagphase corrections, drift corrections, and gain calibrations before fitting the spectral resolution. The baseline correction uses the mean of the pre-trigger as the baseline value for each trace and applies a linear correction to a scatter plot of the OF energy of the K-K$\upalpha$ line vs baseline value to calibrate out any correlations between baseline and OF; this correction is not complex enough to account for the payload tumbling in the Earth's magnetic field during flight. The lagphase correction fixes any biases in the energy estimate due to the arrival phase of photons relative to the digital sampling~\cite{fowler2016,adams2009,ceballos2019}. The drift correction calibrates out variations in gain by fitting a spline to the measured energy of the K-K$\upalpha$ photons in time. The limited statistics from flight precluded the drift correction from being able to account for the changing magnetic field. Finally, the gain calibration fits each calibration line that is visible in the data (Cl-K$\upalpha$, Cl-K$\upbeta$, K-K$\upalpha$, K-K$\upbeta$, and Mn-K$\upalpha$) to get the observed energy and then calibrates the gain as a function of expected energy with an n-degree polynomial, where n is the number of calibration lines fit minus one. The co-added flight spectrum from all pixels during the stable science observation is shown in Figure~\ref{fig:flight_spectrum}. 

\begin{figure}[htbp]
\centering
\begin{tabular}{ l|l|l|l| } 
    & Laboratory Data & Integration Data & Flight Data \\
    \hline
    \rotatebox{90}{Baseline noise} & \includegraphics[width=0.3\textwidth]{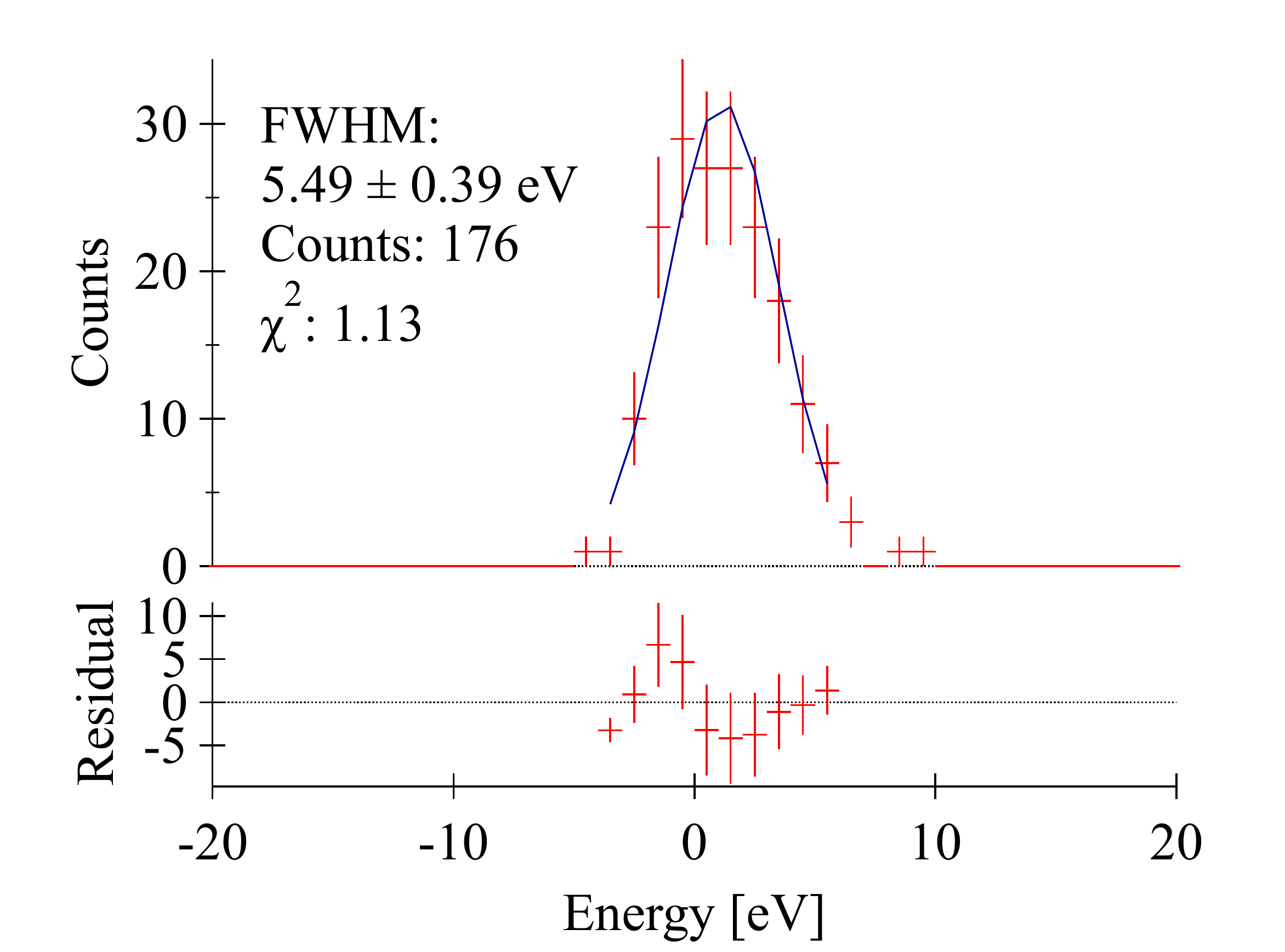} & \includegraphics[width=0.3\textwidth]{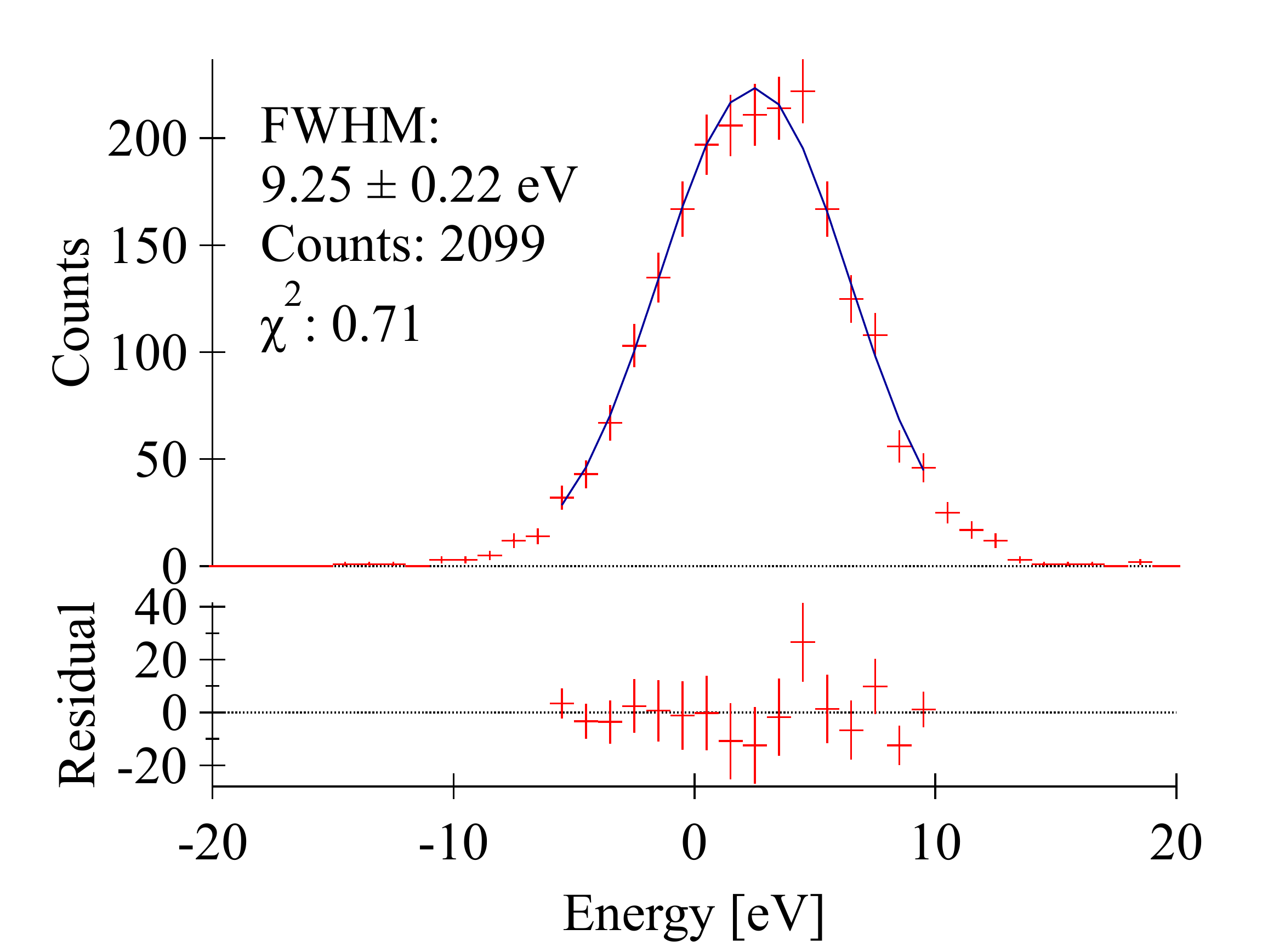} & \includegraphics[width=0.3\textwidth]{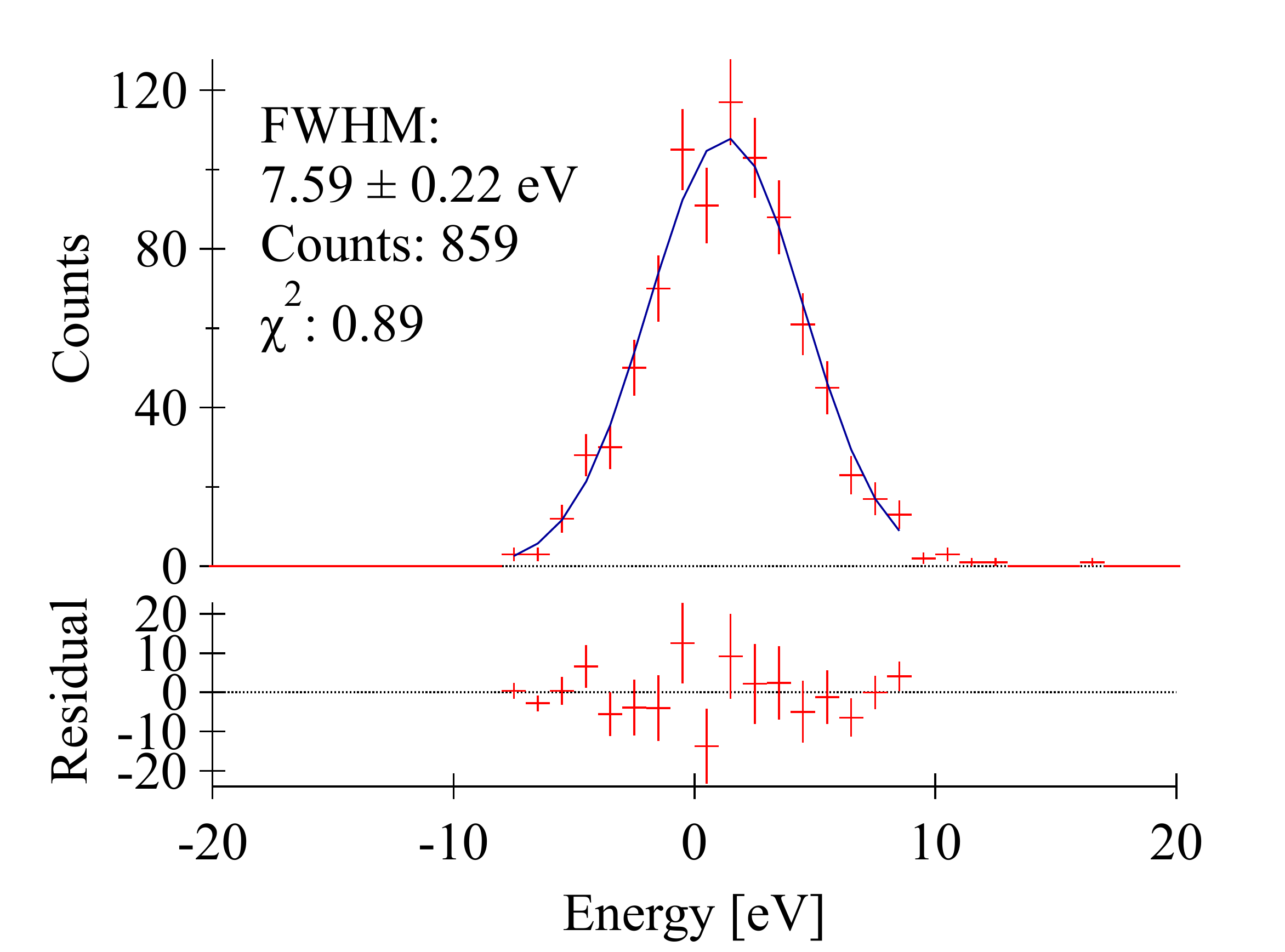} \\ 
    \hline
    \rotatebox{90}{Cl-K$\upalpha$ line} & \includegraphics[width=0.3\textwidth]{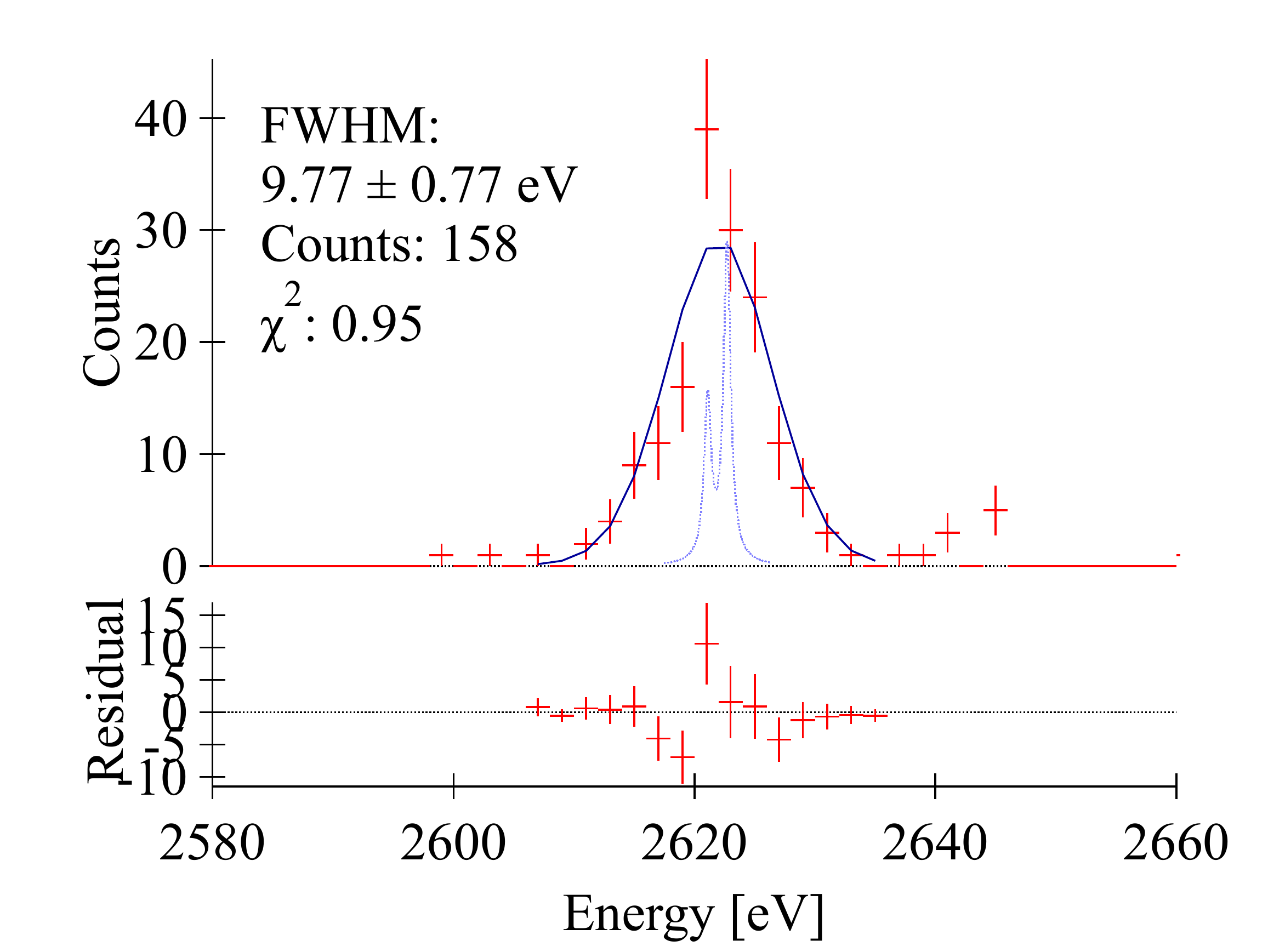} & \includegraphics[width=0.3\textwidth]{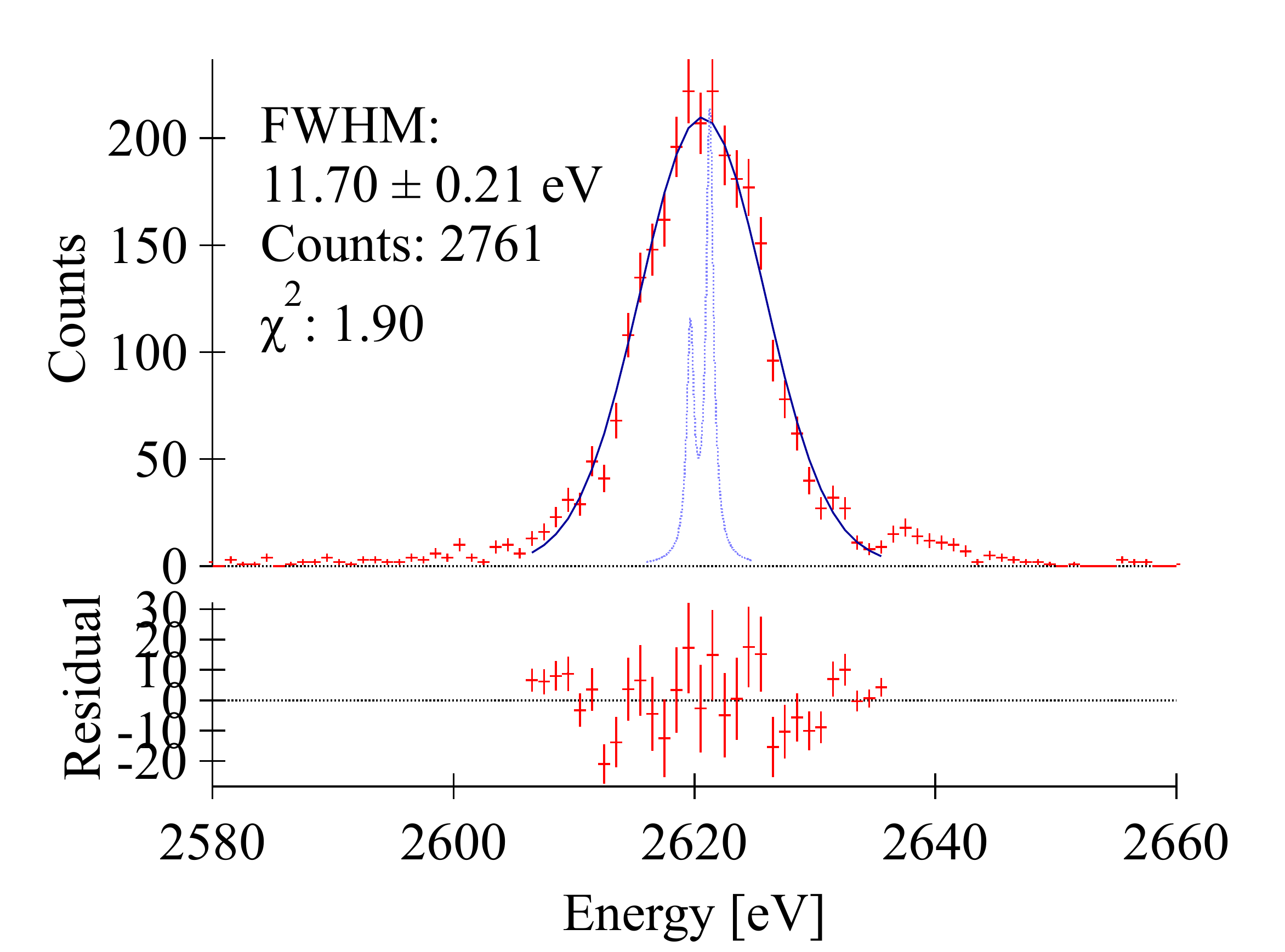} & \includegraphics[width=0.3\textwidth]{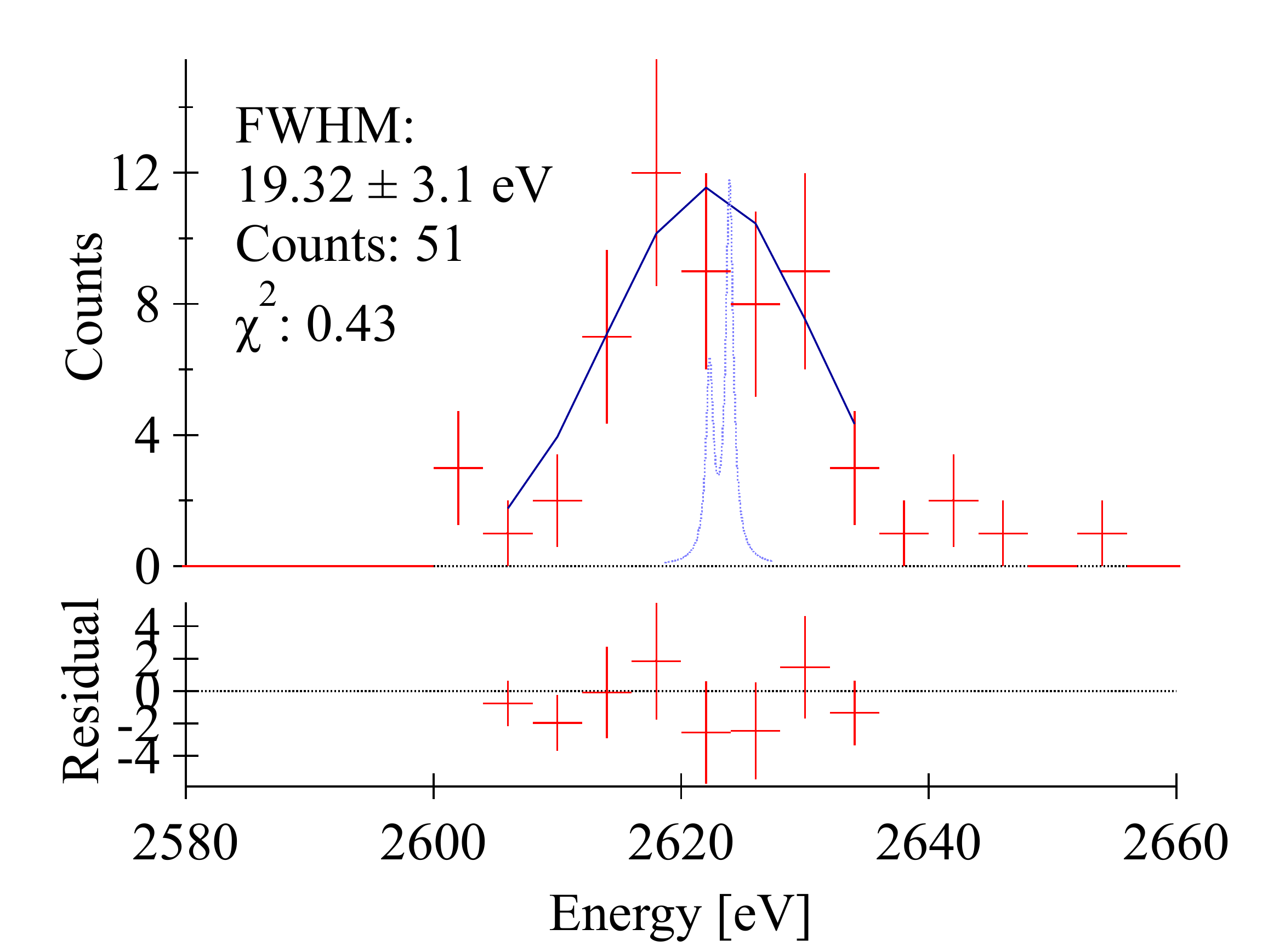} \\ 
     \hline
     \rotatebox{90}{K-K$\upalpha$ line} & \includegraphics[width=0.3\textwidth]{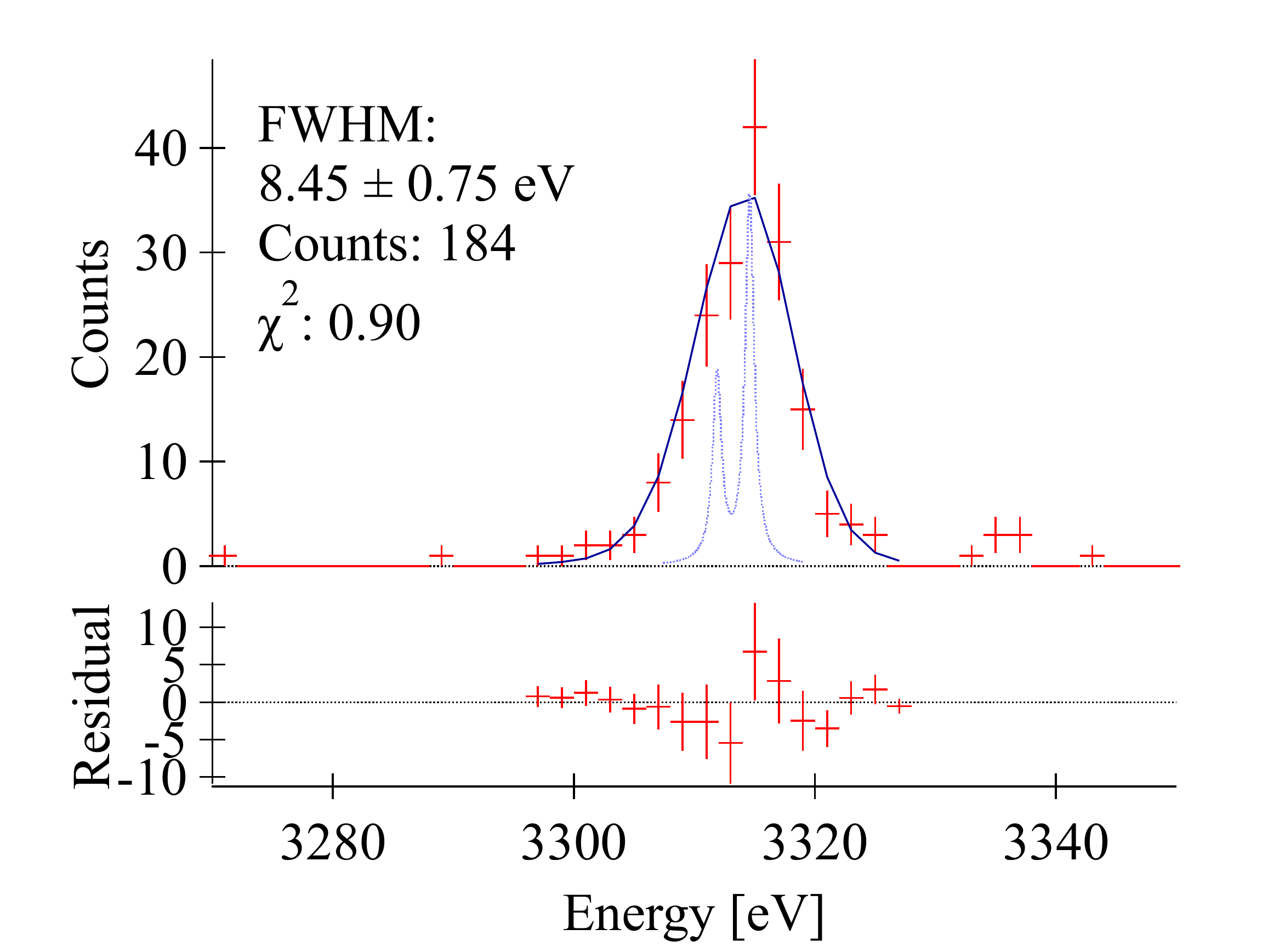} & \includegraphics[width=0.3\textwidth]{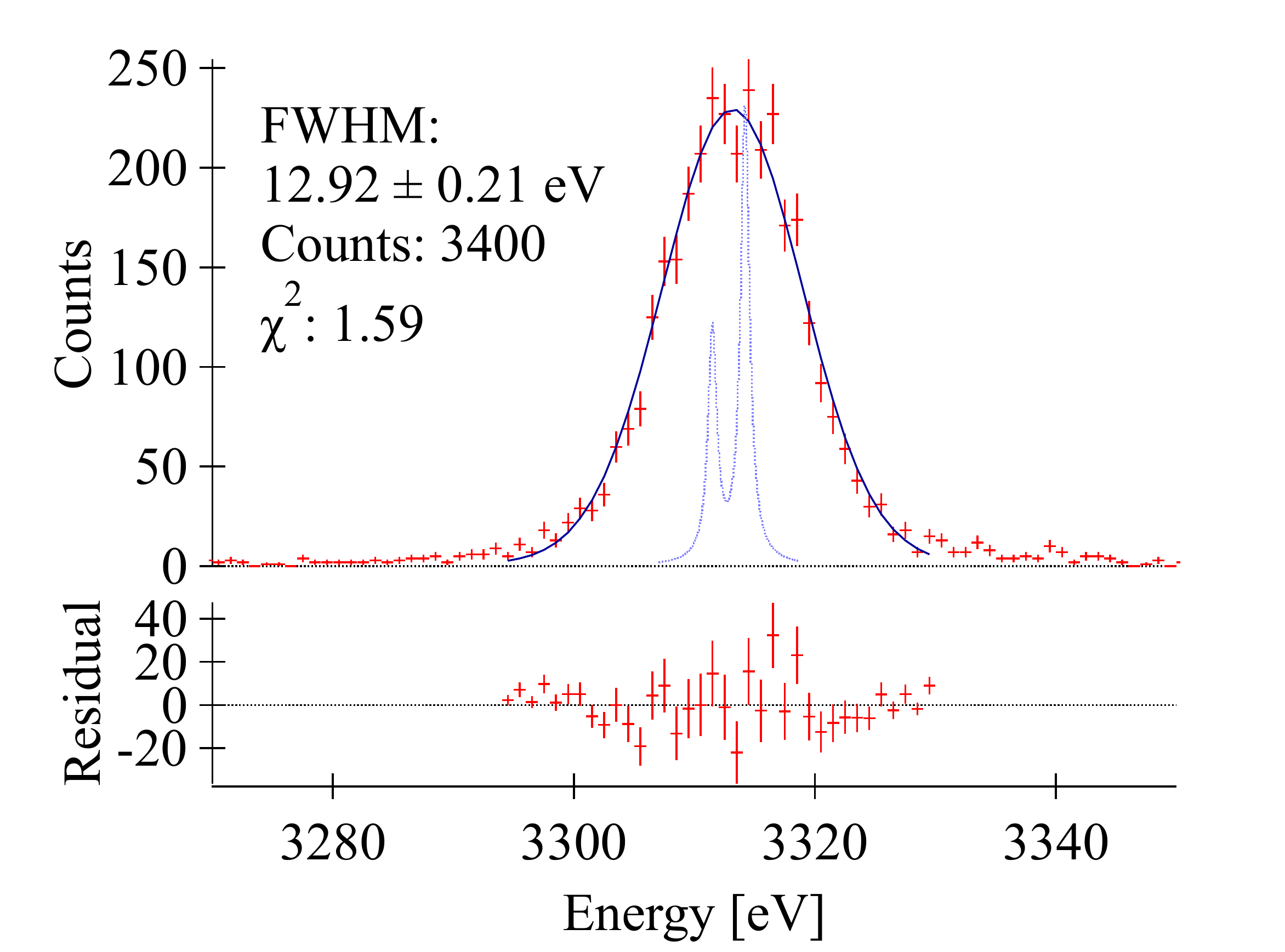} & \includegraphics[width=0.3\textwidth]{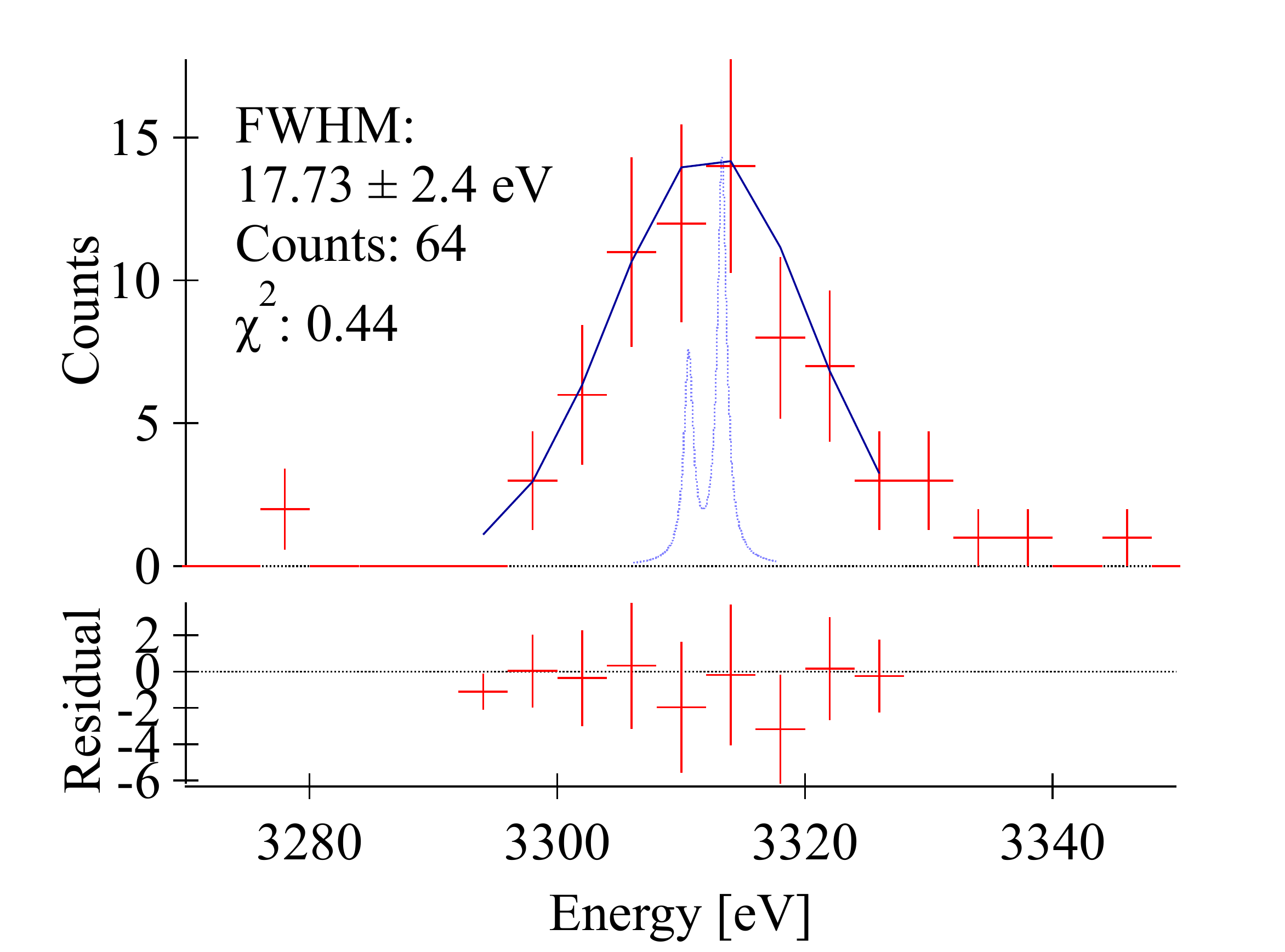} \\ 
     \hline
\end{tabular}
    \caption{Spectral line fits for a single example pixel (YD09) that was one of the best performing pixels -- good resolution with more statistics than most pixels -- for all three datasets. In all figures, the data is in red, the fit to the data is in dark blue, and the line profile is in light blue. The left column is laboratory data, the center column is integration data, and the right column is flight data, as detailed in the beginning of \ref{sec:flight_resolution_methods}. The top row is a fit to the baseline noise events centered on 0~keV, the middle row is a fit to the 2.6~keV Cl-K$\upalpha$ line, and the bottom row is a fit to the 3.3~keV K-K$\upalpha$ line. The residual between the data and the fit is shown at the bottom of each plot.}
    \label{tab:spectral_fits} 
\end{figure}

When fitting the spectral features for each pixel, the baseline noise was fit with a Gaussian function, while the 2.62~keV Cl-K$\upalpha$~\cite{Ca_Ka} and 3.31~keV K-K$\upalpha$~\cite{K_Ka_1,K_Ka_2,K_Ka_3} peaks were fit with double Lorentzian functions. Figure~\ref{tab:spectral_fits} shows these fits for a sample pixel across all three datasets. Lower statistics for the flight data impact the ability to get reliable fits for the Cl-K$\upalpha$ and K-K$\upalpha$ lines for some pixels. The Cl-K$\upbeta$, K-K$\upbeta$, and Mn-K$\upalpha$ peaks (at 2.82~keV, 3.59~keV, and 5.89~keV) did not have enough statistics from the calibration source in flight to get reliable fits. 

%%%%%%%%%%%%%%%%%%%%
\subsubsection{Impact of Instrument Configuration and Readout Mode on Spectral Resolution}
\label{sec:resolution_ground}

\begin{figure} [htb]
    \centering
    \includegraphics[width=\textwidth]{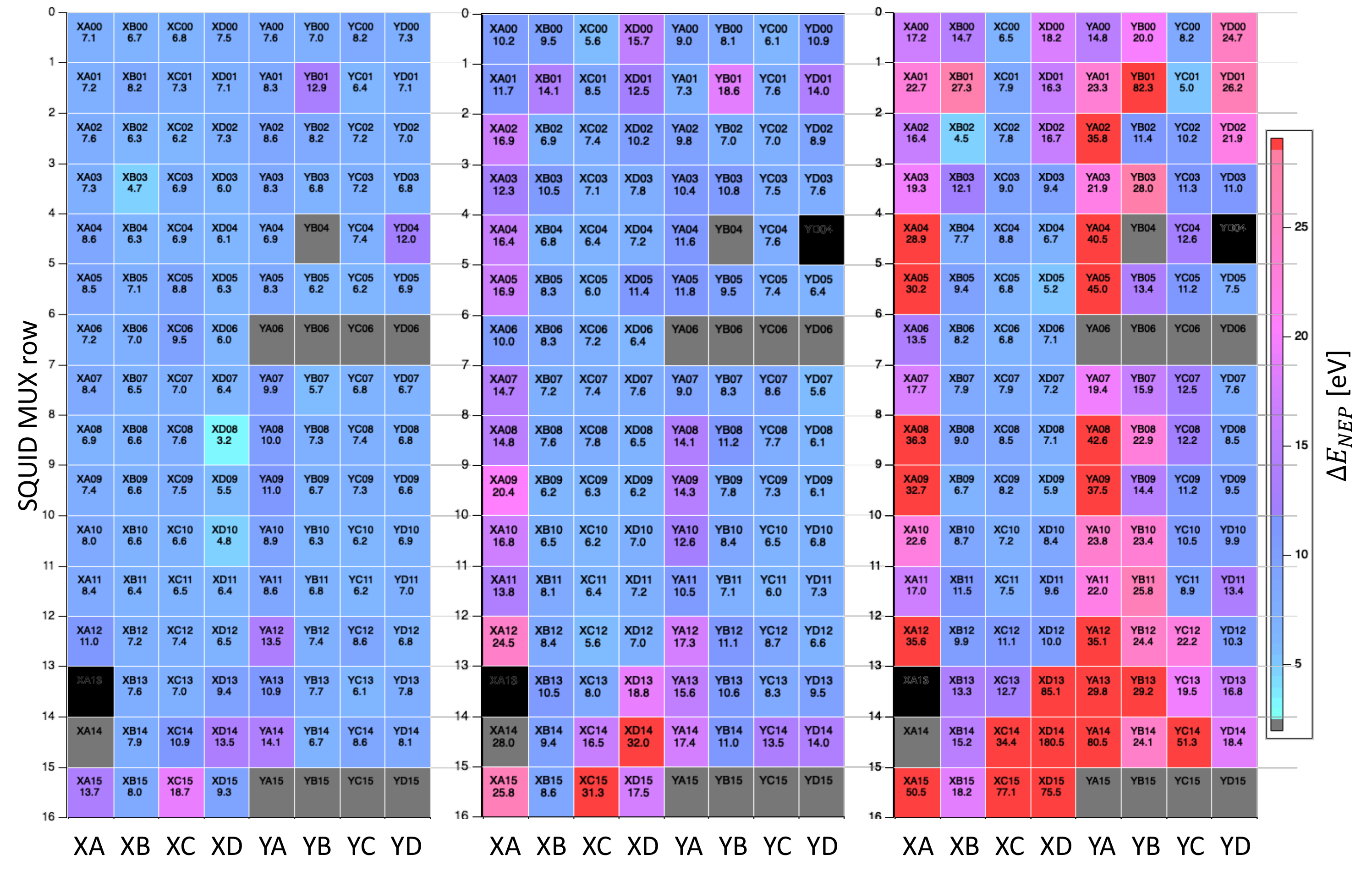}
    \caption{Heat map of $\Delta E_{NEP}$ from the laboratory configuration (left), the integration configuration taken with only one science chain powered and read out at a time (center), and the integration configuration taken with both science chains running (right). The dark gray pixels are known non-operational pixels, the black pixels had insufficient statistics for calculating the $\Delta E_{NEP}$, and the red pixels have $\Delta E_{NEP}$ over 30~eV and are considered pathological.} 
    \label{fig:NEP_readout_config}
\end{figure}

The spectral resolution of the detectors depended both on the configuration of the instrument as a whole and the readout mode (one-sided vs. two-sided). 
Figure~\ref{fig:NEP_readout_config} shows $\Delta E_{NEP}$ heat maps for every pixel in the one-sided readout laboratory data (left), one-sided readout integration data (center), and two-sided readout integration data (right). The pixels are arranged in the SQUID TDM readout order as introduced in Figure~\ref{fig:array_naming}, with each column sharing a single SQ2 and SA.

Increased system noise appeared once the instrument was built up for flight. This noise slightly degraded the spectral resolution, as seen in the comparison between the left and center panels of Figure~\ref{fig:NEP_readout_config}.
The exact mechanism of this degradation is not fully understood, but it is believed to be related to the spatial layout of the instrument electronics, which changed the EMI shielding efficacy. Integrating the instrument brought the electronics boxes and their harnesses physically closer together, which may have introduced pickup on wires or circuit traces. The ADR controller housekeeping signals introduced switching noise at 106~Hz, and the science chain housekeeping introduced noise at 77~Hz.

\begin{figure}[htb]
    \includegraphics[width=0.45\textwidth]{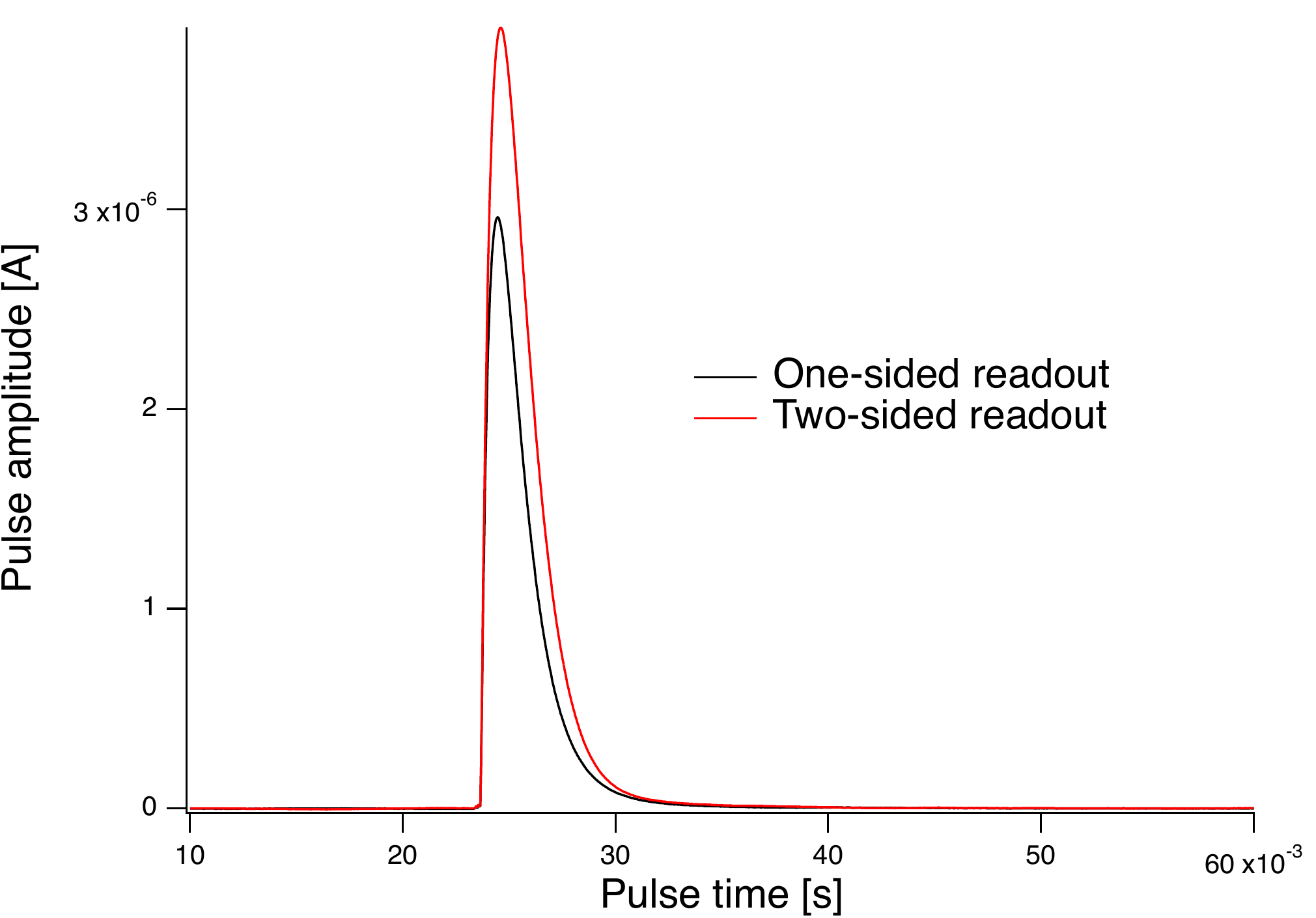}
    \includegraphics[width=0.45\textwidth]{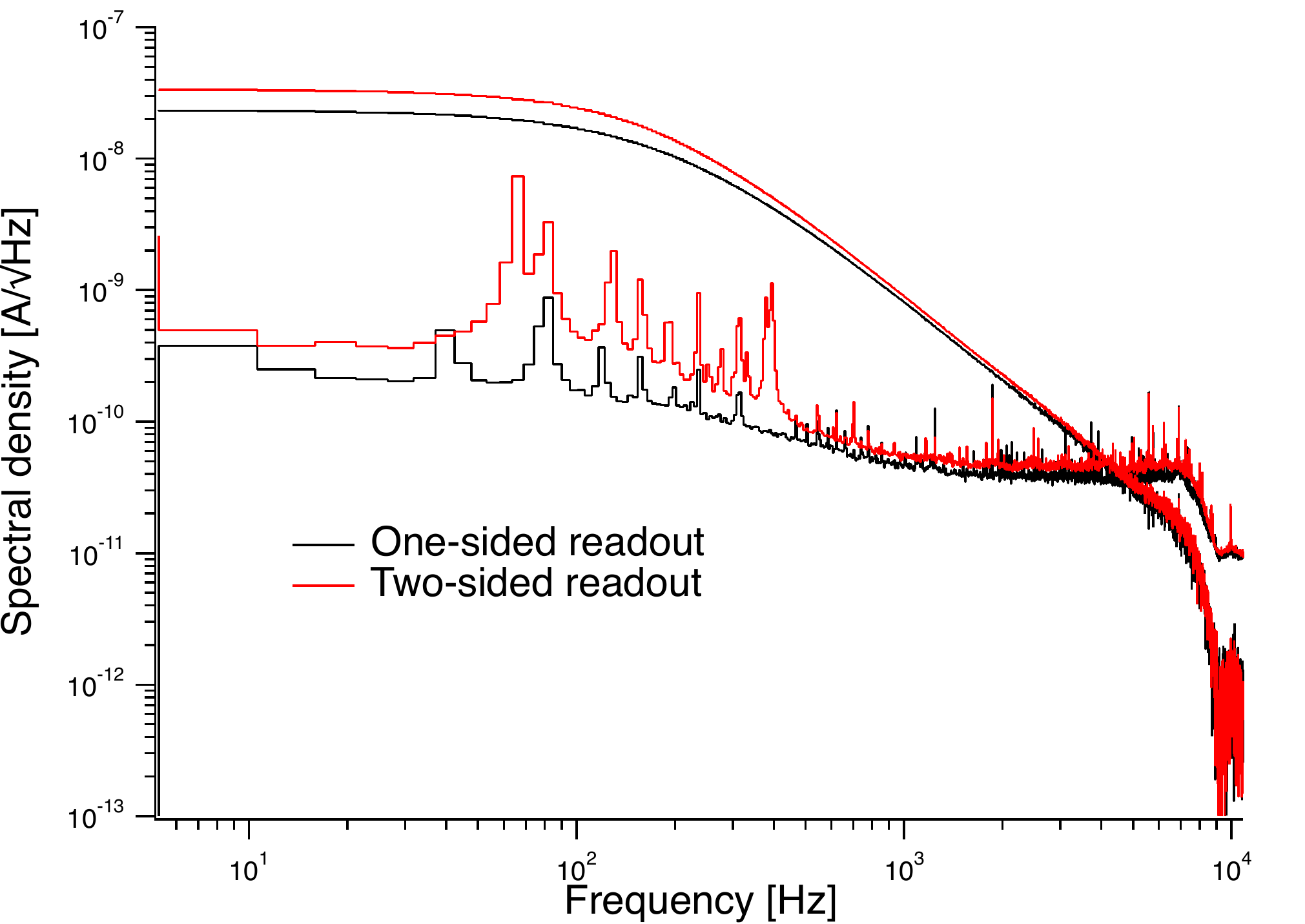}
    \caption{Average K-K$\upalpha$ pulse (left) and amplitude spectral density of the average noise and pulse (right) from one-sided and two-sided integration data for a single pixel (YD03). The data sets used for these plots are the same as those used to make the center and right panels of Figure~\ref{fig:NEP_readout_config}. This shows that the degradation in resolution came mostly from increased noise pickup below 500~Hz. The difference in pulse heights is from biasing the TES with different voltages in each data set; this puts the detector in a different part of the TES transition curve and changes the gain.}
    \label{fig:example_different_noise}
\end{figure}

Changing from the one-sided to the two-sided readout had a larger impact on the noise performance than the physical integration of the payload, as seen in the comparison between the center and right panels of Figure~\ref{fig:NEP_readout_config}. 
As described in \ref{sec:Science Chain Electronics}, each Master Control Board receives its own master clock, and there is a separate board for each independent science chain. The clocks for each chain were not synchronized with each other, so running both science chains at the same time introduced beat frequencies that can vary in time. In addition, running two science chains introduces an increased probability of radio frequency (RF) cross-talk between channels. Synchronizing the clocks was demonstrated in post-flight testing to remove the variable-frequency beating, and this was implemented for the second flight (\S\ref{sec:Reflight}).
Figure~\ref{fig:example_different_noise} shows this effect for one pixel, with a clear increase in noise when going from one-sided to two-sided integration data. The right panel of the figure shows an increase in noise pickup below 500~Hz in the two-sided data that resulted in a degradation of $\Delta E_{NEP}$ from 7.6~eV to 11.0~eV for this particular pixel. The change in the pulse heights of the average pulse are due to a slightly lower TES bias used in the two-sided data set, which increased the gain of the TES.

A third trend, visible in Figure~\ref{fig:NEP_readout_config}, is the degradation of performance when moving down a column to the higher-numbered rows. This is partially attributed to tuning limitations from each column sharing common tuning parameters that were optimized for the lower-numbered rows. The row-dependent degradation was exacerbated by the increased noise seen when the instrument was integrated for flight. This effect is still present after making the post-flight modifications described in \S\ref{sec:Reflight}, so it is independent of the installed SQUID version.

\begin{figure}[htbp]
\centering
    \includegraphics[width=0.7\textwidth]{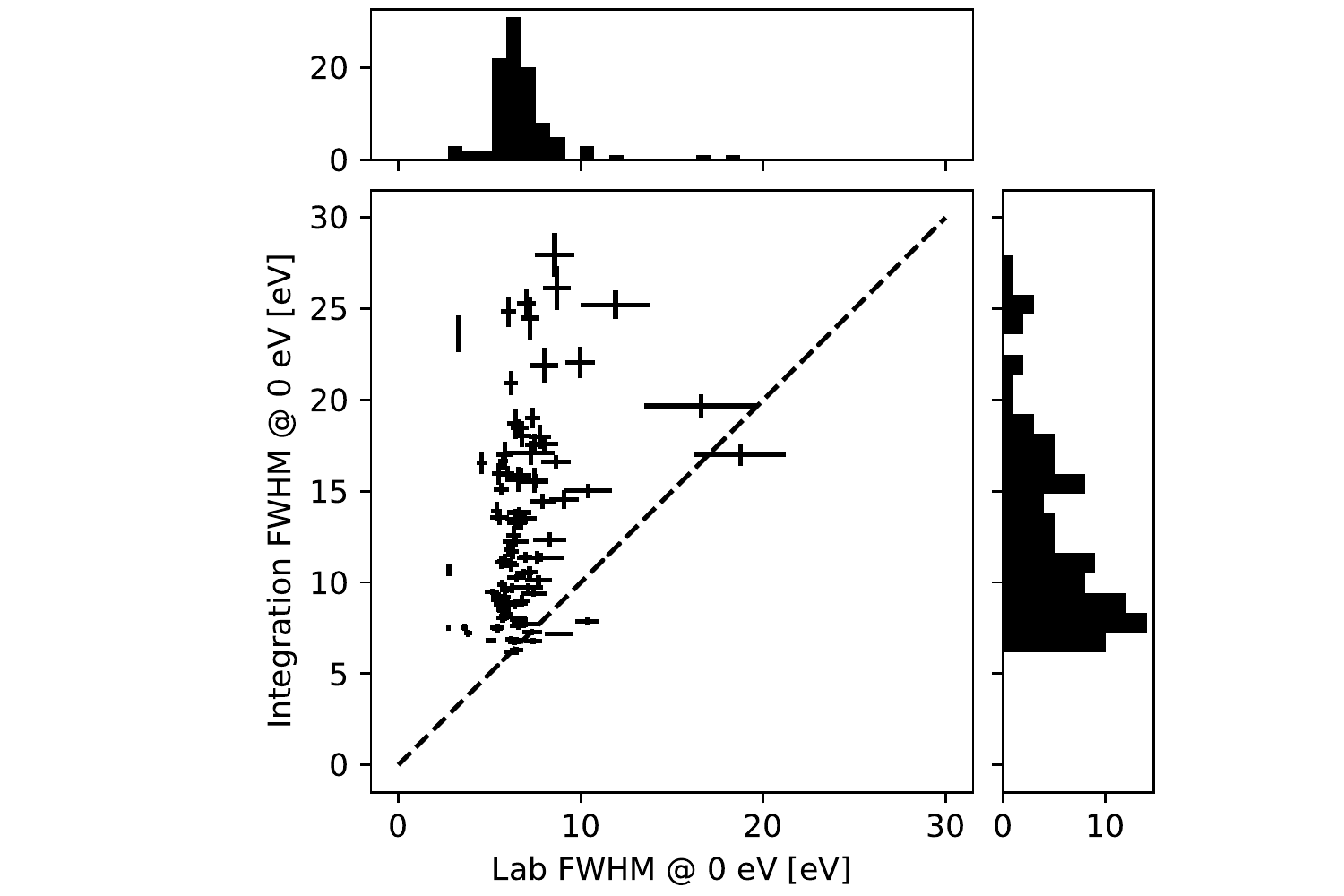}
    \caption{Comparison of the baseline resolution for each pixel in the one-sided laboratory configuration and two-sided integrated configuration. The deviation above the dashed line (slope $=$ 1) indicates that the resolution of 103 out of 110 pixels degraded after integration.}
    \label{fig:scatter_lab_int}
\end{figure}

Figure~\ref{fig:scatter_lab_int} shows the baseline resolution of each pixel between one-sided readout laboratory data and two-sided readout integration data. These are the same data sets as the left and right plots of Figure~\ref{fig:NEP_readout_config}, respectively, but now showing FWHM instead of $\Delta E_{NEP}$. The system had consistently good resolution in the single-sided laboratory mode, yet there is a significant degradation of spectral performance for a large fraction of the pixels in the array when the system was fully integrated.
The amount of resolution degradation correlates to the level of increased broadband noise on a given pixel, which varied across the array.
In the laboratory configuration, 103~pixels had baseline resolution better than 10~eV; in the integrated configuration this number fell to 41~pixels.

%%%%%%%%%%%%%%%%%%%%
\subsubsection{Impact of Flight Environment on Spectral Resolution}
\label{sec:resolution_flight}

\begin{figure}[htbp]
    \centering
    \includegraphics[width=0.7\textwidth]{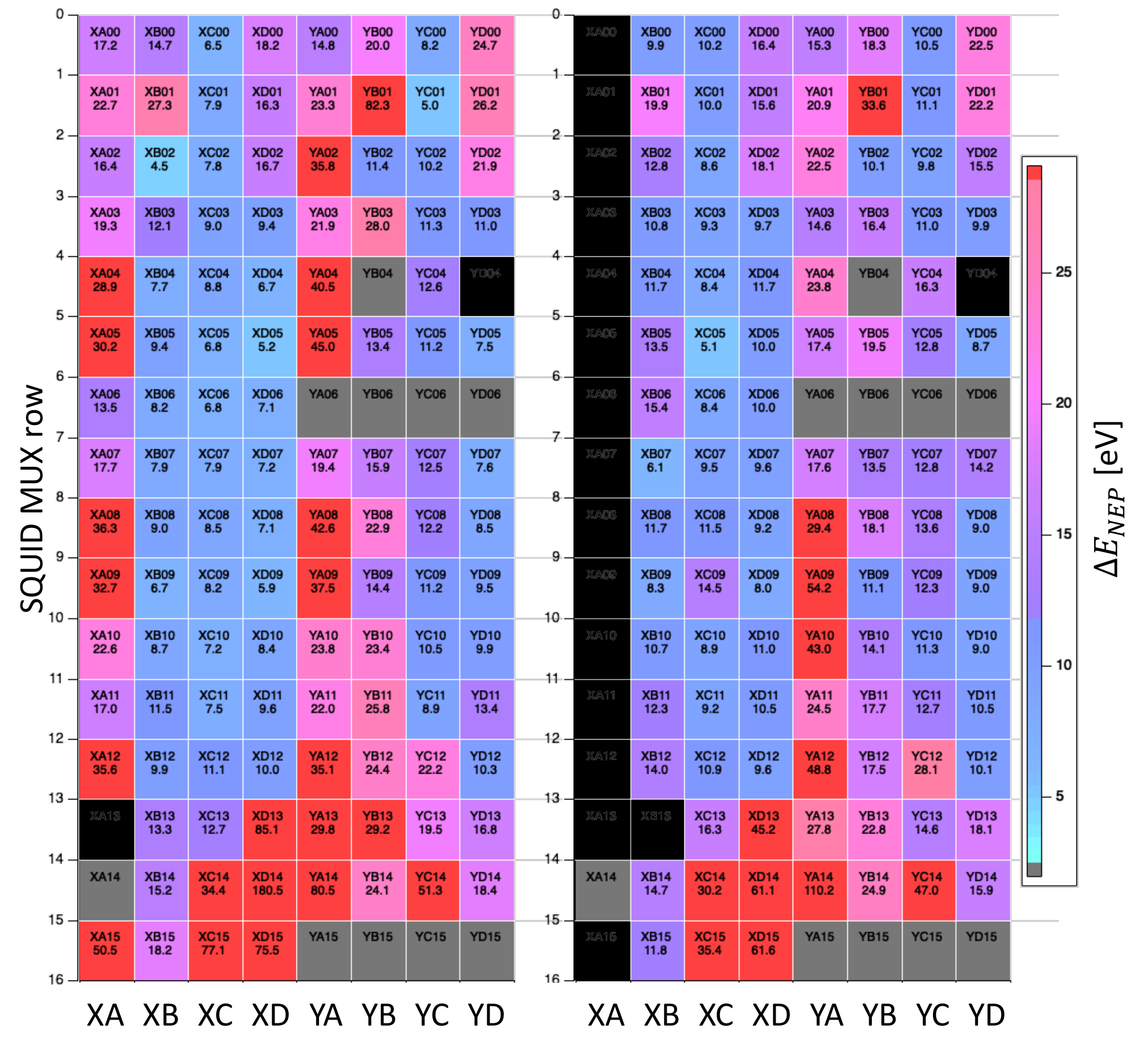}
    \caption{Heat map of $\Delta E_{NEP}$ from the two-sided integration data (left) and flight data (right). The difference in performance between the two data sets is consistent with the difference observed between operating runs of the instrument. The dark gray pixels are known non-operational pixels, the black pixels had insufficient statistics for calculating the $\Delta E_{NEP}$, and the red pixels have $\Delta E_{NEP}$ over 30~eV and are considered pathological. The left plot in this Figure is the same as the right plot from Figure~\ref{fig:NEP_readout_config}.}
    \label{fig:NEP_readout_flight}
\end{figure}

\begin{figure}[htbp]
\centering
    \includegraphics[width=0.7\textwidth]{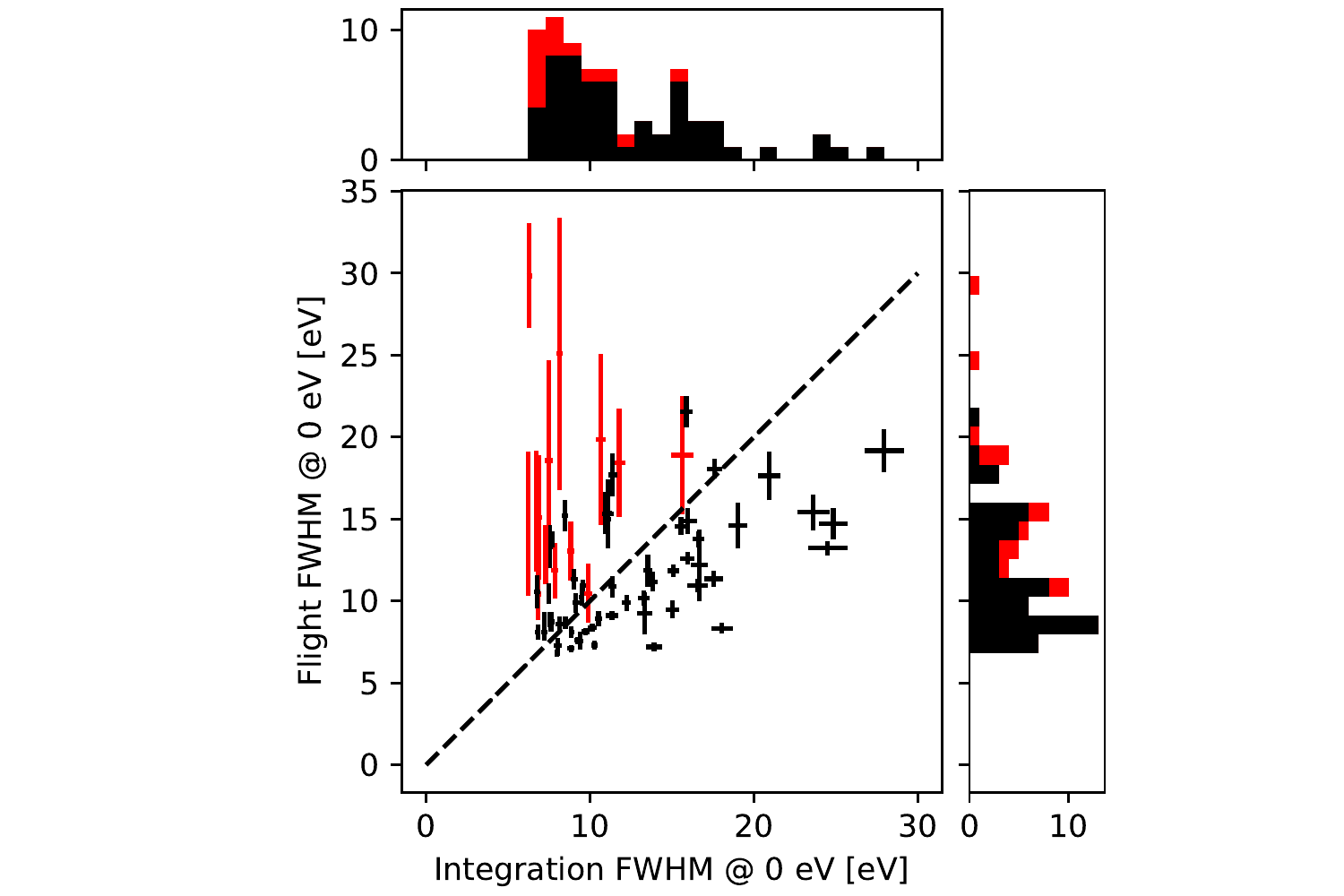}
    \caption{Comparison of the baseline resolution from every pixel in the integration and flight data. The red points have lower statistics ($<150$ counts) in the baseline noise peak, hence the large error bars.}
    \label{fig:scatter_int_flight}
\end{figure}

The energy resolution of the instrument during flight was consistent with the performance from integration data taken prior to flight, as shown in Figures~\ref{fig:NEP_readout_flight} and Figure~\ref{fig:scatter_int_flight}. As highlighted in the beginning of Section~\ref{sec:flight_resolution_methods}, the integration and flight data were taken during different cryogenic cycles of the instrument. The difference in resolutions between the two data sets is within the expected variation from those cycles. Figure~\ref{fig:NEP_readout_flight} shows the $\Delta E_{NEP}$ from two-sided integration and flight data. The $\Delta E_{NEP}$ was measured for only 101 of 118 total live pixels in flight due to the limited livetime of some pixels (see Figure~\ref{fig:Pulse_times}). Figure~\ref{fig:scatter_int_flight} shows the baseline resolution FWHM for the same two data sets shown in Figure~\ref{fig:NEP_readout_flight}.
Of the 101 pixels with measurable $\Delta E_{NEP}$, 77~pixels provided the required statistics to determine their baseline resolution. Focusing on pixels with decent statistics ($>$150~traces, black points in Figure~\ref{fig:scatter_int_flight}), on average the data is comparable between integration and flight environments. Most pixels that showed significant noise pickup with larger FWHM on the ground had better performance in flight, possibly due to the quieter RF environment in space. The pixels with good resolution ($<15$~eV) in integration data had mixed responses in flight, but did not see large degradation from pre-flight to flight data. Selecting those pixels with both decent statistics ($>$150~flight traces) and good resolution ($<15$~eV in integration), the flight baseline resolution of 18 pixels improved compared to integration data, while 20 pixels degraded.

\begin{figure}[htbp]
    \centering
    \includegraphics[width=0.45\textwidth]{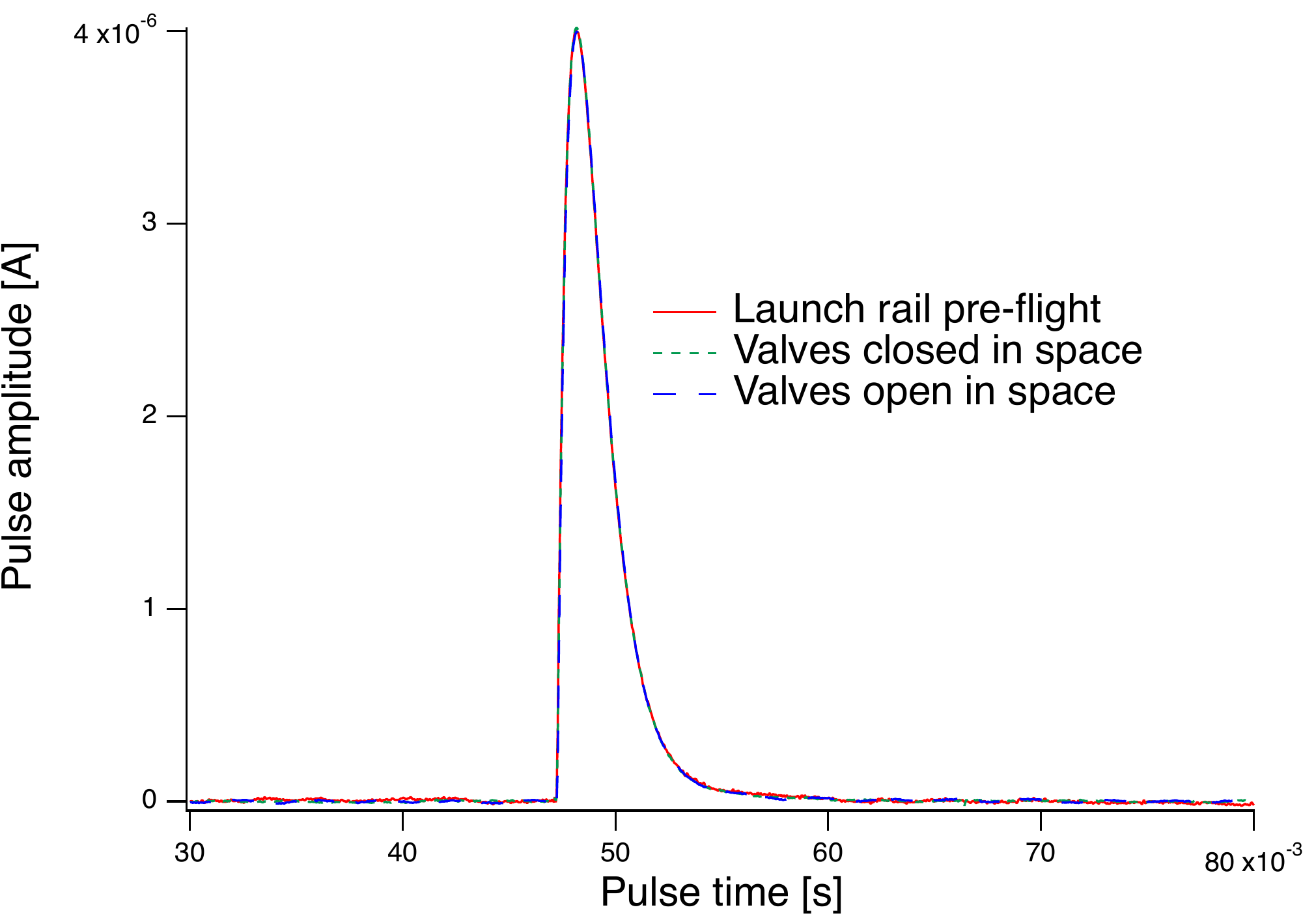}
    \includegraphics[width=0.45\textwidth]{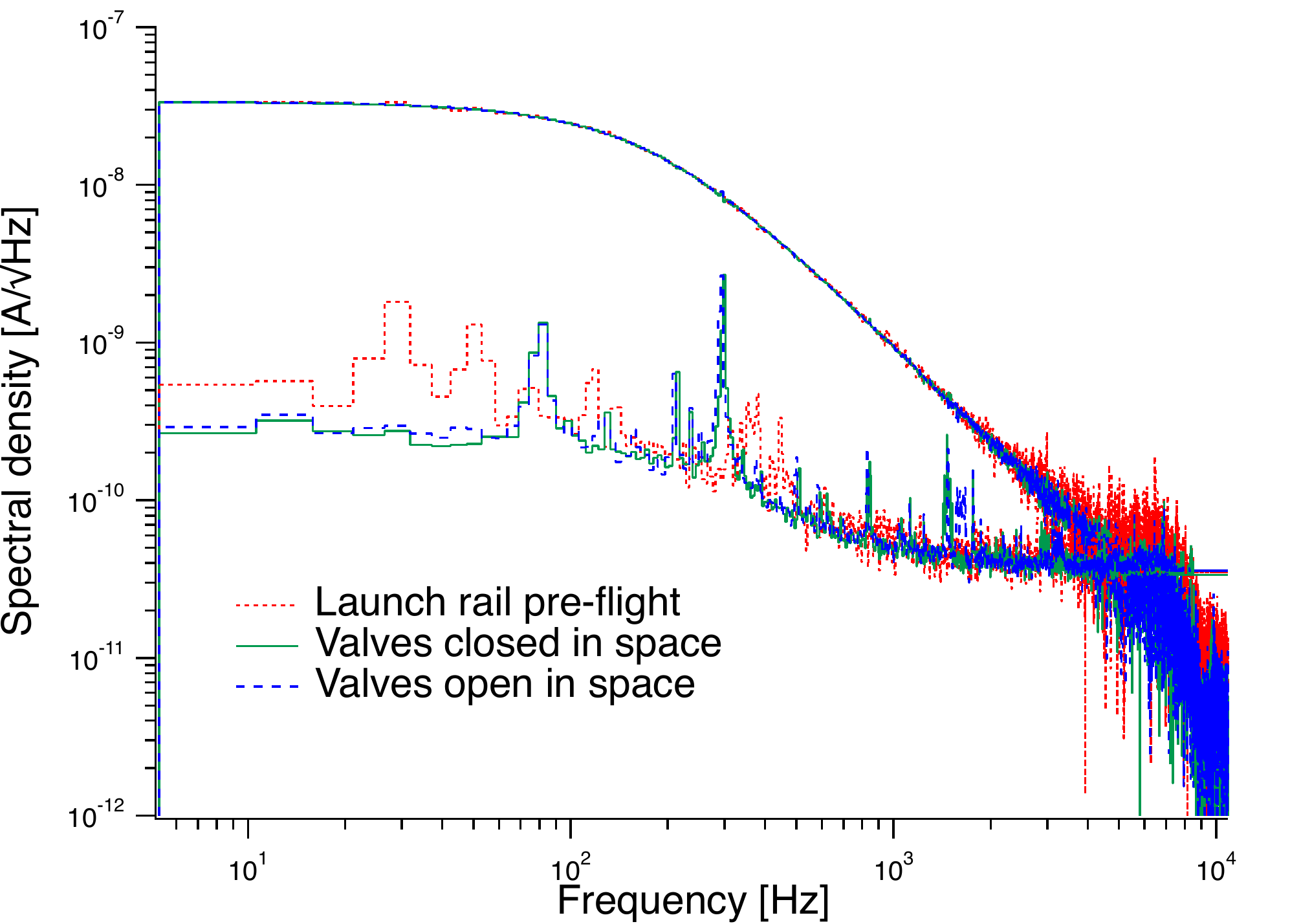}
    \caption{Average K-K$\upalpha$ pulse (left) and amplitude spectral density (right) for an example pixel (YD03) from pre-flight data on the launch rail (red), the last 22~s of the science observation (blue), and the 22~s after the aperture valve was closed at the end of the science observation, when the detector was closed off to space (green). The consistency between the data with the valve open and closed demonstrates that no change in noise or pulse shape was observed when the detectors were exposed to space. }
    \label{fig:shutter_door_open}
\end{figure}

No change was observed in pulse shape or noise when the detectors were exposed to space, as shown in Figure~\ref{fig:shutter_door_open}. No discernible pickup was observed from external sources (e.g., telemetry antennae, transponders, or ambient fields in space) with the cryostat open to space. The comparison was made using data from the last 22~s of the science observation (aperture valve open) and the first 22~s after the observation (aperture valve closed). The same comparison with the opening of the aperture valve at the beginning of the science observation could not be performed because the detectors had not yet reached a stable temperature.

%%%%%%%%%%%%%%%%%%%%%%%%%%%%%%%%%
%%%%%%%%%%%%%%%%%%%%%%%%%%%%%%%%%
%\input{sections/flight-optics.tex}
%%%%%%%%%%%%%%%%%%%%%%%%%%%%%%%%%

%%%%%%%%%%%%%%%%%%%%%%%%%%%%%%%%%
% !TEX root = ../instrument_paper_revB.tex
\section{Status and second flight modifications}
\label{sec:Reflight}

After the first flight, the instrument was recovered in good condition. Upon return to the laboratory, the payload was run in its flight configuration, and no new failures were detected. The post-flight run examined issues that appeared in flight, notably testing sensitivity to magnetic fields and noise susceptibility. Six modifications, motivated by observations from the first flight, were implemented for the second launch, which successfully flew in August 2022 and will be the subject of future publications. These modifications included:  
	\paragraph{The SQUIDs} were upgraded from NIST MUX06a~\cite{Stiehl_2011} to NIST MUX18b~\cite{Reintsema:2019}. The newer SQUIDS have several improvements over the 2006 design, including using flux actuated switches to select rows during TDM operation and a two-stage design, as opposed to the MUX06a's three-stage design. NIST measurements demonstrated that this generation of SQUIDs has a greatly reduced response to magnetic fields~\cite{Reintsema:2019}, and laboratory testing with Micro-X has confirmed that these SQUIDs are less sensitive to external magnetic fields. The Helmoltz coil testing was repeated to mimic spinning in the Earth's field. Unlike the first flight SQUIDs, this new generation remained reliably locked, and the MUX18b SQUIDs exhibited no hysteretic response or dependence on dB/dt. The SQUID series arrays were also upgraded from SAv4 to SA13ax to better match the output of the MUX18b design. 
	
	\paragraph{The science chain clocks,} supplied by the MV encoders in telemetry, were synchronized to eliminate the beat frequencies that degraded the first flight performance. Post-flight laboratory testing with encoder simulators demonstrated that synchronizing the clocks eliminates this beating. The synchronized encoders showed no impact on detector performance. With its second flight, Micro-X became the first program to fly synchronized MV encoders. 
	
	\paragraph{The superconducting Nb shield} uses a spacer at the base of the can that was converted from stainless steel to superconducting lead. The lead ring is superconducting at LHe temperatures, providing a more robust magnetic shield around the FEA in the case where there are undetected gaps in the lead zipper. 
	
	\paragraph{The kinematic mounts} that support the TES and IF chips on the FEA were replaced. The original tungsten carbide balls, measured to be magnetic, were replaced with non-magnetic sapphire balls. Although there is no direct evidence that the tungsten carbide balls degraded performance, they were changed on the best-practice basis of not having any magnetic materials near the TES array.
	
	\paragraph{The wire rope isolators} were replaced by a higher-temperature-rated version of the same design. This change is to prevent the glue failure that was observed in flight and reproduced in laboratory testing. The new isolators retain their structural integrity up to the maximum tested temperature of 170$^\circ$C, which is significantly hotter than the skin temperature reached at any point during flight. 
	
	\paragraph{Science chain data recording} to on-board flash memory was modified to be mechanically initiated by lanyards when the rocket leaves the ground, rather than using timers. This is a robust improvement that mitigates the risk of dropping the command to write data in flight.

\bigskip

%%%%%%%%%%%%%%%%%%%%%%%%%%%%%%%%%%%%%%%
%\subsubsection{New SQUIDs}
%\label{new_squids}
%
%\com{Include or leave for future papers? How much to describe here?}
%
%A number of the SQUID readouts in the first flight shifted to a non-operational state in flight, then returned to an operational state. This is not expected, and as science data cannot be read out when the SQUIDs are non-operational, this is a risk for mission success if it were to occur again. 
%
%Three mitigation strategies have been developed to resolve this. The SQUIDs are magnetically sensitive, and an unexpectedly large change in the local magnetic field could have changed their response and pushed them into the non-operational regime. If this effect is due to a varying magnetic field, we expect that stable rocket pointing would mitigate this problem. Additionally, the SQUIDs have been upgraded to a new design that has been shown to be less magnetically susceptible. \com{Expand depending on what we see this run.}
%%%%%%%%%%%%%%%%%%%%%%%%%%%%%%%%%
% !TEX root = ../instrument_paper_revB.tex
\section{Conclusions}
\label{sec:Conclusions}

Micro-X is a sounding rocket mission designed to take high-spectral-resolution, imaged spectra of extended X-ray sources. Its first flight on July 22, 2018 saw the first operation of TES microcalorimeters and multiplexed SQUID readouts in space, marking an important milestone for the field as it continues driving towards high-spectral-resolution X-ray instruments in space. While an ACS pointing failure led the payload to have no time on target, the flight provided an invaluable test of the instrument design, using X-rays from the on-board calibration source to analyze detector performance. 

The fundamental design of the instrument proved to be sound, including operating TESs and SQUIDs in space, and using a SQUID TDM system for readout. As predicted from pre-flight testing, thermal and mechanical isolation allowed the cryostat and ADR to achieve and maintain a stable 75~mK environment in flight. The vibrational load during a sounding rocket flight is significantly more extreme than that experienced by satellite missions, and the survival of the detectors and readout through launch, atmospheric re-entry, and impact demonstrates their mechanical compatibility with space missions. The system noise increased as the instrument was built up for flight, but was shown to be consistent between the ground and flight, which is a key measurement for using these detectors in space. The performance achieved on the ground is therefore predictive of flight performance and provides confidence in the predictive strength of future measurements in preparation for future flights. This includes identifying the sources of performance degradation between the laboratory and integration configurations when operating on the ground. 

The modifications made for future flights were focused on improving the detector resolution and mitigating risks identified in the first flight. The only fundamental issue observed in flight that was not identified in pre-flight testing was the magnetic sensitivity of the system in the presence of changes in the Earth's magnetic field (in the rocket frame) as the payload spun in flight. The ACS failure kept the payload spinning through the science observation, which would not have occurred in a normal flight trajectory. The magnetic sensitivity was attributed to the a combination of the superconducting magnetic shield's position- and direction-dependent shield function, the unexpected tumbling of the payload, and SQUID's effective pickup loop area. This magnetic field driven response led the SQUIDs to unlock and limited the detector live time during the observation period. Post-flight testing and modifications identified and resolved this issue, so it is no longer considered a risk and indeed was not seen in the second flight of the Micro-X payload in 2022, which furthered our understanding of how best to fly these detectors and will be the focus of future publications. 

%Future flights -- and the instrument upgrades between them -- will continue to further our understanding both of the science targets and how best to fly these detectors. 

%\com{CR: Last paragraph (starting line 904): You probably expected me to say this, but it seems like too much blame is being assigned to the SQUIDs. The reason this problem was not identified in pre-flight testing was because we didn’t look for it. It was a known vulnerability (as described in Stiehl’s paper) and perhaps I should have been more vigilant calling attention to this. But no one expected the ACS to fail either. Why would we have looked for an oscillating field effect on the bench pre-flight? Who’s to say that if the shielding factor was closer to the simulated value (10-40X improvement) we wouldn’t have stayed locked (maybe that could be backed out of Joel’s model?). }
%%%%%%%%%%%%%%%%%%%%%%%%%%%%%%%%%%
\acknowledgments
We gratefully acknowledge the technical support of Ernie Buchanan, John Bussan, Travis Coffroad, Sam Gabelt, Rob Hamersma, Kurt Jaehnig, Frank Lantz, Ken Simms, Tomomi Watanabe, George Winkert, and Mike Witthoeft. We equally gratefully acknowledge the support of the WFF team, whose significant efforts quite literally got us off the ground: Max King, Adam Blake, Bob Camiano, Eric Taylor, John Peterson, Cliff Murphy, Freddy Ayala, Tom Shockley, Rob Marshall, Tom Russell, Jim Diehl, Nate Wroblewski, Charlie Kupelian, and countless others who stepped up to provide their time and expertise. Micro-X has been supported by NASA Grants NNX10AE25G, NNX13AD02G, NNX14AM57H, NNX16AL94G, NNX17AC96G, 80NSSC18K1445, 80NSSC20K0430, 80NSSC21K1856, and the MIT Marble Foundation. Part of this work was performed under the auspices of the U.S. Department of Energy by Lawrence Livermore National Laboratory under Contract DE-AC52-07NA27344. 
%%%%%%%%%%%%%%%%%%%%%%%%%%%

\bibliographystyle{ieeetr}
\bibliography{instrument_bib}
\end{document}